\documentclass[10pt,journal]{IEEEtran}

\usepackage{graphicx}
\usepackage{amsmath, amsthm, amssymb}
\usepackage{lipsum} % to generate some filler text
\usepackage{cite}
\usepackage{subfigure}
\usepackage{rotating}
\usepackage{fancyhdr}
\usepackage{gensymb}

\usepackage{caption} %[belowskip=-10pt,aboveskip=5pt]
\usepackage{float}
\usepackage{multicol}
\usepackage{algorithm}
\usepackage{algorithmic}
\usepackage{breqn}
\usepackage{framed}
\usepackage{color}

\newtheorem{lemma}{Lemma}
\newtheorem{theorem}{Theorem}

\newtheorem{remark}{Remark}
\newtheorem{definition}{Definition}

\newtheorem{assumption}{Assumption}
\newtheorem{proposition}{Proposition}

\newcommand\at[2]{\left.#1\right|_{#2}} 

\DeclareMathOperator{\argmax}{arg\,max}

\normalsize

\ifCLASSINFOpdf
\else
\fi

\begin{document}
\title{Optimal Hierarchical Radio Resource Management for HetNets with Flexible Backhaul}
%
%

\iffalse ***
\author{\IEEEauthorblockN{Naeimeh Omidvar\IEEEauthorrefmark{1}\IEEEauthorrefmark{2}, An Liu\IEEEauthorrefmark{1}, Vincent Lau\IEEEauthorrefmark{1}, Fan Zhang\IEEEauthorrefmark{1}, Danny Tsang\IEEEauthorrefmark{1} and Mohammad Reza Pakravan\IEEEauthorrefmark{2}} \\
\IEEEauthorblockA{\IEEEauthorrefmark{1}{Department of Electronic and Computer Engineering, Hong Kong University of Science and Technology, Hong Kong,}}\\
\IEEEauthorblockA{\IEEEauthorrefmark{2}{Department of Electrical Engineering, Sharif University of Technology, Iran.\footnote{An early version of this work can be found in \cite{omidvar2015Globecom}.}}}
*** \fi

\author{
		Naeimeh~Omidvar,~\IEEEmembership{Student Member,~IEEE,}
		An~Liu,~\IEEEmembership{Senior Member,~IEEE,}
        Vincent~Lau,~\IEEEmembership{Fellow,~IEEE,}
        Fan~Zhang,~\IEEEmembership{Member,~IEEE,}
        Danny~Tsang,~\IEEEmembership{Fellow,~IEEE,}
        and~Mohammad Reza~Pakravan,~\IEEEmembership{Member,~IEEE.}% <-this % stops a space
%        \iffalse      
\thanks{An early version of this work can be found in \cite{omidvar2015Globecom}.}
\thanks{Naeimeh Omidvar was with the Department of ECE, The Hong Kong University of Science and Technology. She is now with the Department of EE, Sharif University of Technology (e-mail: nomidvar@connect.ust.hk).}
%\thanks{M. Shell is with the Department of Electrical and Computer Engineering, Georgia Institute of Technology, Atlanta, GA, 30332 USA e-mail: (see http://www.michaelshell.org/contact.html).}% <-this % stops a space
\thanks{An Liu was with the Department of ECE, The Hong Kong University of Science and Technology. He is now with the College of Information Science and Electronic Engineering, Zhejiang University (e-mail: wendaolstr@gmail.com).}
\thanks{Vincent Lau and Danny Tsang are with the Department of ECE, The Hong Kong University of Science and Technology (e-mails: \{eeknlau, eetsang\}@ece.ust.hk).}
\thanks{Fan Zhang was with the Department of ECE, The Hong Kong University of Science and Technology (e-mail: fzhangee@connect.ust.hk).}% <-this % stops a space
\thanks{Mohammad Reza Pakravan is with the Department of EE, Sharif University of Technology (e-mail: pakravan@sharif.edu).}
%\thanks{TCOM version based on Michael Shell's bare{\textunderscore}jrnl.tex version 1.3.}
%\fi
}

%\markboth{IEEE Transactions on Wireless Communications}%
%{Submitted paper}

\maketitle

% *****************************************************************

\begin{abstract}
%\boldmath

Providing backhaul connectivity for macro and pico base stations (BSs) constitutes a significant share of infrastructure costs in future heterogeneous networks (HetNets). To address this issue, the emerging idea of \textit{flexible backhaul} is proposed. Under this architecture, not all the pico BSs are connected to the backhaul, resulting in a significant reduction in the infrastructure costs. In this regard, pico BSs without backhaul connectivity need to communicate with their nearby BSs in order to have indirect accessibility to the backhaul. This makes the radio resource management (RRM) in such networks more complex and challenging. In this paper, we address the problem of cross-layer RRM in HetNets with flexible backhaul. We formulate this problem as a two-timescale non-convex stochastic optimization which jointly optimizes flow control, routing, interference mitigation and link scheduling in order to maximize a generic network utility. By exploiting a \textit{hidden convexity} of this non-convex problem, we propose an iterative algorithm which converges
to the global optimal solution. The proposed algorithm benefits from low complexity and low signalling, which makes it scalable. Moreover, due to the proposed two-timescale design, it is robust to the backhaul signalling latency as well. Simulation results demonstrate the significant performance gain of the proposed solution over various baselines.

\end{abstract}


\IEEEpeerreviewmaketitle

\section{Introduction}

\IEEEPARstart{}{H}{eterogeneous} networks (HetNets) are a promising network architecture in future mobile access networks \cite{damnjanovic2011survey}.
In current HetNet designs, many pico cells are
deployed across the network, each with fixed high-capacity backhaul
connectivity. %*****{\color{red}{that guarantees the fulfilment of the capacity demand of each hotspot.}} % \cite{QualcommLTEHetNet}. 
However, %*****these designs suffers from high CAPEX and OPEX \cite{paolini2011crucial}, {\color{red}{since in order to guarantee the fulfilment of the capacity demands, the statically allocated resource and deployment should be highly excessive, which leads to low utilization efficiency when traffic is low.}} It has been shown that 
due to the large number of pico base stations (BSs) in future HetNet deployments, 
providing the fixed backhaul for all BSs will lead to high capital and operating expenditures %CAPEX and OPEX 
\cite{paolini2011crucial, tombaz2011impact,farias2013green}.  
%*****huge increase in the the network power consumption \cite{tombaz2011impact,farias2013green}. 
%
%To address this problem, the authors in \cite{shi2014flexible} propose to statically deploy one relaying BS between each pico-BS and the macro BS, called Type-A Relay. Although the proposed structure can improve the network capacity, it will lead to significant increase in CAPEX and OPEX as well. Since for each pico BS one extra type-A relay is statically deployed as well, which do not serve any mobile user (MU) and just relays the backhaul traffic of the associated pico BS.
%
%%To address this problem, the authors in \cite{shi2014flexible} propose static deployment of one relay BS (named as Type-A relay) between the macro BS and each pico BS, which relays the backhaul traffic of a pico cell. They study the capacity analysis of the proposed structure. However, the proposed architecture still cannot address the low utilization problem due to the static deployment of Type-A relay nodes. Moreover, it requires deploying a large number of relays (as many as the number of pico eNBs) that do not serve any mobile users (MUs) and just do the relaying job for the pico eNBs. Accordingly, although their proposed structure will increase the network capacity, it will significantly increase the CAPEX and OPEX, as well.
%
In addition, future networks should provide better support of
emergency communications as well as fast recovery, in case of the unpredicted crash of access points.
%To achieve this, the emergency communication vehicles are currently being utilized to replace the crashed point in order to rebuild the network. 
However, in a fixed backhaul deployment, the
crash of one point requires a physical replacement (e.g., using an emergency communication
vehicle), which causes slow network recovery and low
reliability.

{\color{black}{
To overcome the aforementioned problems of fixed backhaul deployment,
the idea of \textit{flexible backhaul} has been proposed \cite{tornatore2017Fiber-WirelessConvergenceinNext-GenerationCommunicationNetworks}. The concept behind
flexible backhaul is to flexibly utilize any idle network resources
to maximize end-to-end experience with minimal cost. Under flexible backhaul deployment, not all of the pico BSs are provided with backhaul connectivity, and those BSs without backhaul connectivity communicate with the other BSs and utilise their idle system resources to reach the backhaul. Such a flexible backhaul architecture results in a significant reduction in infrastructure
costs, and lower capital and operational expenditures 
%a lower CAPEX and OPEX 
can be gained through aggregation and reuse of idle system resources. %, under flexible backhaul architectures, not all of the pico BSs are connected to the backhaul, resulting in a significant reduction in the infrastructure costs.
Moreover, as the data flows are routed in a multi-hop manner between the backhaul and the mobile users (MUs), dynamically adjusting the flow routings and flexibly allocating the associated system resources in such networks will provide dynamic topology, %intelligent topology establishment
 %and flexible resource allocation that bring 
 reliability and a better end-to-end user experience. %For example, unlike conventional HetNets, in HetNets with flexible backhaul, the crash of any access point would not lead to an inability to reach the backhaul. Because under the flexible backhaul deployment, the BSs can communicate and route each others' data flows, the network could dynamically reroute the traffic to other access points. % and change the topology. 
For example, %unlike conventional HetNets, in HetNets with flexible backhaul, 
when an access point (AP) crashes, the users associated with the crashed AP may be redirected to other APs, and the APs which used this crashed AP as a relay to fetch the backhaul data may still find another route to fetch the backhaul data. Therefore, with flexible backhaul deployment, the network can dynamically reroute the traffic to other access points. Such capability provides resilience of the data path in the network.

%Because under the flexible backhaul deployment, the BSs can communicate and route each others' data flows and hence, the network could dynamically reroute the traffic to other access points. Such advantage provides resilience of the data path in the network.

%if an AP crashed, the APs which used this AP as a relay to fetch the backhaul data may still nd another route to fetch the backhaul data, using dynamic routing in the radio resource management scheme.

{\color{black}{

%as well as better user experience through flexible resource allocation 
Considering the above discussion, the key features of flexible backhaul technology can be identified as follows:
%In this way, multi-hop routing 

%{\color{red}{The key features of flexible backhaul technology can be identified as follows: (Not BOLD, REF)!!!}}

\begin{itemize}
\item Flexible utilization of system resources: It flexibly utilizes any idle system resources in the network in order
to increase the resource utilization efficiency.

\item Dynamic resource scheduling: It fully exploits the degree
of freedom of network resources (in terms of time, frequency, space, etc.), in order to maximize the transmission capability of
backhaul.

%Flexibly utilizing any system resources including wired (e.g., optical fiber, cable), or wireless (e.g., microwave, idle nodes) and hybrid resources (e.g., combined optical fiber and idle nodes), in order to increase the resource utilization efficiency.

\item Dynamic network topology: %{\color{red}{\cite{flexible_network_deployment_in_5G_2015} (Is this ref suitable?) :}} 
It intelligently adjusts network
topology and backhaul transmission strategies in order to match the
traffic variation and meet the transmission/reliability requirements.
\end{itemize}

\iffalse **** LONGER VERSION ****
{\color{red}{(Maybe deleting the following paragraph:)}}

Considering these features, the expected benefits of flexible backhaul
can be categorized into three different aspects:
\begin{enumerate}
\item Cost: Lower CAPEX and OPEX will be gained through
aggregation and reuse of idle system resources. Under
flexible backhaul architectures, not all of the pico BSs are connected
to the backhaul, resulting in a significant reduction in the infrastructure
costs.
\item End-to-end experience: Flexible backhaul provides a better
user experience through flexible resource allocation.
\item Reliability: Intelligent topology establishment would bring higher system reliability. For example, the crash of an access
point would not lead to an inability to reach the backhaul since, the topology could dynamically reroute the traffic to other access points.
%\item \textbf{Interoperability:} It also provides easier inter-radio access technology (inter-RAT) interoperability as data can be routed through different RATs under flexible backhaul.
\end{enumerate}

\fi

To achieve the above benefits of flexible backhaul, it is important to design an efficient  dynamic cross-layer radio resource management (RRM) control  that can adapt to the changing environment, %(e.g., when some BSs have crashed), 
while fully exploiting the resources of different layers in order to match the
traffic variation and meet the transmission requirements. Moreover, it should dynamically and jointly allocate the system resources in order to fully and flexibly utilise them in the network, as well as provide reliability and robustness.

}}

%{\color{blue}{

We would like to note that the idea of flexible backhaul %*****for future HetNets 
%*****{\color{red}{is a technical idea learnt from Huawei,}} that 
extends the idea of cellular relay networks \cite{relay2001integrated, sreng2003relayer, relay2011flexible, survey2004relay}. In a cellular relay network, 
%where 
relaying stations are employed to enhance the network coverage at the cell edges \cite{relay2011flexible} or to divert traffic from possibly congested areas of a cellular system to cells with a lower traffic load \cite{relay2001integrated}. Therefore, HetNets with flexible backhaul and cellular relay networks share some similarities as they are both multihop-augmented and infrastructure-based, and they reduce infrastructure deployment costs. %*****There are many works in the literature on RRM in cellular relay networks. 
%***** For example, \cite{sreng2003relayer} investigates the effects of relay selection strategies and maximum relay transmit power level on the coverage of a two-hop cellular network and performs relay selection and power allocation, separately. The authors of \cite{relay2011flexible} consider a cellular network where relays are deployed at cell borders to enhance the cell coverage. They schedule the frequency-time resource blocks to the mobile users (MUs) served by either macro BSs or relays, within each cell. 
A complete overview of relay-based deployments for wireless networks can be found in  \cite{survey2004relay}. 
%
%
%***** {\color{red}{However, the flexible backhaul considered in this paper has several new features such as dynamic multi-path routing \cite{hasan2017survey}, joint optimization of network layer controls (flow and routing control) and physical layer controls (interference control and link scheduling), and hierarchical (i.e., two-timescale) RRM design. These new features are the key to achieve the aforementioned benefits of flexible and they are not considered in existing works on relay networks.}}
%
\iffalse ***
 Although HetNets with flexible backhaul and cellular relay networks share some similar features such as
\begin{itemize}
\item infrastructure based deployment/communication,
\item multi-hop transmission from source to destination, and
\item enhanced network coverage with low deployment costs,
\end{itemize}

\fi
%*****However, there are several differences between {\color{red}{the considered work in this paper (i.e., HetNets with flexible backhaul)}} and the existing works on relay networks, as follows:
%
However, the flexible backhaul considered in this paper has several new features that are the key to achieve the aforementioned benefits of flexible backhaul and are not considered in existing works on relay networks, as follows:

\begin{itemize}
\item[(1)] In HetNets with flexible backhaul, dynamic multi-path routing is employed for intelligently adjusting the network topology and backhaul access strategies %in order to better exploiting the degree of freedom in the network 
and flexibly utilising idle system resources. However, the existing works on relay networks only consider static single-path routing, where each node has one fixed path to the backhaul \cite{survey2004relay}. There are a few works on relay selection only, but the routing is still fixed under each relay choice  \cite{sreng2003relayer}.

%*****\item[(2)] In HetNets with flexible backhaul, multi-path routing is allowed in order to better utilising the capacity of the links and being able to route higher data flow rates. Other than efficient resource utilisation, multi-path routing brings more advantages including better load balancing, and even improving security for HetNets with flexible backhaul. However, in the existing works on relay networks, multi-path routing scheme is not considered and they only consider one path for each flow  (see e.g., [REF???]).

\iffalse ***
Moreover, in the existing works on relay networks, multi-path routing scheme is not considered  [REF???]. While in flexible backhaul deployment, multi-path routing is allowed in order to better utilising of the capacity of the links and route more data flows with higher rates. Other than efficient resource utilisation, multi-path routing brings more advantages including better load balancing, and even improving security for HetNets with flexible backhaul.
\fi

%@@@ \item[(3)] In relay networks, the inter-relay interference mitigation is not proposed, while in this paper, we propose DTX control policy which tries to coordinate the interference between BSs.

\item[(2)] The existing works on relay networks do not consider joint optimization of network layer controls (flow and routing control) and physical layer controls %dynamic resource block allocation for 
(interference control and link scheduling), which is the key to flexible utilization of system resources. Our considered joint resource optimization for flexible backhaul enables important benefits of flexible backhaul, including improved resource utilization efficiency,  %reliability and fast recovery ability, 
intelligent and dynamic network topology flexibility and improved dynamic backhaul transmission capability. 

\item[(3)] In order to fully utilise the available mixed-timescale CSI knowledge in the network and provide scalability and robustness (which are important features for future HetNets with flexible backhaul), we consider a hierarchical (i.e., two-timescale) design for RRM in HetNets with flexible backhaul. While there are some works considering hierarchical RRM (i.e., multi-timescale design, as will be discussed later), their approaches are heuristic and the solution is not derived from a single optimization problem. The optimization of the long-term (e.g., relay placement) and short-term variables (e.g., power control) are considered separately, and hence, there is no guarantee of optimality in the overall network utility \cite{niyato2009relay}. %, [REF???]. 

\iffalse ***
Most existing works on relay networks do not consider a hierarchical radio resource management design (i.e., multi-timescale design as will be discussed in the following). %***, and those hierarchical works are just heuristic approaches. 
%However, However, in order to provide scalability and robustness (which are important features for flexible backhaul), in this paper we have 
However, in order to fully utilise the available mixed timescale CSI knowledge and provide scalability and robustness %*****(which are important features for flexible backhaul) 
for future HetNets, in this paper we  consider a two-timescale hierarchical design for RRM in HetNets with flexible backhaul. %, that efficiently exploits the available mixed timescale CSI knowledge in the HetNet. 
In addition, the existing hierarchical RRM designs in relay networks are mainly heuristic approaches, where the RRM solution is not derived from a single optimization problem, but the optimization of the long-term and short-term variables are considered separately and accordingly, some heuristic algorithms are proposed without optimality guarantee (e.g., see \cite{niyato2009relay}, [REF???]). 
\fi

\end{itemize}

In this paper, we focus on the problem of dynamic joint resource control for HetNets with flexible backhaul. We model the problem
as a two-timescale stochastic optimization problem and propose a hierarchical two-timescale control structure that can provide scalability and robustness to signalling latency. %In the proposed hierarchical model, the long-term controls are adaptive to the large scale fading, and the short-term control is adaptive to the local channel state information (CSI) within a pico or macro BS. In such a hierarchical RRM design, the long-term controls can be implemented centrally on an RRM server (RRMS), and the short-term controls can be updated locally at each pico or macro BS.
The main contributions of this paper can be summarized as follows:

%\begin{itemize}
$ \bullet $ \textbf{Two-timescale hierarchical formulation for RRM control in HetNets with flexible backhaul:}
This problem modelling and formulation is highly important for the following reasons:

\begin{enumerate}

%*****\item RRM for \textit{flexible backhaul} is much more challenging than traditional RRM for \textit{fixed backhaul} since, in HetNets with flexible backhaul, the data flows associated with a BS without backhaul connectivity may ride on the idle backhaul resources of other BSs to reach the backhaul network. Hence, some other BSs can relay the data flows of that BS. Accordingly, multi-hop routing will be required among the different BSs. This leads to higher complexity in resource allocation as well as more complex signalling and message passing, which need to be efficiently dealt with. Furthermore, a BS relaying another BSs' data flows may increase the interference in the HetNet, and this should be addressed as well.

%RRM for \textit{flexible backhaul} is much more challenging than traditional RRM for \textit{fixed backhaul}, due to the multi-hop transmission among BSs and higher imposed interference in the network. This leads to higher complexity in resource allocation as well as more complex signalling and message passing that need to be efficiently dealt with.

\item Most existing works on RRM in HetNets only consider short-term (instantaneous) CSI adaptation or long-term (statistical) CSI adaptation. However, in most practical cases, mixed-timescale CSI knowledge is available in the HetNet: The local instantaneous CSI is available at each BS, while the global statistical CSI can be available at a central network controller. Therefore, in order to fully utilize the available mixed timescale CSI knowledge and provide scalability and robustness to signalling latency, a \textit{two-timescale hierarchical RRM} should be considered. % in the HetNet.
 While fast-timescale RRM design imposes huge signalling overhead and slow-timescale RRM design does not have good performance (as it cannot achieve multi-user diversity gain), the proposed two-timescale RRM design benefits from better performance as well as low signalling overhead, as will be verified by the simulation results presented later.

\item The existing works on two-timescale RRM for HetNets are mostly based on \textit{heuristic} approaches \cite{wang2012performance},
\cite{pang2012optimized}, i.e., the RRM solution is not derived from a single optimization problem. In \cite{liu2014hierarchical} the authors formulate
a two-timescale RRM problem for HetNets with enhanced
inter-cell interference coordination and propose an asymptotically
optimal solution in high SNR regimes. Yet none of these works consider multi-hop routing, which is an intrinsic characteristic of HetNets with flexible backhaul. Recently, works \cite{omidvar2016cross} and \cite{omidvar2015PIMRC} consider multi-hop routing in RRM as well. However, their objective is to minimize the the total average transmit power subject to instantaneous rate constraints, which is different from the goal of this paper. This problem formulation is very restrictive, since it is only suitable for a narrow class of applications that require a fixed end-to-end data rate for each user. Moreover, their solution is also heuristic.

%Furthermore, the existing works on two timescale RRM for HetNets are mostly based on heuristic approaches \cite{wang2012performance}, \cite{pang2012optimized}, i.e., the RRM solution is not derived from a single optimization problem. In \cite{liu2014hierarchical} authors formulate a two-timescale hierarchical RRM problem for HetNets with enhanced inter-cell interference coordination and propose an asymptotically optimal solution for high SNRs. Yet, none of these works consider multi-hop routing which is an intrinsic characteristic of HetNets with flexible backhaul. 

\end{enumerate}

{\color{black}{

$ \bullet $ \textbf{Global optimal solution for the two-timescale hierarchical RRM control problem:}
In general, the problem of two-timescale hierarchical RRM in HetNets with flexible backhaul is highly non-trivial and includes several challenges that need to be tackled properly:

\begin{enumerate} 

%\item \textbf{Non-convexity:}
%The overall radio resource optimization problem is non-convex. Hence, the conventional convex optimization techniques cannot be used to solve the problem.

%\item \textbf{Lack of Closed-form Expectation:}
%The stochastic optimization problem involves expectation operation in the constraints related to the average data rate of the links, which do not have closed form expression.

\item \textbf{Mixed time-scale non-convex %*****and combinatorial 
optimization problem:}
Due to the mixed-timescale hierarchical RRM structure, the problem is a two-timescale stochastic optimization, where the optimization variables change at different timescales and the constraints involve expectation operations related to the average data rate of the links, which do not have a closed-form expression. 
%***** Moreover, the optimization of discontinuous transmission (DTX) control\footnote{This refers to a dynamic resource block allocation for interference control that will be defined in Section \ref{sec: control variables}, later.} and links scheduling control variables belongs to non-convex and combinatorial optimization problems. Therefore, the current stochastic optimization techniques, such as stochastic cutting plane, cannot be applied in such problems.

\item \textbf{Complex coupling between long-term and short-term control variables:}
Since the average data rate constraint of the links involves both long-term and short-term control variables, there is a strong coupling between the variables. Therefore, the short-term and long-term control variables cannot be solved independently.
\end{enumerate}

To address the above challenges, we first apply the primal-dual decomposition
method to decouple the optimization problem into 
an inner and an outer subproblem. %, which are coupled with the constraint on average data rate of each link. The
The inner problem involves data flow control and routing control, while
the outer problem involves long-term interference mitigation among
BSs (DTX control) and short-term link scheduling control. %\textcolor{red}{{[}From your algorithm, it seems that the data flow control and routing control is the inner problem, right? If so, please change all.{]}} 
The inner
problem is convex and can be solved by standard convex optimization
methods, while the outer problem is non-convex and involves
combinatorial optimization. Using a hidden convexity in the outer
problem, we propose a sufficient condition for global optimality in this problem. Then, based on the derived global optimality condition, we propose an iterative algorithm that converges to the global optimal
solution. Finally, we simulate and compare our proposed solution with
various baselines to illustrate the significant performance gain of
our proposed solution.

{\color{blue}{
%The rest of the paper is organised as follows. Section \ref{sec: system_model} introduces system model and the control variables. The problem formulation is presented in Section \ref{sec: prob_formulation}. Problem transformation and decomposition is presented in Section \ref{prob_transform_decompos}. Section \ref{sec: solution} presents the proposed solution and iterative algorithm. Implementation considerations and signalling flow have been discussed in Section \ref{sec: sig flow}, and Simulation results have been brought in Section \ref{sec: sim}. Finally Section \ref{sec: conclusion} concludes the paper.
}}

}}

{\color{black}{

\section{System Model} \label{sec: system_model}
%System model, long timescale/short timescale variables, problem formulation, etc go here.

\subsection{Heterogeneous Network Topology}
Consider the downlink of a two-tier multi-cell heterogeneous network, as illustrated in Fig. \ref{fig:HetNet Topology_2cells}. Within each cell, there is one macro BS, several pico BSs and several MUs. Moreover, there exists a radio resource management server (RRMS) in the network which coordinates the resource allocation among BSs. All the BSs are connected to and controlled by this central RRMS via a low-cost signalling backhaul. %{\color{red}{(e.g., X2 interface in LTE \cite{khan2009LTE}) in ref ro ye ja dige cite konam}}. 
%The gray lines between the RRMS and each BS indicate the control plane, which represents the signalling interface between the macro/pico BSs and RRMS. 

In order to reduce the backhaul cost, only the macro BSs and a portion of the pico BSs are connected to the high-speed payload backhaul. The other pico BSs do not have direct access to the backhaul and hence need to communicate with the other BSs in order to reach the backhaul. It is assumed that the set of BSs with a backhaul connection is known.

%A radio resource management server (RRMS) is connected to the backhaul network and is used to coordinate the resource allocation among BSs. All the BSs are connected to and controlled by a radio resource management server (RRMS).

\begin{figure}[]
        \centering
        \includegraphics[scale=0.33]{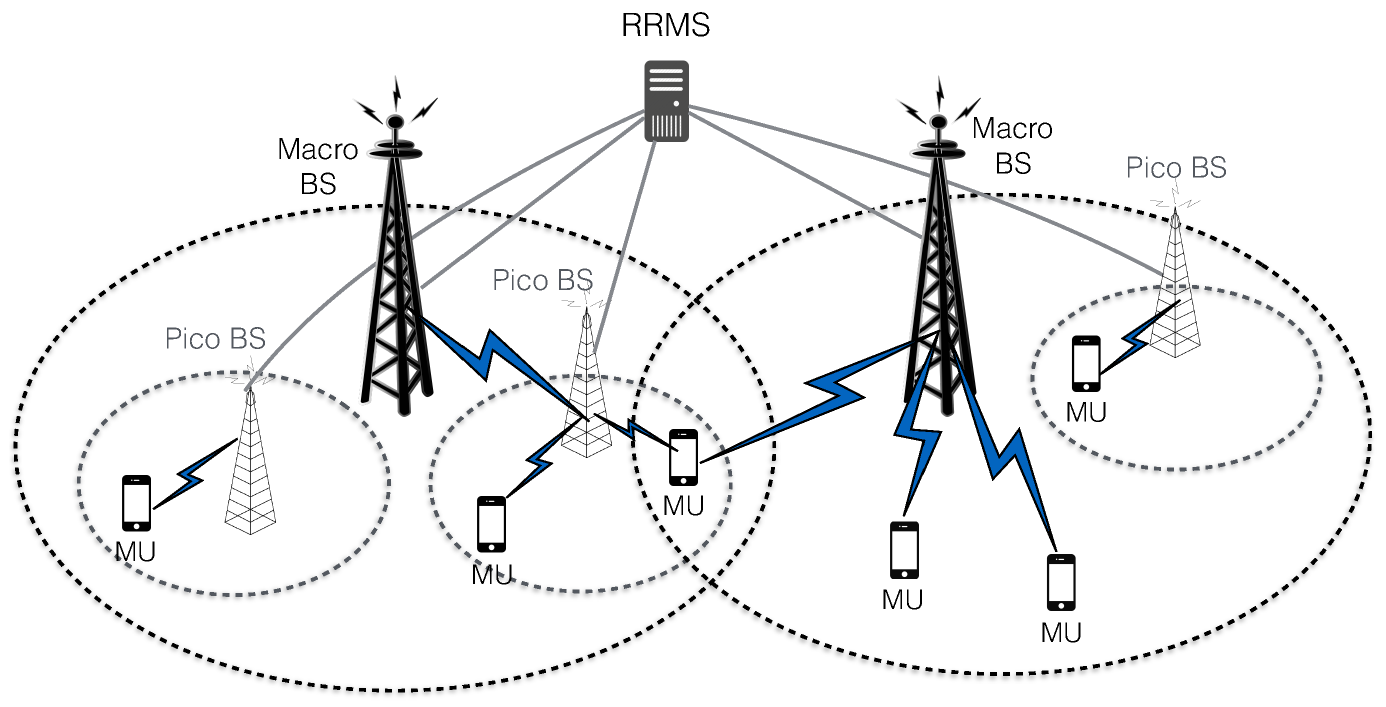} %[width=0.53\textwidth, height=0.27\textwidth]{pics/HetNet_Topology_2cells.pdf}
        \caption{\label{fig:HetNet Topology_2cells}{\small{A two-tier HetNet with macro and pico BSs. The gray lines represents the signalling interface between RRMS and the BSs.}}
%The signalling interface between the RRMS and each BS is shown by the grey lines. Note that the payload backhaul is not sketched in the figure, for clarity.
 }
 \end{figure}

There are $ K $ data flows that are to be routed from some source BSs, which have connections to the backhaul, to some destination MUs in the network. The BSs need to communicate with each other to transfer these flows from the sources to the end-users in a multi-hop mode. In this regard, some BSs need to relay the data of other source BSs in order to help their data flows reach the associated mobile users. Moreover, the total available bandwidth is divided into $ M $ subbands, which are shared by the BSs for the data transmission (BS to MU or BS to BS transmission). The HetNet topology is represented by a topology graph defined below.

\begin{definition}{(HetNet Topology Graph)}
Define the \textit{topology graph} of the HetNet as a directed graph $ G=\{\mathcal{N},\mathcal{L}\} $, where $ \mathcal{N} $ is the set of all BSs and MUs and $ \mathcal{L} $ is the set of all directed edges (BS-to-BS or BS-to-MU links). Each edge $ l\in \mathcal{L} $ is a directed link connecting its head node to its tail node, and is associated with its CSI label $ \{ h_{l,m}, \forall m \} $, where $ h_{l,m} $ represents the channel coefficient between the head and tail nodes of the $ l^{th} $ link on subband $ m $.
\end{definition}

The set of edges that are outgoing from node $ n $ is denoted by $ \mathcal{T}(n) $. Note that since the coverage area of the macro BS includes the whole cell, there is a direct link between the macro BS and each MU, and hence all the MUs are also associated with the macro BS as well. Moreover, let $ N=|\mathcal{N}| $, $ L=|\mathcal{L}| $ and $ N_{BS} \leq N $ be the number of all nodes, links, and BSs in the network topology graph, respectively.
The topology of the network can be summarized using its \textit{node-link incidence matrix} $ \boldsymbol{G} $, defined below.

\begin{definition}{(Node-link Incidence Matrix)}\label{def_G}
The node-link incidence matrix $ \boldsymbol{G} $ for the HetNet is an $ N \times L $ matrix, with a row for each node and a column for each link, in which its $ (n,l)^{th}$ element is associated with node $ n $ and link $ l $ and is given by
%{\small{
\begin{equation}\label{A_n,l}
G_{n,l} \triangleq \left\lbrace
	\begin{array}{rrr}
		1  & \mbox{if \textit{n} is the head node of link \textit{l},} \\
		-1 & \mbox{if \textit{n} is the tail node of link \textit{l},} \\
		0 & \mbox{otherwise.}
	\end{array}
\right.
\end{equation}
%}}
%*****Moreover, we define $ \boldsymbol{g}_{n} $ as the $ n^{th} $ row in the node-link incident matrix (associated with node $ n $ in the network topology graph).
\end{definition}

The time domain is divided into time slots of fixed length called a \textit{superframe}, each consisting of $ T_{s} $ \textit{subframes}, as illustrated in Fig. \ref{fig:frame}. 
%, and each superframe consists of $ T_{s} $ \textit{subframes}. 
Moreover, there are $ M $ available subbands that can be used in each subframe. For clarity, we refer to the superframe, subframe and subband by different indices $ i $, $ t $ and $ m $, respectively.

%***There are $ M $ resource blocks (RBs) in each subframe $ t $, which are associated with $ M $ subbands. Each resource block is indexed by $ (t,m) $, where $ t $ determines the subframe number and $ m $ is the subband of that RB. For clarity, we refer to the superframe, subframe and RB by the different indices $ i $, $ t $ and $ (t,m) $, respectively.

\begin{figure}
        \centering
        \includegraphics[width=0.4\textwidth, height=0.13\textwidth]%[scale=0.2]
        {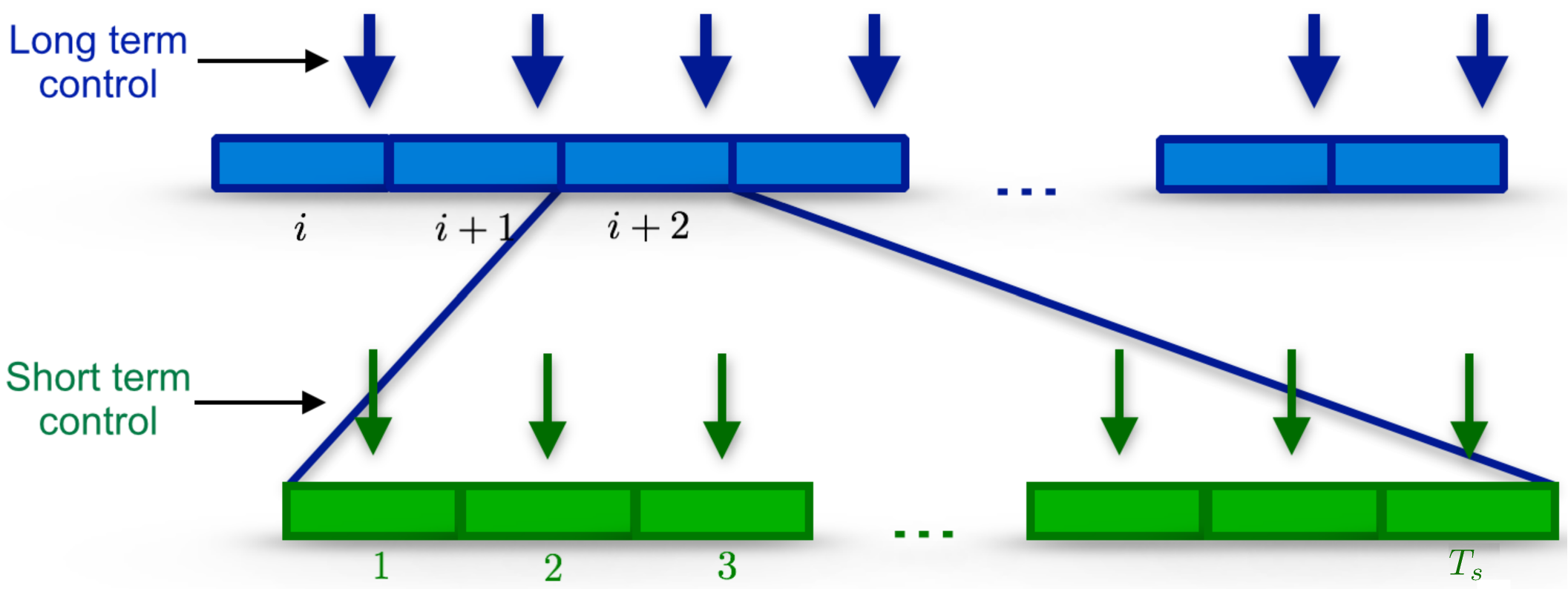}
        \caption{\label{fig:frame}{Superframes (the blue timeslots) and subframes (the green timeslots within each superframe) structure.}}
    %\caption{Caption place holder}
\end{figure}

As in many standard channel models, a two-timescale fading model has been assumed for the wireless channels between different nodes (BS-to-BS or BS-to-MU channels). Accordingly, the channel fading coefficient is given by $ h_{l,m}(t)= h^{small}_{l,m}(t) h^{large}_{l}, ~\forall l,m $,
%\begin{equation}\label{ch_fading}
%h_{l,m}(t)= h^{small}_{l,m}(t) h^{large}_{l}(t), \hspace{20 pt} \forall l,m,
%\end{equation}
where the small-scale fading $  h^{small}_{l,m}(t), ~ \forall l  \in \mathcal{L} $ remains constant in each subframe and changes over different subframes. On the other hand, the large-scale fading process $ h^{large}_{l} > 0, ~ \forall l  \in \mathcal{L} $ is caused by path loss and shadow fading (which are almost the same for all subbands, as long as the bandwidth is small compared to the carrier frequency \cite{khaled2007interpolation}, \cite{sadek2008transmit}) and is assumed to be a slow ergodic process, i.e., it remains constant for many superframes.

%$ h^{large} $ denotes the large-scale fading caused by path loss and shadowing effects which are almost the same for all sub-bands, as long as the bandwidth is small compared to the carrier frequency. Therefore, the large-scale fading is considered to be the same for all sub-bands 

\subsection{Two-Timescale Hierarchical Radio Resource Control Variables}\label{sec: control variables}

The radio resource management control variables are categorized into two groups of \textit{long-term} and \textit{short-term} control variables. 
The long-term control variables are determined centrally at the RRMS in the longer timescale (i.e., in each superframe, as shown in Fig. \ref{fig:frame}) and are adaptive to the large-scale fading process $ \boldsymbol{H}^{large}=\{ h^{large}_{l}, ~ \forall l \} $. On the other hand, the short-term control variables are adaptive to the instantaneous CSI $ \boldsymbol{H}=\{ h_{l,m}, ~ \forall l,m \} $ and are determined locally at each BS in the shorter timescale (i.e., in each subframe, as shown in Fig. \ref{fig:frame}).
%SHORTEN TWC 
%Table \ref{table: variables} summarizes the long-term and short-term control variables of our model.

\vspace{5pt} 

 \textit{a) Long-term Control Variables}
 
%Here is the list of long-term control variables which have been considered in our system model.

 %\vspace{5pt} 
 
\subsubsection{Flow Control}
 %\textit{\textbf{1) Flow Control}}

As mentioned before, there are $ K $ data flows that are to be routed in the downlink of the HetNet. We define flow control vector, $ \boldsymbol{d}=\left[ d_{1}, \ldots , d_{K}\right] $, in which each element $ d_{k} $ indicates the average allowed traffic rate for the $ k^{th} $ data flow.

As the flow control is an end-to-end control variable in the network, it should be determined centrally at the RRMS. Moreover, as $ d_{k} $'s are average values, they do not need to be adapted to the short-term realizations of the channels. Hence, it is more appropriate to adapt them to the long-term statistical information of the channels (i.e., $ \boldsymbol{H}^{large} $). Therefore, flow control is categorized as a long-term control variable, which is centrally determined at the RRMS.
 
 %\vspace{5pt} 
 
 \subsubsection{Routing Control}
 %\textbf{\textit{2) Routing Control}}
 
As the destination users may not be in the coverage area of their source BSs, the other BSs should help in between in order to provide an accessible routing path between the source and destination. 
 
}}

To better utilise the capacity of the links and provide more flexibility for the network to efficiently route more data flows and with higher rates, we allow multi-path routing in the network. It
 brings many advantages, including efficient resource utilisation, better load balancing, and even improved security, for various applications \cite{hasan2017survey}. Moreover, it is compatible with the current LTE systems that have fixed routing %*****\cite{astely2009lte}, 
 \cite{khandekar2010LTE}. Due to these advantages, it has emerged as the technology of choice for future wireless networks and a variety of incrementally deployable techniques have been proposed in the literature to implement it in practice \cite{he2008toward, kandula2007dynamic, xu2006miro, singh2015survey}. 
% \cite{he2008toward}, \cite{kandula2007dynamic}, \cite{xu2006miro}, \cite{singh2015survey}. 
In particular, software-defined networking (SDN) can be used to implement multi-path routing \cite{SDN2015programming}. Through network programmability, SDN enables the network controller to tell a network node how to split a single fow into sub-fows and route the sub-flows among different paths, using some traffic splitting approaches such as round robin \cite{singh2015survey, he2008toward, SDN2015programming}. %The interested readers  %and combining them back together into a single flow, programmatically.
%There are many traffic splitting algorithms in the literature including round robin, WFQ, PFS, etc [6, 7].

 %
\iffalse *****
Other than efficient resource utilisation, multi-path routing brings more advantages including better load balancing, and even improving security, for various applications \cite{hasan2017survey}.%*** \cite{he2008toward}, [4]
\footnote{{\color{blue}{Due to all these advantages, multipath routing strategy has emerged as the technology of choice for future wireless networks, which can fulfil QoS metrics and networks constraints in most real-time applications \cite{hasan2017survey}. %***Due to the aforementioned advantages of multi-path routing, 
Accordingly, %***many works have already addressed the practical implementation issues of multi-path routing and there exists 
a variety of incrementally deployable techniques have been proposed in the literature to implement multipath routing in practice \cite{he2008toward}, \cite{kandula2007dynamic}, \cite{xu2006miro}, \cite{singh2015survey}. %***Consequently, it is considered as the emerging the technology of choice in designing future networks. 
For more details on practical implementations of multipath routing, interested readers may refer to {\color{red}{[5] and [4]}}.
}}} 
\fi
Under multi-path routing, the traffic corresponding to each data flow $ k $ can be split arbitrarily across multiple paths between the source and destination nodes in the network. As a result, each link $ l $ may carry some part of the data flow $ d_{k} $, $ \forall k $. This is determined in the routing vector  defined as follows:  For each data flow $ k $, the corresponding routing vector is denoted by $ \boldsymbol{x}_{k}=[x_{k,1},\ldots,x_{k,L}]^{T} \in R_{+}^{L}$, in which each element $ x_{k,l} $ indicates the average carried traffic of demand $ k $ over link $ l $. Moreover, the overall $ (KL) $-dimensional routing vector is defined as $ \boldsymbol{x}=[\boldsymbol{x}_{1}^{T},\ldots,\boldsymbol{x}_{K}^{T}]^{T} $.

 {\color{black}{
The routing vectors $ \{ \boldsymbol{x}_{k}, k=1,\ldots,K \} $ are adaptive to the global network topology, which is a function of the mobility of MUs, and hence does not change in the short timescale (e.g., during several subframes in LTE \cite{astely2009lte, damnjanovic2011survey}%***** {\color{red}{OMIT if space limitation: \cite{damnjanovic2011survey}, \cite{astely2009lte}}}
). Therefore these control variables are also regarded as long-term control variables and they can be determined according to statistical CSI in the longer timescale. Moreover, since they are end-to-end control variables, they are implemented centrally in the RRMS.
  
  }}
  
%The following example show the

For better illustration of the multi-path routing scheme for a data flow, consider the simple HetNet example depicted in Fig. \ref{fig:Example_multipath_routing_2}, where a data flow is going to be routed from macro BS1 to the MU. There are three different paths available from the source to the destination, as shown by different colors in the figure. The flow control variable (i.e., $ d_1 $) determines the traffic rate that will be carried in the network for this data flow, while the routing control $ \boldsymbol{x} $ determines how this data flow is going to be routed in the network from the source to the destination. Assume the rate of the data flow is equal to 1 (i.e., $ d_1=1 $) and the capacity of all the links is equal to $ 0.5 $. As such, a single-path routing solution is not feasible for this data flow, while using multi-path routing, the network will still be able to route such a data flow. 
A possible multi-path routing solution is shown in the figure, where the blue path carries an average rate of $ 0.3 $, the green path carries $ 0.2 $ and the red path carries $ 0.5 $. Hence, the total data flow $ d_1=1 $ is fractioned into three different paths, each carrying $ 0.3$, $ 0.2 $ and $ 0.5 $, respectively. For each link indexed by $ l $, the value of $ x_{1,l} $ shown in the figure is the amount of traffic flow $ d_1 $ that is routed via that link. %For example, since link $ 3 $ carries a traffic equal to $ 0.5 $, the associated routing variable is set as $ x_{1,3}=0.5 $. 

%For better illustration of the difference between these two and their relationship, consider the following simple example. Fig. \ref{fig:Example_multipath_routing_2} shows a simple scenario where there is only one data flow that needs to be routed from macro BS1 to the mobile user (MU). The number on each link shows the link index. 

\iffalse

\begin{figure}
\centering
\begin{minipage}{.47\textwidth}
  \centering
  \includegraphics[width=0.8\textwidth, height=0.6\textwidth] %[scale=0.4]
{Replies_to_Reviewers_First_round/Example_multipath_routing.pdf}
%\includegraphics[scale=0.40]{pics/HetNet_Topology_Graph_2cell.eps}
%\par\end{centering}
\caption{\label{fig:Example_multipath_routing_2} \small{A simple example to illustrate multi-path routing for a data flow. %A simple illustration to show the difference between data flow and multi-path routing. 
Note that the bold number on each link shows the link's index. 
}%\vspace{-10 pt}
}
\end{minipage}
\begin{minipage}{.47\textwidth}
  \centering
	\includegraphics[width=0.6\textwidth, height=0.3\textwidth]{pics/toy_example1.pdf}
	%\includegraphics[width=0.45\textwidth]{pics/Signaling_SuperFrame_v4.eps}
	\caption{A simple HetNet example to illustrate \\DTX control.}
	\label{fig:toy_example1}
\end{minipage}%
\end{figure}

\fi

%\iffalse 
\begin{figure}%[h]
\begin{centering}
\includegraphics[width=0.4\textwidth, height=0.3\textwidth] %[scale=0.4]
%{Replies_to_Reviewers_First_round/Example_multipath_routing.pdf}
{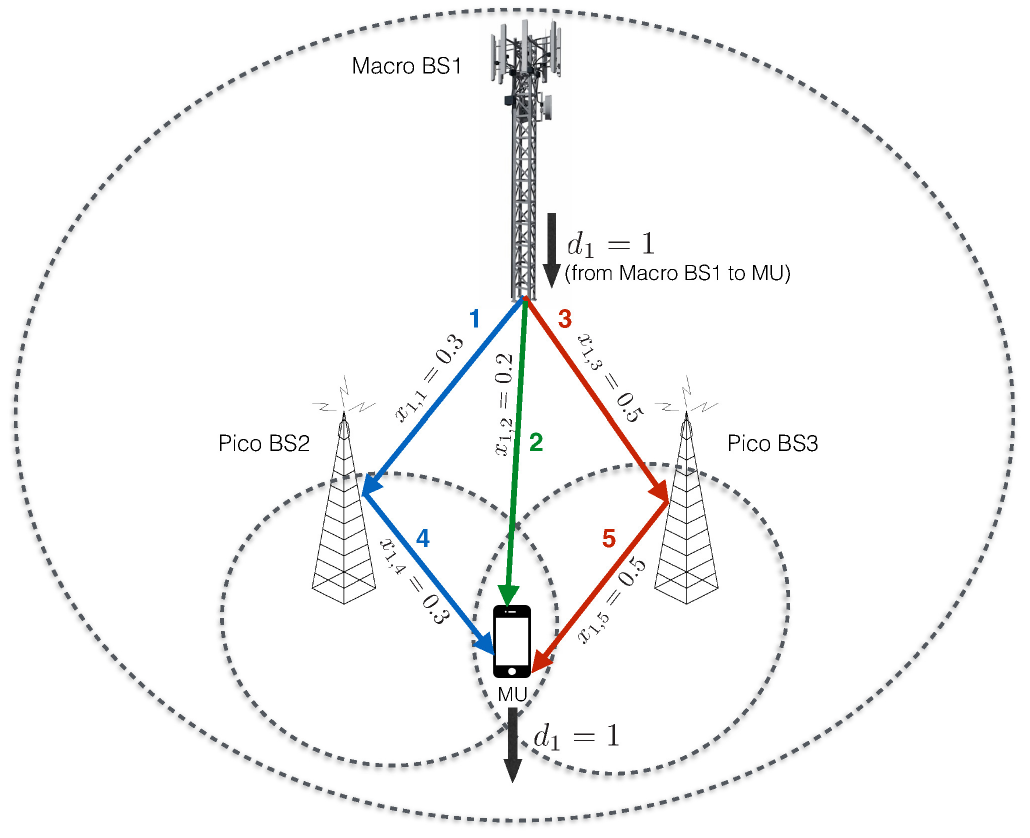}
\par\end{centering}
\caption{\label{fig:Example_multipath_routing_2} A simple example to illustrate multi-path routing for a data flow. %A simple illustration to show the difference between data flow and multi-path routing. 
Note that the bold number on each link shows the link's index. 
}
\end{figure}

  {\color{black}{
   {\color{black}{
 \subsubsection{Discontinuous Transmission (DTX) Control} 

As a macro BS covers all the other nodes (pico BSs and MUs) in its cell, it can cause strong interference to them. This is
called \textit{cross-tier interference}. Moreover, pico BSs may also
suffer from \textit{co-tier interference} caused by their neighbouring
pico BSs. To control both cross-tier and co-tier interference, we propose \textit{discontinuous 
transmission (DTX)} to mitigate interference in the
HetNet. When a DTX is scheduled in a macro or pico BS, the BS will shut
down the transmission on the current subframe. This eliminates the
interference from this BS to the other BSs. Hence, scheduling DTX
over the time domain allows us to control interference. Specifically, at each subframe, a \textit{DTX pattern} indicates which BSs are allowed to transmit, and which are not.}} %***Therefore, in order to design an efficient RRM scheme, DTX is very important in the RRM design for coordinating the interference between BSs. 
Note that a similar concept has been adopted in LTE-A systems as almost blank subframe (ABS), %\footnote{Almost Blank Subframe}
 which is a method for enhanced inter-cell interference coordination (eICIC) \cite{deb2014eICIC}. In fact, the ABS scheme in LTE-A is designed for coordinating the interference of macro BSs to pico BSs in traditional HetNets (i.e., with fixed backhaul) by determining the macro BSs that are allowed to transmit in each subframe. Our proposed DTX control scheme extends the existing ABS scheme from considering only macro BSs to considering all the BSs (macro and pico) in the network. 
 %*****In fact, unlike the conventional HetNets, in flexible backhaul HetNets with dynamic routing, it is necessary for the pico BSs as well to adopt interference coordination. Because in the downlink of conventional HetNets, apart from the interference of macro BSs to the pico BSs, there is also interference between pico BSs that needs to be coordinated as well, via a proper DTX control as will be elaborated in the following.
This is necessary because with flexible backhaul, %apart from the interference of macro BSs to the pico BSs, 
there is also interference between pico BSs that needs to be coordinated as well. %, via a proper DTX control as will be elaborated in the following.

%***The ABS scheme in LTE-A is designed for interference mitigation in traditional HetNets (i.e., with fixed backhaul) by determining which macro BSs are allowed to transmit and which are not, in each subframe, while the proposed DTX scheme {\color{red}{generalises/extends}} ABS scheme from the macro BSs to all the BSs in the HetNet, as will be elaborated in the following. %In fact, our proposed DTX control scheme extends the existing ABS scheme. 

A DTX pattern is denoted by $\boldsymbol{a}=[a_{1},...,a_{N_{BS}} ]\in\mathcal{A}$,
where $a_{n}$ indicates whether a DTX is scheduled for $BS_{n}$
on all subbands of subframe ($a_{n}=1$) or not ($a_{n}=0$),
and $\mathcal{A}= \left\lbrace \boldsymbol{a}^{(1)}, \ldots, \boldsymbol{a}^{( |\mathcal{A}| )} \right\rbrace $ is the set of all %interference-free (i.e., feasible)
feasible DTX patterns. %Note that we may drop the index $t$ when there is no ambiguity. 
%*****In each subframe, the pattern of transmitting BSs ischosen according to the time-sharing of the DTX patterns in $\mathcal{A}$. At each superframe, 
The DTX control determines the probability of using each DTX pattern in the subframes of that superframe, and is defined as follows. %In other words, at any superframe, the DTX time-sharing control, denoted as $ \boldsymbol{q}= [ q_{1},\ldots,q_{| \mathcal{A} | } ]^{T} $, determines that each DTX pattern $ \boldsymbol{a}_{j} $ will be used in a $ q_{j} $ percentage of the subframes. 

%***by the DTX control policy from the set of all feasible DTX patterns $ \mathcal{A}$. %which guarantees that no interference would be caused in the HetNet.
%{\color{blue}{This set may include DTX patterns with any desired level of interference. For example, apart from weak-interference patterns (patterns that cause weak interference, we may allow DTX patterns with more aggressive reuse factor (i.e., with more percentage of BSs turned on) as well to improve the spectrum efficiency at the cost of higher interference level in the network, or DTX patterns with all macro BSs being turned on/off (this is the Almost blank subframe (ABS) scheme in LTE-A for enhanced inter-cell interference coordination (eICIC) \ref{eICIC2014}. In this way, our interference coordination scheme includes all the existing state of the art designs, such as universal frequency reuse (i.e., all BSs are turned on all the time), ABS in LTE-Advanced and dynamic ABS, as special cases of ours.}}

%As the optimal DTX control policy depends on the HetNet topology graph as well as the global channel state information of the network, we control the DTX dynamically with respect to the large scale fading. Therefore, DTX control is categorized as long-term control variables.

\begin{definition}{(DTX Control)}\label{def: DTX}
For a given set of DTX patterns $ \mathcal{A} $, the DTX (time-sharing) control variable $ \boldsymbol{q}= [ q_{1},\ldots,q_{| \mathcal{A} | } ]^{T} $ %, where $ \boldsymbol{A} \subseteq \mathcal{A} $ and the time sharing vector $ \boldsymbol{q} $ satisfies $ q_{j} \in [0,1],  \forall j=1,\ldots,| \boldsymbol{A} |$ and $ \sum_{j=1}^{|\boldsymbol{A}| } q_{j} = 1 $. 
at any superframe 
determines that each DTX pattern $ \boldsymbol{a}^{(j)} \in \mathcal{A} $ will be used in a $ q_{j} $ percentage of the subframes, where $ q_j \geq 0, ~ \forall j $ and $ \sum_{j=1}^{| \mathcal{A} |} q_j =1 $. Hence, the feasibility set for the DTX control is defined as %\vspace{-3 pt}
%{\small{ 
\begin{align}\label{equ: q feas set}
\Lambda_{\boldsymbol{q}} \triangleq \left\lbrace \boldsymbol{q} \left| \forall j= 1,\ldots, | \mathcal{A} | : q_{j} \geq 0, ~\sum_{j=1}^{|\mathcal{A} |} q_{j} =1
 \right. \right\rbrace.
\end{align} 
%}}
\end{definition}

%*****Note that the optimal DTX control depends on the HetNet topology graph as well as the global coordination %channel state information of the network. Hence, in practice, it is not scalable to perform DTX control in the short timescale; otherwise, high signalling latency and the large required signalling overhead would significantly degrade the efficiency and optimality of the DTX control. Accordingly, for improving the robustness of the RRM design against CSI delay, we adapt the DTX control to the long-term CSI and regard it as a long-term control variable that is updated at each superframe. %Therefore, it is updated at each superframe.
%and regard it at longer timescale (i.e., each superframe).

 {\color{black}{
To avoid excessive signalling overhead/latency, we consider long-term DTX control which does not directly determine the DTX control at each subframe based on instantaneous global CSI, but determines the probability (percentage) of using each DTX pattern over the subframes within a superframe.

For better understanding of the DTX control, consider the simple network in Fig. \ref{fig:toy_example1}. %***It includes two pico BSs and three MUs, where $ \mathrm{MU}_3 $ is associated with $ \mathrm{pico~BS}_1 $ but will receive very strong interference from $ \mathrm{pico~BS}_2 $ if $ \mathrm{pico~BS}_2 $ is allowed to transmit.
 Assume that the set of all DTX patterns $ \mathcal{A} $ consists of two DTX patterns, $ \boldsymbol{a}^{(1)} = \left[ 1, 0 \right] $ (i.e., only $ \mathrm{pico~BS}_1 $ is allowed to transmit) and $ \boldsymbol{a}^{(2)} = \left[ 1,1 \right] $ (i.e., both BSs are allowed to transmit), with the associated time-sharing of $ q_1=0.3 $ and $ q_2=0.7 $, respectively. 
%DTX pattern $  \boldsymbol{a}_1 $ means that $ \mathrm{Pico~BS}_1 $ is allowed to transmit, but not $ \mathrm{Pico~BS}_2 $. On the other hand, under DTX pattern $  \boldsymbol{a}_2 $, both the pico BSs are allowed to transmit. 
Note that DTX pattern $  \boldsymbol{a}^{(1)} $ covers only $ \mathrm{MU}_1 $ and $ \mathrm{MU}_3 $, and it does not cover $ \mathrm{MU}_2 $. Furthermore, under DTX pattern $  \boldsymbol{a}^{(2)} $, $ \mathrm{MU}_3 $ usually cannot be scheduled due to the strong interference caused by $ \mathrm{pico~BS}_2 $. %It can be seen that there exists no DTX pattern that can cover all the users, due to the consideration of mitigating strong interference.
%Hence, DTX pattern $  \boldsymbol{a}_2 $ covers only $ \mathrm{MU}_1 $ and $ \mathrm{MU}_2 $, and it does not cover $ \mathrm{MU}_3 $. 
Therefore, time-sharing between DTX patterns is necessary to ensure fairness among different users. 
%SHORTEN TWC
%\iffalse
\begin{figure}[]
\centering
\includegraphics[width=0.25\textwidth, height=0.12\textwidth]{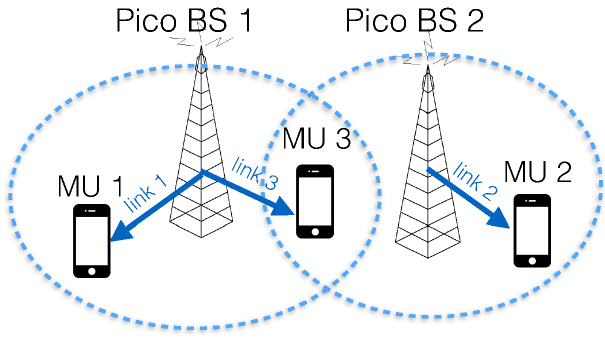}
\caption{{\small{A simple HetNet example to illustrate DTX control.}}\vspace{-10 pt}}
\label{fig:toy_example1}
\end{figure}

Assume that a superframe consists of $ T_s=10 $ subframes. At each of these ten subframes of a superframe, one of the two DTX patterns will be chosen by the pico BSs, randomly and according to their time-sharing $ \boldsymbol{q}=\left[ 0.3, 0.7 \right] $ of the current superframe, as its probability profile. 
%*****This is equivalent to randomly generate a sequence of 10 indices (corresponding to the DTX patterns indices of the 10 subframes) using $ \boldsymbol{q} $ as the probability profile of the DTX pattern, by the pseudo random generator. Note that all the BSs use the same pseudo-random generator, i.e., the seed of their random generator is the same (e.g., they may use the subframe index to generate the same seed). Therefore, since they also use the same probability profile (i.e., $ \boldsymbol{q} $), their generated pseudo-random sequence will be the same. In our example with the assumption of $ T_s=10 $ subframes in each superframe and $ \boldsymbol{q}=\left[ 0.3, 0.7 \right] $ for the current superframe, on average three of the subframes of that superframe will use DTX pattern $  \boldsymbol{a}^{(1)} $, while seven subframes will use DTX pattern $  \boldsymbol{a}^{(2)} $. 
For example, one possible realization of DTX patterns within a superframe is shown in Fig. \ref{fig:toy_example_DTX_patterns}. Note that all BSs generate the same sequence of DTX patterns by using identical pseudo-random generators with the same seed (e.g., using the subframe index as the seed).

}}

%***At each subframe of a superframe, one of the two DTX patterns will be chosen by the pico BSs, randomly and according to their time-sharing $ \boldsymbol{q}=\left[ 0.3, 0.7 \right] $ of the current superframe, as their probability profile. Note that all the BSs use the same pseudo-random generator, and hence, the DTX patterns generated at different BSs will be the same. Therefore, assuming that each superframe consists of $ T_s=10 $ subframes, on average three of the subframes of the current superframe will use DTX pattern $  \boldsymbol{a}^{(1)} $, while seven subframes will use DTX pattern $  \boldsymbol{a}^{(2)} $. For example, one possible realization of DTX patterns within a superframe is listed in Fig. \ref{fig:toy_example_DTX_patterns}. 

%SHORTEN TWC
%\iffalse
\begin{figure}[]
\centering
\includegraphics[width=0.5\textwidth]{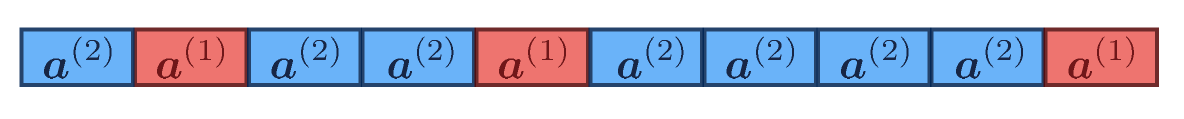}
\caption{A possible realization of DTX patterns within a superframe consisting of 10 subframes.  }
\label{fig:toy_example_DTX_patterns}
\end{figure}
%\fi

%Moreover, if we just pick one of the patterns and use them in the entire subframes, for example if we only use pattern $  \boldsymbol{a}_1 = \left[ 1,0 \right] $, then $ \mathrm{MU}_2 $ will not be served.%%/has very poor performance because it suffers from strong interference under pattern 1. 
%So time-sharing between DTX patterns ensures fairness between users. 

%According to the above explanations and illustrations, the time-sharing among different patterns has been considered for \textit{fairness} consideration of DTX control design. Specifically, if we only use a fixed DTX pattern, some MUs can hardly be scheduled for transmission either because the serving BS is not allowed to transmit or because these MUs are strongly interfered under this fixed DTX pattern (e.g., in Fig. \ref{fig:toy_example1}, $ \mathrm{MU}_3 $ is always strongly interfered by pico $ \mathrm{BS}_2 $ if we use fixed DTX pattern $ \left[ 1,1 \right]  $).
%%Specifically, if the time-sharing is not done, some feasible patterns may never be used and hence some BSs may not be chosen, which contradicts the fairness of the RRM design. 
%Therefore, if some utility functions with fairness requirements (such as PFS) have been considered, we need to somehow maintain fairness in the design of our RRM algorithm. It should be noted that if the fairness is not considered in the utility function (e.g., sum throughput utility function), it is possible that under the proposed algorithm, only one DTX will be chosen and used for all subframes of the current superframe. %However,  

Although the DTX pattern changes at each subframe, the DTX time-sharing control $ \boldsymbol{q} $ is fixed for all subframes of each superframe, as a long-term control variable. Moreover, the set of all feasible patterns $ \mathcal{A} $ may include DTX patterns with any desired level of interference. For example, apart from patterns with low interference, it may also allow DTX patterns with a more aggressive reuse factor (i.e., with a higher percentage of BSs turned on) to improve the spectrum efficiency at the cost of a higher interference level in the network. It may also include DTX patterns with all macro BSs turned on/off, which is the ABS scheme in LTE-A for eICIC  \cite{deb2014eICIC}. 
Consequently, our interference coordination scheme includes all the existing state-of-the-art designs, such as universal frequency reuse (i.e., all BSs are turned on all the time), ABS in LTE-Advanced \cite{khandekar2010LTE} and dynamic ABS \cite{liu2014hierarchical}, as special cases.

\begin{remark}\label{remark: feasible DTX set}
%The set of all feasible patterns $ \mathcal{A} $ includes all the feasible DTX patterns. This set is determined off-line and prior to the RRM problem, and is out of scope of this paper. The size of this set provides trade-off between performance and complexity. Specifically, if the size of this set is small, it contains less DTX patterns and hence the complexity of RRM will be less; while allowing/considering more DTX patterns in $ \mathcal{A} $ may increase the performance of RRM, while its complexity will also increase. 

It should be noted that determining the set of all feasible DTX patterns $ \mathcal{A} $ is an off-line problem that will be solved at the network planning level, prior to our RRM problem, which is out of the scope of this paper. Determining this set for a network topology depends on the level of interference that the network planner would like to allow in the network. 
\iffalse ***
In fact, the size of this set provides a trade-off between performance %(due to the level of interference mitigation) 
and complexity. %(due to the size of the time-sharing vector $ \boldsymbol{q} $). 
Specifically, if it contains fewer DTX patterns, the complexity of RRM will be less, while {\color{red}{allowing/considering}} more DTX patterns in $ \mathcal{A} $ may improve the performance of RRM and at the same time its complexity will also increase.  
\fi
In general, a good DTX profile set $ \mathcal{A} $ should include the DTX patterns that can maximize the spatial reuse efficiency (i.e., turn on as many links as possible subject to the constraint that the interference is below the desired level), and meanwhile, this set needs to cover all users. The size of this set is usually small in practice. For example, in the simulations in Section \ref{sec: sim}, $ | \mathcal{A} | $ is no more than $ 7 $, which can already achieve a good performance. The details are out of the scope of this paper and interested readers may refer to \cite[Section V.C]{An_eICIC_Arxiv} for more details.

\end{remark}

}}

  \vspace{5 pt}
  
  \textit{b) Short-term Control Variables (Dynamic Link Scheduling)}
 
% \vspace{5 pt}

%*****\subsubsection{Dynamic Link Scheduling}
  
%***As mentioned before, there are  $ M $ available subbands for transmission by the BSs. If the DTX pattern on the $ t $-th subframe satisfies $ \boldsymbol{a}_{n}(t)=1 $, $ BS_{n} $ is allowed to transmit on the $ (t,m) $-th RB, $ \forall m $. To avoid interference on each subband $ m $, only one of the associated links of $ BS_{n} $ is scheduled for transmission on the $ (t,m) $-th RB.

{\color{blue}{
%As mentioned before, there are  $ M $ available subbands for transmission by the BSs. Under the DTX pattern at each subframe, if a BS is scheduled for transmission, it is allowed to transmit on the $ (t,m) $-th RB, $ \forall m $. 

%As mentioned before, DTX control mitigates the interference between BSs (cross/co-tier interference). However, when a BS is scheduled for transmission under a DTX pattern in a subframe, its outgoing links may still interfere with each other if they use the same subbands.  This is called  inter-link interference. 
 {\color{black}{
To avoid inter-link interference at the outgoing links of a BS and exploit multi-user diversity, we apply link scheduling control. At each BS, link scheduling 
allocates the outgoing links of an active BS to different subbands, allowing at most one link to be scheduled over each subband so that the transmissions over different outgoing links will not interfere with each other.

%***To exploit multi-user diversity and avoid the inter-link interference at the outgoing links of a BS, we apply link scheduling. At each BS, link scheduling allocates the outgoing links of an active BS to different subbands, allowing at most one link to be scheduled over each subband. % $ m $. 

%If the DTX pattern on the (t,m)-th RB satisfies a_n(t,m) = 1, BS n is allowed to transmit on the (t,m)-th RB. To avoid interference, only one of the associated links of BS n is scheduled for transmission on the (t,m)-th RB.

\begin{definition}{(Link Scheduling)}
The links schedule at each subframe is represented by a set of functions $ \rho_{l,m} \left(\boldsymbol{a}, \boldsymbol{H} \right), \forall l,m $. Specifically, under DTX pattern $  \boldsymbol{a}$ and instantaneous CSI $ \boldsymbol{H} $, $ \rho_{l,m} \left( \boldsymbol{a}, \boldsymbol{H} \right) =1 $ means that link $ l $ is scheduled over subband $ m $ and $ \rho_{l,m}\left( \boldsymbol{a}, \boldsymbol{H} \right) =0 $ means the opposite.  Moreover, we define link scheduling policy as $ \boldsymbol\rho \triangleq \left\lbrace   \rho_{l,m} \left( \boldsymbol{a}, \boldsymbol{H} \right), ~ \forall \boldsymbol{a}, \boldsymbol{H}, l, m \right\rbrace $,  which is the collection of scheduling of all links under all DTX patterns and CSI realisations.
\iffalse
For any link $ l $, $ \rho_{l,m} \left(\boldsymbol{a}, \boldsymbol{H} \right) $ denotes its link scheduling over subband $ m $ under DTX pattern $ \boldsymbol{a} $ and instantaneous CSI $ \boldsymbol{H} $. 
Under DTX pattern $ \boldsymbol{a} $ and instantaneous CSI $ \boldsymbol{H} $, the subbands that are used over link $ l $ are determined by $ \rho_{l,m} \left(\boldsymbol{a}, \boldsymbol{H} \right), ~ \forall m $. If $ \rho_{l,m} =1 $ it means that link $ l $ is scheduled over subband $ m $, otherwise, $ \rho_{l,m} =0 $. %Note that each link can be scheduled over more than one subbands. %However, each subban
Moreover, we define link scheduling policy $ \boldsymbol\rho \triangleq \left\lbrace   \rho_{l,m} \left( \boldsymbol{a}, \boldsymbol{H} \right), ~ \forall \boldsymbol{a}, \boldsymbol{H}, l, m \right\rbrace $  which is the collection of link scheduling of all links under all DTX patterns and CSI realisations.
\fi
\end{definition}

%*** For any link $ l \in \mathcal{L} $, let $ \rho_{l,m} \in \{0,1\} $ be the link scheduling variable of link $ l $ over subband $ m $. If the $ m^{th} $ subband is used on link $ l $, then $ \rho_{l,m}=1 $;  otherwise, $ \rho_{l,m}=0 $ indicates that the $ m^{th} $ subband is not scheduled on link $ l $. For example, if the head BS of link $ l $ decides to use the subbands $ m_{1} $ and $ m_{2} $ for transmitting over link $ l $, we would have $ \rho_{l,m_{1}}=1 $ and $ \rho_{l,m_{2}}=1 $. 

It should be noted that if a $ \mathrm{BS}_n $ is not scheduled under the current DTX pattern (i.e, $ a_n=0 $), then none of its outgoing links can be scheduled over any subband (i.e., the link scheduling variables of all its outgoing links should set to zero). Moreover, each  subband $ m $  can only be scheduled to at most one of its outgoing links. These constraints are reflected in the link scheduling feasibility set: %\vspace{-5 pt}
%{\color{red}{ Clearly, the set of all feasible link scheduling policies is given by}}
%
%\vspace{-15 pt}
%{\small{
\begin{align}\label{equ: rho feas set}\small
\hspace{-7 pt}\Lambda_{\boldsymbol{\rho}} \triangleq \Bigg\lbrace \boldsymbol\rho \Bigg|  &
\forall \boldsymbol{a} ,\boldsymbol{H}, l, m, \forall n=1,\ldots, N_{BS}  : \notag\\
&
\rho_{l,m} ( \boldsymbol{a} , \boldsymbol{H}  )   \in \left\lbrace 0,1 \right\rbrace,\hspace{-3 pt} \sum_{l \in \mathcal{T}(n) } \rho_{l,m} ( \boldsymbol{a} , \boldsymbol{H}  )  \leq a_{n}   \Bigg\rbrace.
\end{align}
%}}
%*** Furthermore, we define  $ \boldsymbol{\rho_{l}}=[ \rho_{l,1},\ldots,\rho_{l,M} ]^{T} $ as the associated subband selection vector for link $ l $ and  $ \boldsymbol{\rho}=[\boldsymbol{\rho_{1}},\ldots,\boldsymbol{\rho_{L}}]^{T} $ as the overall link scheduling vector for all the links in the network. %Therefore, under the DTX pattern $ \boldsymbol{a} $ and the channel realization $ \boldsymbol{H} $, the subband(s) that are used on link $ l $ is determined by $ \boldsymbol\rho_{l}=\boldsymbol\rho_{l}(\boldsymbol{a},\boldsymbol{H}) = [ \rho_{l,1},\ldots,\rho_{l,M} ]^{T} , ~ \forall l=1,\cdots,L $. 

Note that link scheduling is decided locally by each BS %(to avoid inter-link interference as explained above) 
and is adapted to the instantaneous CSI. %to exploit multi-user diversity. 
Therefore, we consider this variable $ \boldsymbol\rho $  as a short-time control variable which is updated at each subframe.

Before introducing the problem formulation in the next section, we summarise the key notations in Table \ref{table: list of notations}.

\begin{table}[]%\small
	\begin{center}
\begin{tabular}{| c | p{5cm} |}	
\hline %[-3ex]
		 {\textbf{Notation}}  & {\textbf{Description}} \\%[1ex] 
\hline	%\\ [-3ex]	
{$ K $}  & {The number of data flows }  \\%[1ex] 
	\hline	%\\ [-2ex]	
	{$ L, N, N_{BS} $}  & {The number of links, nodes and BSs in the network topology graph }  \\%[1ex] 
	\hline	%\\ [-2ex]		
	%***{$ N $}  & {The number of nodes in the network topology graph }  \\[1ex] 
	%***\hline	\\ [-2ex]
	%***{$ N_{BS} $}  & {The number of BSs in the network }  \\[1ex] 
	%***\hline	\\ [-2ex]
	{$ \boldsymbol{G} $}  & {The node-link incident matrix of the network topology graph }  \\%[1ex] 
	\hline	%\\ [-2ex]
		 {$ d_k $}  & {Average traffic rate of the $ k^{th} $ data flow }  \\%[1ex] 
	\hline	%\\ [-2ex]		 
			 {$ x_{k,l} $}  & {The amount of traffic carried on link $ l $ for the $ k^{th} $ data flow }  \\%[1ex] 
	\hline	%\\ [-2ex]	
	 {$ \mathcal{A} = [ \boldsymbol{a}^1, \ldots,  \boldsymbol{a}^{|\mathcal{A}|}  ] $}  & {The set of all admissible DTX patterns }  \\%[1ex] 
	\hline	%\\ [-2ex]
	 {$ \boldsymbol{q} = [q_1, \ldots, q_{|\mathcal{A}|}]$}  & {The DTX patterns, time-sharing vector}  \\%[1ex] 
	\hline	%\\ [-2ex]
	 {$ \rho_{l,m} $}  & {The link scheduling variable for link $ l $ over subband $ m $ }  \\%[1ex] 
	\hline	%\\ [-2ex]
	% {$ \boldsymbol{a}_j $}  & {the  $ j^{th} $ DTX pattern in $ \mathcal {A} $}  \\[1ex] 
	%\hline	\\ [-2ex]
	{$ p_{macro},~ p_{pico} $}  & {The  power level of a transmitting macro/pico BS over each subband}  \\%[1ex] 
	\hline	%\\ [-2ex]
	%{$ p_{pico} $}  & {the  power level of a transmitting pico BS over each subband}  \\[1ex] 
	%\hline	\\ [-2ex]
%\hline  
{$ \mathcal{T}(n) $}  & {The  set of outgoing links from $ \mathrm{BS}_n $}  \\%[1ex] 
	\hline
	{$ n(l) $}  & {The  head BS of link $ l $}  \\%[1ex] 
	\hline
	{$ tail(l) $}  & {The  tail node of link $ l $}  \\%[1ex] 
	\hline
\end{tabular}
	\end{center}
	\caption{{\small{List of the key notations used in the paper.}}\vspace{-10 pt}}	
		\label{table: list of notations}
		
\end{table}

%$  tail(l), ~\forall l \in \mathcal{L} $ denotes the tail node of link $ l $

%}}

\section{Two-Timescale Hierarchical RRM Problem Formulation} \label{sec: prob_formulation}
% CHALLENGES

%{\color{blue}{

For given DTX control $ \boldsymbol{q} $, and link scheduling policy $ \boldsymbol{\rho} $ and under large-scale channel fading state $ \boldsymbol{H}^{large}=\{ h^{large}_{l} \} $, %{\color{red}{the conditional average data rate}} 
the average data rate of link $ l $ %conditional on given DTX pattern  $ \boldsymbol{a} $ 
is given by %Note that all the expectations are with respect to the random instantaneous CSI $ \boldsymbol{H} $ and conditional on the large-scale channel fading state $ \boldsymbol{H}^{large} $.
\iffalse ***
\begin{equation}\label{ave_rate_DTXpolicy}\small
\overline{r}_{l} ( \boldsymbol{q}, \boldsymbol\rho %| \boldsymbol{H}^{large}
 ) = \sum_{j=1}^{ | \mathcal{A} |} q_{j}     \sum_{m=1}^{M} \mathbb{E} \left[ \rho_{l,m}(\boldsymbol{a}^{(j)},\boldsymbol{H}) ~ \log \left(1+ \dfrac{| h_{l,m} |^{2} p_{n(l)} }{ 1 + \sum_{l^\prime \in I_l^{(j)} } | h_{l^\prime,m} |^2 ~p^{ (j)}_{n(l^\prime),m}   } \right) \right] .
 %  \left| \boldsymbol{H}^{large} \right.   \right] ,
\end{equation}
\fi
\begin{equation}\label{ave_rate_DTXpolicy}%\small
\overline{r}_{l} ( \boldsymbol{q}, \boldsymbol\rho %| \boldsymbol{H}^{large}
 ) = \sum_{j=1}^{ | \mathcal{A} |} q_{j}    r_{l} ( \boldsymbol{a}^{(j)}, \boldsymbol\rho %| \boldsymbol{H}^{large}
 ),
\end{equation}\vspace{-10 pt}
where
\begin{align}\label{ave_rate_DTXpattern}\small
r_{l} ( \boldsymbol{a}^{(j)}, \boldsymbol\rho %| \boldsymbol{H}^{large}
 ) = & \sum_{m=1}^{M} \mathbb{E} \Bigg[ \rho_{l,m}\left(\boldsymbol{a}^{(j)},\boldsymbol{H} \right) \notag \\
& \times \log \bigg( 1 + \dfrac{| h_{l,m} |^{2}  p_{n(l)} }{ 1 + \sum_{l^\prime \in I_{l,m}^{(j)} } | h_{l^\prime,m} |^2 ~p_{n(l^\prime)}   } \bigg) \Bigg],
 %  \left| \boldsymbol{H}^{large} \right.   \right] .
\end{align}
%
%where the expectation is with respect to the random instantaneous CSI $ \boldsymbol{H} $, conditional on a fixed long-term CSI $ \boldsymbol{H}^{large} $.
\iffalse***

For given DTX pattern $ \boldsymbol{a} $, link scheduling policy $ \boldsymbol{\rho} $ and under large-scale channel fading state $ \boldsymbol{H}^{large}=\{ h^{large}_{l} \} $, %{\color{red}{the conditional average data rate}} 
the average data rate of link $ l $ %conditional on given DTX pattern  $ \boldsymbol{a} $ 
is given by
% the following formula. 
%\begin{align}\label{ave_rate_DTXpattern}
%r_{l}(\boldsymbol{a},&\boldsymbol\rho | \boldsymbol{H}^{large} ) =\notag
%\\ &\sum_{m=1}^{M} \mathbbm{E} \left[ \rho_{l,m}(\boldsymbol{a},\boldsymbol{H}) ~ \log(1+ | h_{l,m} |^{2} p_{l} ) \left| \boldsymbol{H}^{large} \right. \right]
%\end{align}
\begin{equation}\label{ave_rate_DTXpattern_2}\small
r_{l}(\boldsymbol{a},\boldsymbol\rho | \boldsymbol{H}^{large} ) =
\sum_{m=1}^{M} \mathbb{E} \left[ \rho_{l,m}(\boldsymbol{a},\boldsymbol{H}) ~ \log \left(1+ \dfrac{| h_{l,m} |^{2} {\color{red}{p_{l,m} }} }{ \sigma_{l,m}   } \right)   \left| \boldsymbol{H}^{large} \right.   \right] ,
\end{equation}
\fi
where for any link $ l $, %$ n(l) $ denotes its head BS, 
$ I_{l,m}^{(j)} = \left\lbrace l^\prime \in \mathcal{L} \left| \rho_{l^\prime,m}=1 ,  n(l^\prime) \neq n(l) , tail(l^\prime) = tail (l) \right.  \right\rbrace $ denotes the set of its active interfering links under DTX pattern $ \boldsymbol{a}^{(j)} $ on subband $ m $. 
%, and $  tail(l), ~\forall l \in \mathcal{L} $ denotes the tail node of link $ l $. 
Moreover, $ p_{n} $ is the transmission power level of $ \mathrm{BS}_n %, ~ \forall n = 1,\ldots, N_{BS} 
$ on each subband %over link $ l $ 
and is given by %{\color{red}{(hazfe in age ja nabud)}}
%{\small{
\begin{equation}\notag
\forall n = 1, \ldots, N_{BS}: %\quad 
p_{n} = \left\lbrace
	\begin{array}{ll}
%		a_{n(l)}^{(j)} 
		p_{macro}  & \mbox{if $ BS_{n} $ is a macro BS,} \\
%		a_{n(l)}^{(j)} 
		p_{pico} & \mbox{if $ BS_{n} $ is a pico BS,} 
	\end{array}
\right.
\end{equation}%}}
where $ p_{macro} $ and $ p_{pico} $ are the power levels of the macro BS and pico BS, respectively, on each subband. Note that under DTX control, the inter-cell interference among the active BSs is sufficiently small, even if all active BSs transmit at the maximum power. Therefore, it is near-optimal for each active BS to transmit with its maximum power so as to maximize its desired signal power. Nonetheless,  the considered problem formulation can be easily extended to include power control as well, by treating the power level  of each macro/pico BS as another optimization variable. 
It should be noted that all expectations in \eqref{ave_rate_DTXpattern} are with respect to the random instantaneous CSI $ \boldsymbol{H} $ and conditional on the large-scale channel fading state 
$ \boldsymbol{H}^{large} $.

%***Moreover, as the coverage area of a macro BS is greater than the coverage area of a pico BS \cite{dementev2013machine}, clearly we have $ p_{macro} > p_{pico} $.

\iffalse ***
Moreover, $ \sigma_{l,m}^{(j)}  $ is noise plus the interference from other BSs to link $ l $  under DTX pattern $ \boldsymbol{a}^{(j)} $ and channels realisation $ \boldsymbol{H} $, and is given by % {\color{red}{$ \sigma_{l,m} = 1 + \sum_{l^\prime \in I_l } h_{l^\prime,m} ~p_k $}}, 
\begin{equation}
%\sigma_{l,m} = 1 + \sum_{\forall l^\prime \neq l, o_{l^\prime}=o_{l}, \boldsymbol{a}(n_{l^\prime})=1 } | h_{l^\prime,m} |^2 ~p_k,
\sigma_{l,m}^{(j)} = 1 + \sum_{l^\prime \in I_l } | h_{l^\prime,m} |^2 ~p_{l^\prime,m}, % {\color{red}{ \quad p_{k,m}  ??? }}
\end{equation}
where $ I_l^{(j)} $ is the set of all interfering links with link $ l $ under DTX pattern $ \boldsymbol{a}^{(j)} $. %***, i.e., the set of all links that their head nodes is different from the head node of link $ l $ and is active according to the DTX pattern and their tail node is the same as the tail node of link $ l $.
\fi

}}

 {\color{black}{
The performance of the network is measured by a utility function $ U(\boldsymbol{d}) $, where $ \boldsymbol{d} = \left[ d_{1}, \ldots , d_{K}\right] $ is the average data flow vector, as mentioned before. We make the following assumption on the utility function.%$ U(\boldsymbol{d}) $.

\begin{assumption}{(Utility Function)}\label{assumption: utility func}
The utility function is expressed as $ U(\boldsymbol{d})=\sum_{k=1}^{K} U_{k}(d_{k}) $, where $ U_{k} $ is a twice continuously differentiable, strictly concave and increasing function of the average data flow rate $ d_{k} , $ $\forall k =1,\ldots,K$. %Moreover, $ U_{k}, ~ \forall k $ is L-Lipschitz, i.e.,
%\begin{align}\label{equ: Lipschitz}
%& \forall k=1,\ldots,K: \notag\\ &\left| \frac{ \partial U_{k} (d_{1}) }{\partial d} - \frac{ \partial U_{k} ( d_{2} ) }{\partial d} \right| \leq L \left| d_{1} - d_{2} \right|, ~ \forall d_{1},d_{2} \geq 0
%\end{align}
%for some constant $ L > 0 $.
\end{assumption}

Note that Assumption \ref{assumption: utility func} on the utility function can capture a lot of interesting cases, such as $ \alpha $-fairness \cite{alpha_fair} and proportional fairness\cite{PFS}.

For a given HetNet topology graph $ \mathcal{G} = \{ \mathcal{N},\mathcal{L} \} $ and known source-destination pairs, the hierarchical RRM optimization problem can be formulated as follows: 
\begin{subequations}
  \begin{align}
  \mathcal{P}_{org}: &&&  \max_{\boldsymbol{d}, \boldsymbol{x}, \boldsymbol{q}, \boldsymbol{\rho} }  U \left(  \boldsymbol{d} \right) \label{obj_func} & \\
  & \quad\text{subject to:} &&& \nonumber\\
  &&& \hspace{-20 pt} \boldsymbol{G} \boldsymbol{x}_{k} = \boldsymbol{v}_{k} (d_{k}), &&  \hspace{-10 pt}  \forall k=1,\ldots,K, \label{flow_cons} \\
  &&& \hspace{-20 pt} \sum_{k=1}^{K} x_{k,l} \leq \overline{r}_{l} \left( \boldsymbol{q}, \boldsymbol{\rho} \right),   %( \Omega )
   &&    \forall l=1,\ldots,L, \label{ave_rate} \\
   &&& \hspace{-20 pt} \boldsymbol{q} \in \Lambda_{ \boldsymbol{q}}, ~ \boldsymbol{\rho} \in \Lambda_{\boldsymbol{\rho}}. \label{equ: PHY feasibility set}
  \end{align}
\end{subequations}

Constraint (\ref{flow_cons}) is the flow conservation constraint, in which $ \boldsymbol{G} $ is the node-link incidence matrix of the HetNet (as defined in \eqref{A_n,l}), and $ \boldsymbol{v}_{k} (d_{k}) = [v_{k,1},\ldots,v_{k,N}]^{T} \in \mathbb{R}^{N} $, which specifies the net outgoing data rate for $ d_{k} $ in each node, is defined as 
%
%\vspace{-17 pt}
%{\small{
\begin{equation}\label{v_k}%\small
v_{k,n} (d_{k}) = \left\lbrace
	\begin{array}{rl}
		d_{k}  & \mbox{if \textit{n} is the source node of flow \textit{k},} \\
		-d_{k} & \mbox{if \textit{n} is the destination node of flow \textit{k},} \\
		0 & \mbox{otherwise.}
	\end{array}
\right. \notag
\end{equation}
%}}
%
%Moreover, $ \overline{r_{l}} $ in \eqref{ave_rate} is the conditional data rate of link $ l $ defined in \eqref{ave_rate_DTXpolicy}. 
Furthermore, constraint \eqref{ave_rate} is the capacity constraint on each link that denotes the physical layer limitation on the average data rate of each link. Finally, constraint \eqref{equ: PHY feasibility set} is the feasibility constraint for the DTX control  and link scheduling variables (i.e., $ \boldsymbol{q} $ and $ \boldsymbol{\rho} $).

%*** Moreover, constraint \eqref{equ: rho_l  a_j} guarantees that if under any DTX pattern $ \boldsymbol{a}^{(j)} $, $ \mathrm{BS}_n $ is not activated, non of its outgoing links can be scheduled over any subband. Finally, constraint \eqref{equ: PHY feasibility set} is the feasibility constraint for the DTX control  and link scheduling variables (i.e., $ \boldsymbol{q} $ and $ \boldsymbol{\rho} $).
%*** Finally, constraints \eqref{equ: rho_l 0,1 }-\eqref{equ: q_j >= 0} {\color{red}{are the feasibility constraints OR show the feasibility space}} for the link scheduling and DTX time-sharing variables. For simplicity of notations, we show this feasibility space by $ \Lambda $, i.e., $ \Lambda \triangleq \left\lbrace  \left( \boldsymbol{q}, {\color{red}{\boldsymbol{\rho}}} \right) \mathrm{ satisfying ~ \eqref{equ: rho_l 0,1 }-\eqref{equ: q_j >= 0} } \right\rbrace $.

%***Finally, constraint (\ref{Theta_feas_cond})  guarantees that the chosen physical layer control policy be feasible.

 %{\color{blue}{
Note that in the above problem formulation, there is a tight coupling between different control variables, %which makes it {\color{red}{impossible/difficult}} to optimise them independently, 
and hence, we need to optimise them jointly. %The existing coupling be
For example, the network layer control variables (including flow control and routing) and the physical layer control variables (including the DTX control and link scheduling)  cannot be optimised independently. %, and hence, we need to jointly optimise them. 
Moreover, for fixed network layer controls, there is a coupling between the DTX time-sharing and link scheduling controls, and even for fixed DTX time-sharing control, the link scheduling variables for different subframes of a superframe are still coupled together.

It should be noted that although the application focused on in this paper is flexible backhaul, the proposed model can be utilised in a broader range of scenarios, including conventional HetNets (i.e., with full backhaul connection), relay networks, D2D networks, cellular networks with inbound backhaul and  multi-hop cellular networks.
 Moreover, the proposed two-timescale design and solution techniques (which will be presented later) can also be adapted to various applications to give a highly scalable and practical RRM design.
}} 
 %***The problem formulation covers many other scenarios and deployments as special cases and the formulated problem and the proposed algorithm can be used for modelling and solving the RRM problem in those applications as well.

}}

%Note that in the above problem formulation, there is a tight coupling between different control variables, which makes it {\color{red}{impossible/difficult}} to optimise them independently, and hence, we need to optimise them jointly. First of all, there is a tight coupling between the the network layer and physical layer control variables, as follows:  The network layer control variables (including flow control and routing) cannot be optimised independent of the physical layer control variables (including the DTX control and link scheduling) and visa versa. For example, the routing depends on the DTX control as well. Because under di↵erent DTX controls, the average throughput of each link will be di↵erent and consequently, this will clearly a↵ect the routing choice via the link capacity constraint (12).

{\color{black}{

\section{Problem Transformation and Decomposition} \label{prob_transform_decompos} \label{prob_transform_decompos}
% CONVERTING THE ABBOVE OPT PROBLEM INTO A SIMPLE ONE WITH OMEGA INSTEAD OF CONSTRAINTS:

%The above objective function can be expressed as a function of average data rate $ \overline{\boldsymbol{r}} $. As a result, the above optimization problem can be expressed by the following compact form expression.

%\begin{equation}\label{original_prob}
%P_{org}: \hspace{5 pt} \max_{ \Omega \in \Lambda }  U(\overline{\boldsymbol{r}}(\Omega))
%\end{equation}
%where $ \Omega \triangleq \left\lbrace \boldsymbol{d}, \boldsymbol{x}, \Theta , \boldsymbol{\rho} \right\rbrace $ and $ \Lambda $ represents the feasible set defined by \eqref{flow_cons}-\eqref{Theta_feas_cond}.

% CHALLENGE #1: NO CLOSED FORM EXPRESSION FOR THE RATE CONSTRAINT:

Note that according to \eqref{ave_rate_DTXpolicy} and \eqref{ave_rate_DTXpattern}, the %conditional 
average data rate $ \overline{\boldsymbol{r}}_l , \forall l $ in \eqref{ave_rate} %$ \overline{\boldsymbol{r}} = [ \overline{r}_{1}, \ldots ,\overline{r}_{L} ]^{T} $ 
involves a stochastic expectation over CSI realizations and does not have a closed-form expression. 
Moreover, due to the combinatorial link scheduling variables and the production terms $ q_j \times \rho_{l,m} $ in the average rate expression, $ P_{org} $ is a non-convex stochastic optimization problem which cannot be solved by the common stochastic optimization techniques  such as stochastic subgradient and stochastic cutting plane. 
\iffalse *****
This implies that $ P_{org} $ is a stochastic optimization problem. However, since the problem includes control variables with a combinatorial structure (i.e., DTX control policy), the common stochastic optimization techniques, such as stochastic subgradient and stochastic cutting plane, cannot be applied in this problem.
%Therefore, in order to make the problem tractable, we need to tackle the following challenge.
% CHALLENGE #2: NON-CONVEXITY OF THE PROBLEM:
Moreover, the existence of combinatorial variables (i.e., $ \boldsymbol\rho_{l,m} , \forall l,m $) in the capacity constraints \eqref{ave_rate}, results in the feasibility region of the problem to be non-convex. Consequently, the overall problem is not convex, and hence the conventional convex optimization techniques cannot be used to solve it.  %Therefore, in order to make the problem tractable, we need to tackle the following challenge.
%\begin{framed}
%\noindent\textbf{Challenge 1.} Exploit the specific structure of the problem in order to find a hidden convexity.
%\end{framed}
\fi
In the following, we will show how to tackle this challenge by transforming the original problem into a form which can exploit a hidden convexity.

First, using primal decomposition \cite{palomar2006tutorial, bertsekas1999nonlinear}%\cite{boyd2007notes}%\cite{boyd2008decomposition}
, we decompose the original problem into two sub-problems: the inner (a.k.a slave) problem $ \mathcal{P}_{1} $ and the outer (a.k.a master) problem $ \mathcal{P}_{2} $, as follows. %*****{\color{red}{It should be noted that according to the primal decomposition, solving the master problem includes solving the slave problem as well and  is equivalent to solving the original problem \cite{boyd2007notes}.}}

\textbf{Subproblem 1} (Optimization of the routing control $\boldsymbol{d}$
and flow control $\boldsymbol{x}$ under a fixed DTX control and link scheduling policy): %\vspace{-10 pt}
\begin{subequations}
  \begin{align}
  \mathcal{P}_{1}: &&&\widetilde{U}(\overline{\boldsymbol{r}}\left( \boldsymbol{q}, \boldsymbol{\rho} \right)) =  \max_{\boldsymbol{d}, \boldsymbol{x}}  U \left(  \boldsymbol{d} \right) && \label{def_U_tilda}\\ %\sum_{k=1}^{K} U_{k}(d_{k}) 
  & \text{subject to:} &&& & \nonumber\\
  &&& \hspace{-22 pt} \boldsymbol{G} \boldsymbol{x}_{k} = \boldsymbol{v}_{k} (d_{k}),  && \hspace{-22 pt} \forall k=1,\ldots,K, \label{P2_flow_cons}\\
  &&& \hspace{-22 pt} \sum_{k=1}^{K} x_{k,l} \leq \overline{r}_{l}\left( \boldsymbol{q}, \boldsymbol{\rho} \right),  && \hspace{-22 pt}  \forall l=1,\ldots,L. \label{P2_ave_rate}
  \end{align}
\end{subequations}

\iffalse ******
\begin{equation}\label{def_U_tilda}\small
\mathcal{P}_{1}: ~ \widetilde{U}(\overline{\boldsymbol{r}}\left( \boldsymbol{q}, \boldsymbol{\rho} \right)) =  \max_{\boldsymbol{d}, \boldsymbol{x}}  U \left(  \boldsymbol{d} \right) %\sum_{k=1}^{K} U_{k}(d_{k})
\end{equation}
\textit{\hspace{35 pt} subject to}
\begin{equation}\label{P2_flow_cons}\small
\boldsymbol{G} \boldsymbol{x}_{k} = \boldsymbol{v}_{k} (d_{k}), \hspace{20pt}  \forall k=1,\ldots,K
\end{equation}
\begin{equation}\label{P2_ave_rate}\small
\sum_{k=1}^{K} x_{k,l} \leq \overline{r}_{l}\left( \boldsymbol{q}, \boldsymbol{\rho} \right),   \hspace{15pt}  \forall l=1,\ldots,L.
\end{equation}
******
\fi

\textbf{Subproblem 2} (Optimization of the physical layer controls, i.e, DTX time-sharing and link scheduling):
\begin{align}\label{opt_U_tilda1}%\small 
 \mathcal{P}_{2}: &&&&&&&&&&&&& \max_{  \boldsymbol{q} \in \Lambda_{\boldsymbol{q}} , ~ \boldsymbol{\rho} \in \Lambda_{\boldsymbol{\rho}} }   %_{ \Omega \in \Lambda }
  \widetilde{U}(\overline{\boldsymbol{r}} \left( \boldsymbol{q}, \boldsymbol{\rho} \right) ). &&&&&&&&&&
\end{align}
\iffalse ***
\begin{center}
\textit{\hspace{20 pt} %{35 pt} 
subject to}
\eqref{equ: rho_l  a_j}-\eqref{equ: q_j >= 0}
\end{center}
\fi

%\begin{figure}
%\begin{centering}
%\includegraphics[scale=0.65]{pic/problem-struc.pdf}
%\par\end{centering}
%\caption{\label{fig:trans}Overall problem decomposition and transformation procedure.}
%\end{figure}

%The physical meaning behind this problem decomposition is elaborated below. First by solving problem $ \mathcal{P}_{2} $, the routing paths and the average data rate of the flows are determined. Then, having the routing paths and data flows as input, the DTX control policy and link scheduling policy are updated by solving problem $ \mathcal{P}_{1} $. These problems are solved iteratively until the global optimum solution is achieved.

%The overall problem decomposition and transformation procedure has been illustrated in Fig. \ref{fig:trans}. 
}}

From the primal decomposition \cite{palomar2006tutorial, bertsekas1999nonlinear}, 
%\cite[Chapter 3]{boyd2007notes},   %{\color{red}{[REF???, Chapter???]}},
 we have the following lemma. 
 %***** that shows the optimal solutions to the subproblems are indeed the optimal solution to the original problem. The proof has been omitted due to space limitation and directly follows from \cite[Chapter 3]{boyd2007notes}.

\begin{lemma}\label{lem: primal decomposition}
If $ \left( \boldsymbol{q}^\ast, \boldsymbol{\rho}^\ast \right) $ is the optimal solution of $ \mathcal{P}_2 $ and $ \left( \boldsymbol{d}^\ast, \boldsymbol{x}^\ast \right) $ is the optimal solution of $ \mathcal{P}_1 $ with the DTX control and link scheduling policy fixed as $ \left( \boldsymbol{q}^\ast, \boldsymbol{\rho}^\ast \right) $, then $ \left( \boldsymbol{d}^\ast, \boldsymbol{x}^\ast, \boldsymbol{q}^\ast, \boldsymbol{\rho}^\ast \right) $ is the optimal solution to the original problem $ \mathcal{P}_{org} $.
\end{lemma}

{\color{black}{

The intuition behind this decomposition of the original problem formulation $ \mathcal{P}_{org} $ into the sub-problems corresponds to the network layers structure. In other words, we have decoupled the original cross-layer problem into individual layer sub-problems, namely the network layer sub-problem (flow and routing control) and the physical layer (PHY) sub-problem (DTX and link scheduling).
%Specifically, the sub-problem $ \mathcal{P}_1 $ is the network layer problem, and the sub-problem $ \mathcal{P}_2 $  is the physical layer problem. 
The network layer problem $ \mathcal{P}_1 $ is a long-term problem due to the fact that the network layer control variables (i.e., routing and flow controls) are long-term control variables. However, since the physical layer control variables include both short-term (i.e., link scheduling) and long-term control variables (i.e., DTX control), the physical layer sub-problem $ \mathcal{P}_2 $ is still a mixed-timescale problem. 

It should be noted that under any fixed DTX control and link scheduling policy $ \left( \boldsymbol{q}, \boldsymbol{\rho} \right) $, or equivalently any data rate vector $ \overline{\boldsymbol{r}} $, problem $ \mathcal{P}_{1} $ is a standard convex optimization problem. Hence, using existing convex optimization methods, it can be easily solved in the RRMS at the beginning of each superframe in order to obtain the routing vector $ \boldsymbol{x} $ and data flow rate vector $ \boldsymbol{d} $ for that superframe.

\begin{remark}{(Solving Problem $ \mathcal{P}_{1} $)}\label{remark: solving P1}
%Note that we use primal-dual methods for solving problem $ \mathcal{P}_{1} $, so that when solving problem $ \mathcal{P}_{1} $, we obtain the Lagrangian multipliers or dual variables associated to constraints \eqref{P2_ave_rate}, at the same time as well. The reason for applying these methods is that we will need these dual variables later on, for obtaining the subgradient vector of function $ \widetilde{U} $.
Note that for solving problem $ \mathcal{P}_{1} $, we specifically use primal-dual methods \cite[Chapter 11]{boyd2009convex}. Consequently, by solving problem $ \mathcal{P}_{1} $, we will simultaneously obtain the Lagrangian multipliers or dual variables %$ \boldsymbol\omega = [\omega_{1}, \ldots, \omega_{L} ]^{T} $ 
associated with constraints \eqref{P2_ave_rate} as well. 
Using the conditions in Assumption \ref{assumption: utility func}, \cite[Proposition 3.3.3]{bertsekas1999nonlinear} shows that the weight  %Lagrangian multiplier 
vector $ \boldsymbol\omega $ 
is actually the gradient vector of $ \widetilde{U} \left( \boldsymbol{r} \right) $, i.e., %$ \boldsymbol\lambda = \nabla \widetilde{U} \left( \boldsymbol{r} \right) $, 
$ \boldsymbol\omega = \nabla \widetilde{U} \left( \boldsymbol{r} \right) $, 
which will be used later on in our proposed algorithm. % \boldsymbol{g}_{\widetilde{U}} $.

%Later on, we will show how these Lagrangian multipliers will be used for obtaining the subgradient vector of function $ \widetilde{U} $.

%These Lagrangian multipliers will be used later on, for obtaining the subgradient vector of function $ \widetilde{U} $.
\end{remark}

On the other hand, it is very difficult to find the solution for Problem
$\mathcal{P}_{2}$, because $\mathcal{P}_{2}$
is a non-convex stochastic optimization problem with the objective function being the optimal objective of Problem $\mathcal{P}_{1}$. In the next section, we focus on addressing the following challenge. 
\iffalse *****
On the other hand, it is very difficult to find the solution for problem
$\mathcal{P}_{2}$, due to the following reasons. First, there is no closed-form expression for
the utility function $\widetilde{U}$ of problem $\mathcal{P}_{2}$
because $\widetilde{U}$ is the solution of another constrained optimization
problem (i.e., problem $\mathcal{P}_{1}$). Moreover, the %conditional
average data rates $\overline{r}_{l},~\forall l$ in the objective
function $\widetilde{U}(\overline{\boldsymbol{r}})$ involve stochastic
expectation over CSI realizations and $\overline{r}_{l}$ is non-convex, since it includes combinatorial control variables
$\boldsymbol{\rho}$.
%*** $\overline{r}_{l}$ is a a non-convex function of $\Omega$ which includes combinatorial control variables $\boldsymbol{\rho}$ and $\boldsymbol{A}$. 
As a result, $\mathcal{P}_{2}$
is a non-convex stochastic optimization problem with combinatorial
optimization variables, and hence the conventional stochastic optimization
techniques (such as stochastic subgradient and stochastic cutting
plane) cannot be applied for solving it. 
In the next section, we focus
on addressing the following challenge. %{\color{red}{(kam konam ino)}}
\fi
\begin{framed}%
\noindent \textbf{Challenge 1.} Exploit the specific structure of
Problem $\mathcal{P}_{2}$ to find a global optimal solution for this
non-convex stochastic optimization problem whose objective has no closed-form expression. %and whose optimization variables contain combinatorial variables.
\end{framed}

\section{Solution to Problem $ \mathcal{P}_{2} $} \label{sec: solution}

%HIDDEN CONVEXITY RO INJA NAGU! CHON THE FOLLOWING PROBLEM CONVEX NIST.
%*****As elaborated before, problem $ \mathcal{P}_{2} $ is a non-convex problem which has no closed-form expression for either the objective function or the constraints. Accordingly, it is very difficult to find a necessary and sufficient global optimality condition of this problem. 

%{\color{blue}{

In this section we aim to address Challenge 1.  For this purpose, we first study a hidden convexity of $ \mathcal{P}_{2} $, which can then be exploited to tackle the first challenge and derive the global optimality condition for this problem. Next, based on the global optimality condition, we propose an iterative algorithm to efficiently solve the problem. Note that all the proofs have been provided at the end of the paper in Appendices.

%}}

%solving problem $ \mathcal{P}_{2} $ is highly challenging. The first challenge is that as the utility function of this problem is the solution of another optimization problem, there is no closed form expression for it. The second challenge is that even the constraint of this problem cannot be expressed in any closed form. 

\subsection{Hidden Convexity and Global Optimality Condition of $ \mathcal{P}_{2} $}

The average data rate region can be defined as:
\begin{equation}\label{def_R}%\small
\mathcal{R} \triangleq  \bigcup_{      \boldsymbol{q} \in \Lambda_{\boldsymbol{q}} , ~ \boldsymbol{\rho} \in \Lambda_{\boldsymbol{\rho}}    } \left\{\boldsymbol{r} \in \mathbb{R}_{+}^{L}: \boldsymbol{r} \leq \overline{\boldsymbol{r}} \left( \boldsymbol{q}, \boldsymbol{\rho} \right) \right\} ,
\end{equation}
\iffalse ***
\begin{equation}\label{def_R}\small
\mathcal{R} =  \bigcup_{\Omega \in \Lambda} \left\{\boldsymbol{\nu} \in R_{+}^{L}: \boldsymbol{\nu} \leq \overline{\boldsymbol{r}}(\Omega)  \right\} ,
\end{equation}
\fi
%where $ \overline{\boldsymbol{r}}(\Omega) = [\overline{r}_{1}(\Omega),\ldots,\overline{r}_{L}(\Omega)]^{T} $ in which $ \overline{r}_{l}(\Omega) = \sum\limits_{j=1}^{|\boldsymbol{A}|} q_{j} r_{l}(\boldsymbol{a_{j}} , \boldsymbol\rho) ,  ~ \forall l=1,\ldots,L$, as earlier defined in \eqref{ave_rate_DTXpolicy}.
Using this definition, Lemma \ref{theorem: P_2=P_E}    
shows the relationship between problem $ \mathcal{P}_{2} $ and the following optimization problem:
\begin{equation}\label{opt_1_Equivalent}
\mathcal{P}_{E}: \hspace{5 pt} \max_{  \boldsymbol{r}  \in \mathcal{R} }  \widetilde{U}( \boldsymbol{r} ).
\end{equation}

\begin{lemma} \label{theorem: P_2=P_E}
Suppose that $ \left( \boldsymbol{q}^\ast, \boldsymbol{\rho}^\ast \right)  $ is the global optimal solution of $ \mathcal{P}_{2} $. Then $ \overline{\boldsymbol{r}}\left( \boldsymbol{q}^\ast, \boldsymbol{\rho}^\ast \right) $ is the optimal solution of $ \mathcal{P}_{E} $; and if $ \boldsymbol{r}^{\ast} $ is the optimal solution of $ \mathcal{P}_{E} $, then any $ \left( \boldsymbol{q}^\ast, \boldsymbol{\rho}^\ast \right) $  satisfying $ \overline{\boldsymbol{r}}\left( \boldsymbol{q}^\ast, \boldsymbol{\rho}^\ast \right)=\boldsymbol{r}^{\ast} $ is the global optimal solution of $ \mathcal{P}_{2} $.   
\end{lemma}

%***\proof Refer to Appendix \ref{ap: proof of theorem: P_2=P_E}.

%{\color{red}{(Bring its proof here if it results in saving some space. OR omit this lemma and just mention it in the main text, where (the proof is omitted due to the limited space).)}}

Moreover, the following proposition shows that the equivalent problem $ \mathcal{P}_{E} $ is a convex problem.

\begin{proposition}{(Convexity of Problem $ \mathcal{P}_{E} $)} \label{th: Convexity of P_E}
In problem $ \mathcal{P}_{E} $, the objective function $ \widetilde{U}(\boldsymbol{r}) $ is concave and the feasible set $ \mathcal{R} $ is a convex set. Hence, $ \mathcal{P}_{E} $ is convex.
\end{proposition}

}}

%\textbf{Theorem 2.} (Convexity of Problem $ P_{E} $). The objective function $ \widetilde{U}(\overline{\boldsymbol{r}}) $ and the feasible set $ R $ in problem $ P_{E} $ are convex function and set, respectively. Hence, $ P_{E} $ is a convex optimization problem.

%***\proof The detailed proof has been provided in Appendix \ref{ap: proof of th: Convexity of P_E}.

%\textbf{Proof.} Refer to Appendix A.

{\color{black}{It should be noted that although problem $ \mathcal{P}_{E} $ is convex, it is not trivial to find its solution because its objective function as well as its feasible set %$ \mathcal{R} $ still 
do not have any closed-form representations. However, the convexity of problem $ \mathcal{P}_{E} $ results in a hidden convexity in problem $ \mathcal{P}_{2} $, which will be utilized in order to tackle the aforementioned Challenge 1. For this purpose, using the first-order optimality condition of problem $ \mathcal{P}_{E} $, we %derive/
propose
 a sufficient global optimality condition for $ \mathcal{P}_{2} $, as stated in Theorem \ref{theorem: Global Optimality Condition of P_2}.

\begin{theorem}{(Global Optimality Condition of $ \mathcal{P}_{2} $)}\label{theorem: Global Optimality Condition of P_2}
A point $      (     \boldsymbol{q}^\ast \in \Lambda_{\boldsymbol{q}} , ~ \boldsymbol{\rho}^\ast  \in \Lambda_{\boldsymbol{\rho}}   )     $ 
%***A physical layer control policy $   {\color{red}{    (     \boldsymbol{q}^\ast \in \Lambda_{\boldsymbol{q}} , ~ \boldsymbol{\rho}^\ast  \in \Lambda_{\boldsymbol{\rho}}   )   }}  $
%$ \left( \boldsymbol{q}^\ast , {\color{red}{\boldsymbol{\rho}^\ast}} \right) {\color{red}{\in \Lambda}} $ 
is a global optimal solution of $ \mathcal{P}_{2} $ if it satisfies the following condition: $ \forall \boldsymbol{q} \in \Lambda_{\boldsymbol{q}} , \forall \boldsymbol{\rho} \in \Lambda_{\boldsymbol{\rho}}, $
%*** A physical layer control policy $ \Omega ^{\ast} = \{\Theta ^{\ast},\boldsymbol{ \rho^{\ast} } \} $, where $ \Theta^{\ast}=\left\{  \boldsymbol{A}^{\ast} , \boldsymbol{q}^{\ast}  \right\} $, is a global optimal solution of $ \mathcal{P}_{2} $ if it satisfies the following condition:
% \at{g^{T}_{ \widetilde{U} } \left( \boldsymbol{r} \right) }{\boldsymbol{r}=\boldsymbol{\overline{r} \left( \Omega^{\ast} \right) }}
\begin{equation}\label{opt_cond_P1}%\small
%*** \at{\boldsymbol{g} ^{T} _{ \widetilde{U} } 
\at{ {\nabla  \widetilde{U} } ^T  \left( \boldsymbol{\overline{r}} \right) }{ \boldsymbol{\overline{r}} = \boldsymbol{\overline{r}} \left( \boldsymbol{q}^\ast, \boldsymbol{\rho}^\ast \right)  } \cdot \left( \boldsymbol{\overline{r}} \left( \boldsymbol{q}^{\ast} , \boldsymbol\rho^{\ast} \right) - \boldsymbol{\overline{r}} \left( \boldsymbol{q} , \boldsymbol\rho  \right) \right) \geq 0, % \quad \forall \boldsymbol{q} \in \Lambda_{\boldsymbol{q}} ,  \forall \boldsymbol{\rho} \in \Lambda_{\boldsymbol{\rho}},
%
%\forall  j=1,\ldots,| \mathcal{A}^{\ast} |, \forall \boldsymbol{a} \in \mathcal{A} , \forall \boldsymbol\rho \in \Lambda_{\boldsymbol{\rho}} \left( \boldsymbol{a} \right),
\end{equation}
\iffalse ***
\begin{equation}\label{opt_cond_P1}\small
\at{\boldsymbol{g} ^{T} _{ \widetilde{U} }  \left( \boldsymbol{\overline{r}} \right) }{ \boldsymbol{\overline{r}} = \boldsymbol{\overline{r}} \left( \boldsymbol{ \Omega^{\ast} } \right)  } \cdot \left( \boldsymbol{r} \left( \boldsymbol{a}^{\ast}_{j} , \boldsymbol\rho^{\ast} \right) - \boldsymbol{r} \left( \boldsymbol{a} , \boldsymbol\rho  \right) \right) \geq 0, ~~~~ \forall  j=1,\ldots,| \boldsymbol{A}^{\ast} |, \forall \boldsymbol{a} \in \mathcal{A} , \forall \boldsymbol\rho \in \Lambda_{\boldsymbol{\rho}} \left( \boldsymbol{a} \right),
\end{equation}
\fi
%\begin{align}\label{opt_cond_P1_2}
%\forall &j=1,\ldots,| \boldsymbol{A}^{\ast} |:  \notag \\ & \boldsymbol{\lambda^{\ast}}^{T}  \cdot \left( \boldsymbol{r} \left( \boldsymbol{a}^{\ast}_{j} , \boldsymbol\rho^{\ast} \right) - \boldsymbol{r} \left( \boldsymbol{a} , \boldsymbol\rho  \right) \right) \geq 0, ~ \forall \boldsymbol{a} \in \mathcal{A} , \forall \boldsymbol\rho \in \Lambda_{\boldsymbol{\rho}}
%\end{align}
where %$ {\color{red}{\nabla  \widetilde{U} }} \at{\boldsymbol{g} _{ \widetilde{U} } \left( \boldsymbol{\overline{r}} \right) }{ \boldsymbol{\overline{r}} = \boldsymbol{\overline{r}} \left( \boldsymbol{ \Omega^{\ast} } \right)  } $
$  \at{ \nabla  \widetilde{U} \left( \boldsymbol{\overline{r}} \right) }{ \boldsymbol{\overline{r}} = \boldsymbol{\overline{r}} \left( \boldsymbol{q}^\ast, \boldsymbol{\rho}^\ast \right)  } $ is the gradient of $ \widetilde{U} $ at point $ \boldsymbol{\overline{r}} = \boldsymbol{\overline{r}} \left( \boldsymbol{q}^\ast, \boldsymbol{\rho}^\ast \right)  $.
%$ \boldsymbol{\overline{r}} = \boldsymbol{\overline{r}} \left( \boldsymbol{ \Omega^{\ast} } \right) $.

%the Lagrangian multipliers vector obtained from solving problem $ \mathcal{P}_{1} $, as aforementioned in Remark \ref{remark: solving P1}.

% $ \boldsymbol{\lambda^{\ast}} \triangleq \boldsymbol {\lambda} \left( \boldsymbol{\overline{r} \left( \Omega^{\ast} \right) } \right) $ is the Lagrangian multipliers vector obtained from solving problem $ \mathcal{P}_{1} $, as aforementioned in Remark \ref{remark: solving P1}. %!!!!!
% and $ \boldsymbol{\omega^{\ast}} \triangleq g^{T}_{ \widetilde{U} } \left( \boldsymbol{\overline{r} \left( \Omega^{\ast} \right) } \right) $
\end{theorem}

}}

%\textbf{Theorem 3.} (Global Optimality Condition of $ P_{1} $). \textit{A control policy $ \Omega^{\ast}=\left\{  \boldsymbol{\Gamma^{\ast}} , \boldsymbol{q}^{\ast}  \right\} $ is a global optimal solution of $ P_{1} $ if and only if it satisfies the following condition:    }

%\begin{align}\label{opt_cond_P1}
%\forall j=1&,\ldots,|\boldsymbol{\Gamma^{\ast}} |:  \hspace{15pt}  \notag \\ &\sum_{l=1}^{L} \omega_{k} \at{\frac{d \widetilde{U}(r) }{dr}}{r=\overline{r}_{l}(\Omega^{\ast})} (r_{l}(\Gamma^{\ast}_{j}) - r_{l}(\Gamma) ) \geq 0, \hspace{5pt} \forall \Gamma \in \Lambda_{\Gamma}
%\end{align}

%***\proof The detailed proof can be found in Appendix \ref{ap: prrof of theorem: Global Optimality Condition of P_2}.

{\color{black}{

\subsection{Globally Optimal Solution of $ \mathcal{P}_{2} $}

\iffalse *****
Although Theorem \ref{theorem: Global Optimality Condition of P_2} introduces a sufficient condition for the global optimal solution of problem $ \mathcal{P}_{2} $, achieving a solution that satisfies this condition is still non-trivial and challenging. Therefore,  in order to find the optimal solution of the problem, the following challenge needs to be addressed as well. %we need to address the following challenge 
\begin{framed}
\noindent\textbf{Challenge 2.} Find an algorithm to achieve a solution that satisfies the global optimality condition in Theorem \ref{theorem: Global Optimality Condition of P_2}.
\end{framed}

\fi

In the rest of this section, we propose an iterative algorithm for iteratively updating the optimization variables and the gradient vector such that the global optimality condition stated in Theorem \ref{theorem: Global Optimality Condition of P_2} is achieved. % by iteratively updating the optimization variables and the gradient vector.

\iffalse ***
In the rest of this section, using the global optimality condition in Theorem \ref{theorem: Global Optimality Condition of P_2}, we propose an algorithm that iteratively updates the optimization variables %$ \boldsymbol{A}$, $\boldsymbol{q} $ and $ \boldsymbol\rho $ and a weight vector $ \boldsymbol{\omega}=[\omega_{1},\ldots,\omega_{L} ]^{T} $ 
in  a way that the globally optimal solution of $ \mathcal{P}_{2} $ is achieved. 
\fi

Algorithm \ref{alg 1}  shows the pseudo-code of the proposed solution. The indices $ t $ and $ i $ are for subframes and superframes, respectively. For updating the short-term control variables at each subframe $ t \in [ (i-1)T_{s}+1, iT_{s} ]~  $ of the $ i^{th} $ superframe, each BS randomly selects a DTX pattern from the DTX patterns profile  $ \mathcal{A} $ based on the DTX time-sharing vector %*****$ \boldsymbol{q}^{(i)} $ 
of the current superframe. Then the link scheduling is chosen to maximize the weighted rate as in \eqref{equ: link sched}. It should be noted that the DTX patterns generated at different BSs are the same, as explained before. 
%since they use the same pseudo random generators (i.e., they use the same seed, e.g., they all use the subframe index as the seed, along with the same probability profile  $ \boldsymbol{q}^{(i)} $).
%and thus the DTX patterns generated at different BSs are the same for each subframe.

For updating the long-term control variables at the end of each superframe $ i $, first, all BSs feed back the average rate of their outgoing links %during that superframe 
to the RRMS. Then, the RRMS updates the DTX time sharing vector by Procedure I in section \ref{sec: Proc I}. %(as will be discussed later).
 Finally, for the updated DTX control and link scheduling policy, the routing and flow control variables and the weight vector (Lagrangian multipliers of  $ \mathcal{P}_1 $) %$ \boldsymbol{w} $ 
are updated by solving Problem $ \mathcal{P}_{1} $ %, under current physical layer control policy $ \Omega $ 
using the  primal-dual method. Note that as earlier mentioned in Remark \ref{remark: solving P1}, the weight vector is actually the gradient of the objective function $ \widetilde{U}(\overline{\boldsymbol{r}}) $ at point $\overline{\boldsymbol{r}} \left( \boldsymbol{q}^{(i)}, \boldsymbol{\rho}^{(i)} \right) $. %Note that using primal-dual method, the 

%, which also finds the  Lagrangian multipliers $\boldsymbol\lambda $ of $ \mathcal{P}_{1} $. $\boldsymbol\lambda $ is the gradient of the objective function $ \widetilde{U}(\overline{\boldsymbol{r}}) $ at $\overline{\boldsymbol{r}} \left( \Omega \right) $, and it is input to Procedure I for the next iteration.  

%*****{\color{red}{shayad hazf: }} The iterations in Algorithm \ref{alg 1} will continue until convergence, which is checked by the termination condition at line {\color{red}{???}. OR: These steps are done iteratively until convergence, which will be checked by the termination condition}.

}}

\begin{algorithm}[] \small
\begin{algorithmic}[1]
\STATE { \textbf{Initialization:}}
\STATE {Set $ i=1 $, and let  $ \boldsymbol{q}^{(1)}=[ q^{(1)}_{1},\ldots,q^{(1)}_{\left | \mathcal{A}^{(i)} \right | } ]^{T}  = \dfrac{1}{|\mathcal{A}|} \left[ 1, \ldots, 1 \right]^T %\left[ \begin{smallmatrix} 1 & 1 & \cdots & 1 \end{smallmatrix} \right]^{T}
 $ and $ \boldsymbol{\omega}^{(1)} = \left[ \omega^{(1)}_1, \ldots , \omega^{(1)}_L \right]^{T} = \left[ 1, \ldots, 1 \right]^T   $ %$ \boldsymbol{\omega}^{(1)} = \left[ \begin{smallmatrix} 1 & 1 & \cdots & 1 \end{smallmatrix} \right]^{T} $. 
  }
%*** \STATE {\hspace{10pt}Set $ i=1 $, and let $ \boldsymbol{A}^{(1)} = \{ \boldsymbol{a}_{1} \in \mathcal{A} \} $ and $ \boldsymbol{q}^{(1)}= [ q^{(1)}_{1}] = 1 $.} % ... and let $ \boldsymbol{\omega}^{(1)}=\boldsymbol{1}_{L \times 1} $, ... 

\STATE {Choose proper initial routing control $ \boldsymbol{x}^{(1)} $ and flow control $ \boldsymbol{d}^{(1)} \geq 0 $ such that constraint \eqref{P2_flow_cons} is satisfied. %Let $ \boldsymbol{\omega}^{(1)} = \left[ \begin{smallmatrix} 1 & 1 & \cdots & 1 \end{smallmatrix} \right]^{T} $ .  
 }

%\STATE {\hspace{10pt}Solve problem $ \mathcal{P}_{1} $ to obtain the routing vector $ \boldsymbol{x} $, the \\\hspace{10pt}flow control vector $ \boldsymbol{d} $ and the Lagrangian multipliers $ \boldsymbol\lambda $.}

%\STATE{\hspace{10pt}Let $ \boldsymbol{\omega}^{(1)}=\boldsymbol\lambda $.}

%\STATE{\hspace{10pt}$ i \leftarrow i+1 $.}

\iffalse

\FOR {every $ BS_{n} $}
\STATE{$ l^{\ast}_{m} \triangleq \underset{l \in \mathcal{T}(n) }{\arg\max} ~\omega_{l}^{(i)} \log(1+|h_{l,m}^{(i)}|^{2}p_{l}) $}
\STATE{$ \rho_{l^{\ast}_{m} , m } \leftarrow 1 $}
\ENDFOR

\fi

%\STATE {\textbf{Step 1:}}
%\STATE {\hspace{10pt}Solve problem $ \mathcal{P}_{1} $ to update the routing vector $ \textbf{x} $, the flow control vector $ \boldsymbol{d} $ \\\hspace{10pt}and the Lagrangian multipliers $ \boldsymbol\lambda $.}

%\STATE {\hspace{10pt}Let $ \boldsymbol{\omega}^{(i)} = \boldsymbol\lambda $. }

\STATE {\hspace{10pt}$ i \leftarrow i+1 $.}

\STATE {\textbf{Step 1 (Short-timescale link scheduling update by each BS at each subframe $ t \in [ (i-1)T_{s}+1, iT_{s} ] $): }}

%\STATE {\textbf{At each subframe $ t \in [ (i-1)T_{s}+1, iT_{s} ]~  $}    \\\textit{\textbf{(Short timescale link scheduling optimization at each BS)}}: }
\STATE{Each BS randomly selects a DTX pattern $ \boldsymbol{a} $ according to the current time-sharing vector $ \boldsymbol{q}^{(i-1)} $.  }
\STATE{{\color{black}{For given $ \boldsymbol\omega ^{(i-1)} $, %of that superframe
 the %optimum
 link scheduling policy $ \boldsymbol\rho^{(i)}\left( \boldsymbol{a} , \boldsymbol{H} \right) $ 
%  $ \boldsymbol\rho^{\ast}\left( \boldsymbol{a} , \boldsymbol\omega^{(i)} \right) $ 
at subframe $ t $ is obtained by the following:%}}
\iffalse ***
\begin{equation}\label{equ: link sched} 
\mathrm{At ~ every ~ BS_{n},:~} \rho_{l^{\ast}_{m} , m } \leftarrow \boldsymbol{a}(n) %a_n ^{(i)}
%
%
, \quad \mathrm{where ~} l^{\ast}_{m} \triangleq \arg\max_{l \in \mathcal{T}(n) } ~ \omega_{l}^{(i)} \log \left(1+ \dfrac{| h_{l,m} |^{2} {\color{red}{p_{n} }} }{ 1 + \sum_{l^\prime \in I_l^{(j)} } | h_{l^\prime,m} |^2 ~p^{ (j)}_{n(l^\prime),m}   } \right) %\log \left(1+ \dfrac{| h_{l,m}(t) |^{2}p_{l}}{  \sigma_{l,m} \left( t\right)   } \right)
\end{equation} % \underset{l \in \mathcal{T}(n) }{\arg\max} ~
\fi
\begin{align}\label{equ: link sched} 
&\mathrm{At ~ any ~ BS_{n}:}~ \rho_{l^{\ast}_{m} , m } \leftarrow a_n %\boldsymbol{a}(n) %a_n ^{(i)}
, \quad \mathrm{where ~} l^{\ast}_{m} \triangleq  \notag \\
&\arg\max_{l \in \mathcal{T}(n) } ~ \omega_{l}^{(i-1)} \log \left(1+ \dfrac{| h_{l,m} |^{2} p_{n}  }{ 1 + \sum_{l^\prime \in I_{l,m}^{(j)} } | h_{l^\prime,m} |^2 ~p_{n \left(l^\prime \right)}   } \right). %\log \left(1+ \dfrac{| h_{l,m}(t) |^{2}p_{l}}{  \sigma_{l,m} \left( t\right)   } \right)
\end{align} % \underset{l \in \mathcal{T}(n) }{\arg\max} ~
%}}
%$ \rho_{l^{\ast}_{m} , m } \leftarrow 1 $, where $ l^{\ast}_{m} = \underset{l \in \mathcal{T}(n) }{\arg\max} ~\omega_{l} log(1+|h_{l,m}|^{2}p_{l}) $.  
}}} \vspace{-13 pt}
\STATE {\textbf{Step 2 (Long-term controls update by the RRMS %starting from subframe $ (iT_{s}-T_{d}) $ in 
at the end of superframe $ i $): }}

%\STATE{\textbf{At each superframe $ i $ (Starting from subframe $ (iT_{s}-T_{d}) $)} \\\textit{\textbf{(long timescale DTX control policy optimization):}}}

\iffalse ***
\STATE {\textbf{Step 2a:}}

\STATE {Call \textit{Procedure I} with input $ \boldsymbol{\omega}^{(i)} $ to obtain a new control variable $ \boldsymbol{a}^{\ast}( \boldsymbol{\omega}^{(i)} ) $.}
\STATE {\hspace{10pt}Let  $ \boldsymbol{A}^{(i)} = \left\{ \boldsymbol{a}^{\ast}( \boldsymbol{\omega}^{(i)} ) \right\}  \bigcup  \left\{ \boldsymbol{a}_{j}^{(i-1)} \in \boldsymbol{A}^{(i-1)}: q_{j}^{(i-1)} > 0 \right\} $. }
\fi

\STATE {\textbf{Step 2a (DTX time-sharing update):}}

\STATE {Each BS calculates and feeds back the average rate of its outgoing links under each DTX pattern (i.e., $ r_l \left( \boldsymbol{a}^{(j)} , \boldsymbol{\rho}^{(i)} \right), ~ \forall j=1, \ldots, |\mathcal{A}| $, as determined by \eqref{ave_rate_DTXpattern}) to the RRMS.}

{\color{black}{
\STATE {Call \textit{Procedure I} with the fed back average rates as the input %input  {\color{red}{???}} %$ \mathcal{A}^{(i)} = \left\{ \boldsymbol{a}^{(i)}_{1},\ldots,\boldsymbol{a}^{(i)}_{\left | \mathcal{A}^{(i)} \right | } \right\} $ 
to update the DTX time-sharing vector $ \boldsymbol{q}^{(i)} $. } 
%***$ \boldsymbol{q}^{(i)}=[ q^{(i)}_{1},\ldots,q^{(i)}_{\left | \mathcal{A}^{(i)} \right | } ]^{T} $.}
%\STATE{\hspace{10pt} Let $ \boldsymbol{\Gamma}^{(i)} = \left\{ \Gamma^{\ast}( \boldsymbol{\omega}^{(i)} ) \right\} $ }

{\color{black}{
\STATE {\textbf{Step 2b (routing and flow control update):}}
\iffalse ***
\STATE {Let $ \Theta^{(i)} = \left\lbrace \boldsymbol{A}^{(i)} , \boldsymbol{q}^{(i)} \right\rbrace $ and $ \Omega^{(i)}=\left\{ \Theta^{(i)},\boldsymbol\rho^{(i)} \right\} $,where the link scheduling policy $ \boldsymbol\rho^{(i)} $ is determined by \eqref{equ: link sched}.}
\fi

%\STATE {\hspace{10pt}Call \textit{Procedure III} with input $ \Omega^{(i)}=\left\{ \Theta^{(i)},\boldsymbol\rho^{(i)} \right\} $ to update the gradient vector \\\hspace{10pt}$ \boldsymbol\omega^{(i+1)}=[ \omega^{(i+1)}_{1},\ldots,\omega^{(i+1)}_{L} ]=\at{ \nabla \widetilde{U}(\boldsymbol{r}) }{\boldsymbol{r}=\overline{\boldsymbol{r}}(\Omega^{(i)})} $.} 
\STATE {{Solve problem $ \mathcal{P}_{1} $ with input $ \bar{\boldsymbol{r}} \left( \boldsymbol{q}^{(i)} , \boldsymbol{\rho}^{(i)}\right) $, where $ \bar{r}_l \left( \boldsymbol{q}^{(i)} , \boldsymbol{\rho}^{(i)}\right)  = \sum_{j=1} ^{|\mathcal{A}|} q_j^{(i)} r_l \left( \boldsymbol{a}^{(j)} , \boldsymbol{\rho}^{(i)} \right), ~ \forall j=1, \ldots, |\mathcal{A}|  $ %***which can be calculated using {\color{red}{the conditional average rates fed \\\hspace{10pt}back from the BSs to the RRMS (as mentioned in Line 10)}},
%with input $ \Omega^{(i)}=\left\{ \Theta^{(i)},\boldsymbol\rho^{(i)} \right\} $ 
 to  obtain the updated routing vector $ \boldsymbol{x}^{(i)} $, the flow control vector $ \boldsymbol{d}^{(i)} $ and the weight vector  %multipliers 
$ \boldsymbol{\omega}^{(i)} $. %{\color{red}{ or directly $ \boldsymbol\omega^{(i+1)}  $ ???}}  
}}   
}}
}}
%***\STATE {\hspace{10pt}{\color{blue}{Update the weight vector: $ \boldsymbol\omega^{(i+1)} \leftarrow \boldsymbol\lambda^{(i)} $.
%$ \boldsymbol\omega^{(i+1)}=[ \omega^{(i+1)}_{1},\ldots,\omega^{(i+1)}_{L} ] \leftarrow \boldsymbol\lambda^{(i)} $.  
%***}} }

\STATE {%After updating the long-term control variables based on these average rates, the RRMS 
The RRMS sends the updated long-term control variables $ \boldsymbol{q}^{(i)}, \boldsymbol{x}_l^{(i)}, ~\forall l \in \mathcal{T}(n)  $ and the weights  $ \omega_l^{(i)} , \forall l \in ~\mathcal{T}(n) $ %for all outgoing links of 
to $ \mathrm{BS}_n $  for the next superframe.
}

%, where $ \overline{\boldsymbol{r}}(\Omega^{(i)}) = \sum_{j=1}^{\left | \boldsymbol\Gamma^{(i)} \right |}q_{j}^{(i)} \boldsymbol{r}( \Gamma_{j}^{(i)} )$

\STATE {\textbf{Termination:}}
\STATE {If  $ \left | \widetilde{U}(\Omega^{(i)}) - \widetilde{U}(\Omega^{(i-1)}) \right | \leq \epsilon $, where $ \epsilon >0 $ is a given small number, then the algorithm is terminated. Otherwise, return to line 4. %with the optimum physical layer control policy $ \Omega^{\ast} = \left\{ \Theta^{\ast},\boldsymbol\rho^{\ast} \right\} $, where $\Theta^{\ast} = \left\{ \boldsymbol{A}^{\ast},\boldsymbol{q}^{\ast} \right\} $, $ \boldsymbol{A}^{\ast} = \left\{ \boldsymbol{a}_{j}^{(i)}: q_{j}^{(i)} > 0 \right\} $ and $ \boldsymbol{q}^{\ast} = \left[ q_{j}^{(i)} > 0 \right] $. 
} %Otherwise, return to line 4.}
%\STATE {Return to line 4.}

\iffalse ***
\STATE {\hspace{10pt}If  $ \left | \widetilde{U}(\Omega^{(i)}) - \widetilde{U}(\Omega^{(i-1)}) \right | \leq \epsilon $, where $ \epsilon >0 $ is a small number, then the algorithm is terminated with the optimum physical layer control policy $ \Omega^{\ast} = \left\{ \Theta^{\ast},\boldsymbol\rho^{\ast} \right\} $, where $\Theta^{\ast} = \left\{ \boldsymbol{A}^{\ast},\boldsymbol{q}^{\ast} \right\} $, $ \boldsymbol{A}^{\ast} = \left\{ \boldsymbol{a}_{j}^{(i)}: q_{j}^{(i)} > 0 \right\} $ and $ \boldsymbol{q}^{\ast} = \left[ q_{j}^{(i)} > 0 \right] $. Otherwise, return to line 4.}
\fi

\end{algorithmic}
\caption{Iterative Algorithm for Solving $ \mathcal{P}_{2} $ }
\label{alg 1}
\end{algorithm}

{\color{blue}{

{\color{black}{
%***\vspace{8pt}
\subsubsection{\textbf{Procedure I (Update of time-sharing vector $ \boldsymbol{q} $)}}\label{sec: Proc I}

%This procedure tries to optimize the time sharing among the current control variables in $ \boldsymbol{A}^{(i)} $. Specifically, for each control variable $ \boldsymbol{a}_{j}^{(i)} \in \boldsymbol{A}^{(i)} $, procedure II allocates the optimum fraction of a super frame, or equivalently the optimum number of subframes in a superframe, that will use this control variable $ \boldsymbol{a}_{j}^{(i)} $. %These optimum time sharing 
   
%***Having the set of chosen DTX control variables $ \boldsymbol{A}^{(i)} = \left\lbrace \boldsymbol{a}_{j}^{(i)}, j=1,\cdots, |\boldsymbol{A}^{(i)}| \right\rbrace $, along with the associated data rate of links $ r_{l} \left(\boldsymbol{a}_{j}^{(i)}, \boldsymbol\rho^{\ast}  \left( \boldsymbol{a}^{(i)} , \boldsymbol\omega ^{(i)} \right) \right), ~ \forall l $, which are fed back from the head BSs of the links to the RRMS, 
This procedure obtains the updated time-sharing vector $ \boldsymbol{q}^{(i)} $ by solving the following problem: %for fixed $ \boldsymbol{A}^{(i)} $:
%
%\begin{align}\label{q_Proc}
%& \underset{ \boldsymbol{q}=[ q_{1},\ldots,q_{|\boldsymbol\Gamma |} ]^{T} }{\max} \widetilde{U}(\overline{\boldsymbol{r}})
%\\& s.t.  \notag
%\\& ~ q_{j} \in [0,1],~ \forall j, ~ \sum_{j=1}^{|\boldsymbol\Gamma |} q_{j} = 1 \notag
%\\& \overline{\boldsymbol{r}} = [\overline{r}_{1},\ldots \overline{r}_{L} ]^{T}, where ~ \overline{r}_{l}=\sum_{j=1}^{|\boldsymbol\Gamma |} q_{j} r_{l}(\Gamma_{j}), ~ \forall l=1,\ldots,L \notag
%\end{align}
%
\begin{equation}\label{q_Proc}
\max_{ \boldsymbol{q} \in \Lambda_{\boldsymbol{q}} } \widetilde{U} \left( \overline{\boldsymbol{r}} (\boldsymbol{q}, \boldsymbol{\rho}^{(i)} )  \right),
\end{equation}
%
\iffalse *****
\begin{equation}\label{q_Proc}
\underset{ \boldsymbol{q}=[ q_{1},\ldots,q_{|\boldsymbol{A}^{(i)} |} ]^{T} }{\max} \widetilde{U}( \overline{\boldsymbol{r}} ),
\end{equation}
\hspace{20 pt}\textit{subject to:}
\begin{equation}
q_{j} \geq 0,~ \forall j, ~ \sum_{j=1}^{|\mathcal{A}^{(i)} |} q_{j} = 1, \notag
\end{equation}
***** \fi
where  $ \boldsymbol{\rho}^{(i)}  $ is the link scheduling policy in the $ i^{th} $ superframe as determined by \eqref{equ: link sched}, % in Step 1 of the algorithm.
 and $ \overline{\boldsymbol{r}} (\boldsymbol{q}, \boldsymbol{\rho}^{(i)} )  = \sum_{j=1}^{ | \mathcal{A} | }  q_j r_l \left(\boldsymbol{a}^{(j)}, \rho^{(i)} \right)  $, where $ r_l \left(\boldsymbol{a}^{(j)}, \rho^{(i)} \right)  $ is given in \eqref{ave_rate_DTXpattern}. 
% is determined by the head BS of each link $ l $ using \eqref{ave_rate_DTXpattern} and is fed back to the RRMS as the input to Procedure I. % (as stated {\color{red}{in line 10}} of the algorithm).
%***  $ \overline{\boldsymbol{r}} = [\overline{r}_{1},\ldots\overline{r}_{L} ]^{T} $ and  $  \overline{r}_{l} =  \sum_{j=1}^{|\mathcal{A}^{(i)} |} q_{j} r_{l}( \boldsymbol{a}^{(j)} , \boldsymbol{\rho}^{(i)} ) $ in which, $ r_{l}( \boldsymbol{a}^{(j)} , \boldsymbol{\rho}^{(i)}  )  $ is  the average rate of link $ l $ under  DTX pattern $ \boldsymbol{a}^{(j)} $ and {\color{red}{link scheduling policy in the $ i^{th} $ superframe $ \boldsymbol{\rho}^{(i)}  $.
%$ \boldsymbol{\rho}^{(i)} \left( \boldsymbol{a} , \boldsymbol{H} \right)  $, as is defined as }}
It is easily verified that problem \eqref{q_Proc} is a convex %*****deterministic 
optimization problem. %and is convex with respect to the optimization variable $ \boldsymbol{q} $. 
Therefore, it can be easily solved by the RRMS using the existing convex optimization methods (e.g., primal-dual interior point methods)
 \cite{boyd2009convex}. %as well.

\iffalse *****
\begin{equation}\label{equ: r for optimising q}
r_{l}^{(j)} = 
%\left( \boldsymbol{a_{j}}^{(i)} , \boldsymbol\rho^{\ast}  ( \boldsymbol{a}^{(i)} , \boldsymbol\omega ^{(i)}) \right) 
\sum_{m=1}^{M} \mathbb{E} \left[ {\color{red}{\rho_{l,m}^{(i)} (\boldsymbol{a}^{(j)},\boldsymbol{H}) }} ~ \log \left(1+ \dfrac{| h_{l,m} |^{2} {\color{red}{p_{n(l)} }} }{ 1 + \sum_{l^\prime \in I_l^{(j)} } | h_{l^\prime,m} |^2 ~p^{ (j)}_{n(l^\prime),m}   } \right) \right]
\end{equation}
***** \fi 
%***$ ~ \overline{r}_{l} =  \sum_{j=1}^{|\mathcal{A}^{(i)} |} q_{j} r_{l} \left( \boldsymbol{a_{j}}^{(i)} , \boldsymbol\rho^{\ast}  ( \boldsymbol{a}^{(i)} , \boldsymbol\omega ^{(i)}) \right) $.

%left( \boldsymbol{a_{j}}^{(i)} , \boldsymbol\rho^{\ast}  ( \boldsymbol{a}^{(i)} , \boldsymbol\omega ^{(i)}) \right) 

%\begin{equation}
%\overline{\boldsymbol{r}} = [\overline{r}_{1},\ldots\overline{r}_{L} ]^{T}, \mathrm{~where} ~ \overline{r}_{l} = \hspace{-3 pt} \sum_{j=1}^{|\boldsymbol{A}^{(i)} |} q_{j} r_{l} \left( \boldsymbol{a_{j}}^{(i)} , \boldsymbol\rho^{\ast}  ( \boldsymbol{a}^{(i)} , \boldsymbol\omega ^{(i)}) \right). \notag
%\end{equation}

%$ \overline{\boldsymbol{r}}^{(i)} $ is the conditional average rate vector under 

\begin{remark}
Procedure I requires the conditional average rates $ r_l \left(\boldsymbol{a}^{(j)}, \rho^{(i)} \right), ~ \forall l \in \mathcal{T}(n), \forall j  $ as the input, which can be calculated at the RRMS if the distribution of small-scale fading is known. When the distribution of the small-scale fading is not available, the conditional average rates can be calculated at each BS using the running sample average over the subframes of the $ \mathrm{i}^{th} $ superframe and then fed back to the RRMS.

\end{remark}
}}
}}

%\begin{figure}
%\begin{centering}
%\includegraphics[scale=0.5]{pic/real-struc.pdf}
%\par\end{centering}
%\caption{\label{fig:real}Overall relationship among the main problem and the subproblems.}
%\end{figure}

%{\color{red}{Figure \ref{fig:Signaling_SuperFrame} illustrates the overall relationship between the main problem and the subproblems, as well as the feedback signals between BSs and the RRMS.}}

\subsubsection{\textbf{Convergence and Optimality  of Algorithm 1}}

The following theorem states that Algorithm \ref{alg 1} converges to the global optimal solution of the original problem $ \mathcal{P}_{org} $. The proof is obtained by showing that this algorithm updates the control variables in such a way that the global optimality condition in Theorem \ref{theorem: Global Optimality Condition of P_2} is satisfied. Please refer to Appendix \ref{App: proof of theorem: Global Optimality of Algorithm 1} for the details. %For this purpose, the convergence and optimality of Algorithm \ref{alg 1}.

\begin{theorem}{(Global Optimality of Algorithm 1)}\label{theorem: Global Optimality of Algorithm 1}
Let $ \left( \boldsymbol{x}^{(i)},  \boldsymbol{d}^{(i)},  \boldsymbol{q}^{(i)},  \boldsymbol{\rho}^{(i)} \right) $ be the output of the $ i^{th} $ iteration of Algorithm 1. We will have $ \left( \boldsymbol{x}^{(i)},  \boldsymbol{d}^{(i)},  \boldsymbol{q}^{(i)},  \boldsymbol{\rho}^{(i)} \right), ~ \forall i $, satisfying all the constraints in \eqref{flow_cons}-\eqref{equ: PHY feasibility set} and 
{\small{\begin{equation}\label{equ: global optimality}
\lim_{i \rightarrow \infty} U \left( \boldsymbol{d}^{(i)}  \right)  = U^{\ast},
\end{equation}}}
\iffalse *****
{\small{\begin{equation}\label{equ: global optimality}
\lim_{i \rightarrow \infty} {\color{red}{\widetilde{U} \left( \bar{\boldsymbol{r}} (\boldsymbol{q}^{(i)},  \boldsymbol{\rho}^{(i)} ) \right) }} = \widetilde{U}^{\ast},
\end{equation}}}
\fi
%where $ \widetilde{U}^{\ast} $ is the global optimal value of $ \mathcal{P}_{2} $.
where $U^{\ast} $ is the global optimal value of $ \mathcal{P}_{org} $.

%\begin{equation}\label{equ: opt gap}
%\widetilde{U}^{\ast} - \widetilde{U}^{\ast}_{alg1} \leq \boldsymbol{g}^T _{\widetilde{U}^{\ast}_{alg1}} \cdot \left( \boldsymbol{y}^\ast_{alg1} - \boldsymbol{x}^\ast_{alg1} \right)
%\end{equation}
%in which, $ \boldsymbol{x}^\ast_{alg1} $ and $ \boldsymbol{y}^\ast_{alg1} $ are the convergence points of algorithm 1, where $ \widetilde{U}\left( \boldsymbol{x}^\ast_{alg1} \right) = \widetilde{U}^{\ast}_{alg1} $. Moreover, $ \boldsymbol{g}^T _{\widetilde{U}^{\ast}_{alg1}} $ is the gradient of $ \widetilde{U} $ at the convergence point $ \boldsymbol{x}^\ast_{alg1} $.

% and algorithm 1 monotonically increases the utility $ \widetilde{U} \left( \Omega^{(i)} \right) $ 

\end{theorem}

%***\proof The detailed proof can be found in Appendix \ref{App: proof of theorem: Global Optimality of Algorithm 1}.

%Finally, we present the following special case, in which the optimality gap is zero, i.e., the proposed algorithm converges to the optimal solution.
%
%\begin{corollary}{(Sufficient Conditions for optimality of the Proposed Algorithm)}\label{theorem: Special Case}
%ddd
%\end{corollary}

{\color{black}{
\section{Implementation Considerations} \label{sec: sig flow}

\subsection{Signalling Flow}
According to the description of the proposed algorithm in the previous section, the signalling flow between the RRMS, macro/pico BSs and MUs within each superframe or each subframe can be illustrated as follows. 
%
% has been shown in Figs. \ref{fig:Signaling_SuperFrame} and \ref{fig:Signaling_sub}. 
%

\iffalse
\begin{figure*}[t!]
\centering
\begin{minipage}[t]{.5\textwidth}
  \centering
	\includegraphics[width=0.8\textwidth]{pics/Signaling_Sup.pdf} %Signaling_SuperFrame_v4.pdf
	%\includegraphics[width=0.45\textwidth]{pics/Signaling_SuperFrame_v4.eps}
	\caption{Signalling flow at each superframe (i.e., slow signalling). }
	\label{fig:Signaling_SuperFrame}
\end{minipage}%
\begin{minipage}[t]{.45\textwidth}
  \centering
  \includegraphics[width=0.8\textwidth]{pics/Signaling_sub.pdf} %Signaling_sub_v3.pdf
	%\includegraphics[width=0.35\textwidth]{pics/Signaling_sub_v3.eps}
	\caption{Signalling flow at each subframe (i.e., fast signalling).  }
	\label{fig:Signaling_sub}
\end{minipage}%\vspace{-10 pt}
\end{figure*}%\vspace{-20 pt}
\fi

%
\iffalse
\begin{figure}[t]
\centering
\includegraphics[width=0.45\textwidth]{pics/Signaling_SuperFrame_v4.pdf}
%\includegraphics[width=0.45\textwidth]{pics/Signaling_SuperFrame_v4.eps}
\caption{\small{Signalling flow at each superframe \textit{(slow signalling)}.}}
\label{fig:Signaling_SuperFrame}
\end{figure}
\fi
%
%Figs. \ref{fig:Signaling_beg_sup} and \ref{fig:Signaling_end_sup} show the \textit{slow timescale} signalling flows that are being passed between RRMS and each BS at the beginning and end of each superframe, respectively. 
%
Fig. \ref{fig:Signaling_SuperFrame} shows the \textit{slow-timescale} signalling flows that are being passed between the RRMS and each BS at the end of each superframe $ i $: %As can be seen in this figure, at the end of the $ i^{th} $ super frame:
\begin{enumerate}
\item Each $ BS_{n} $ calculates the average rates of its outgoing links,  $ r_{l} \left(\boldsymbol{a}, \boldsymbol\rho^{(i)}  \right), ~ \forall \boldsymbol{a} \in \mathcal{A} $,  %***{\color{red}{under each DTX pattern}} $ r_{l} \left(\boldsymbol{a}, \boldsymbol\rho^{\ast}  \left( \boldsymbol\omega \right) \right), ~ \forall \boldsymbol{a} \in \mathcal{A} $, 
and reports them to the RRMS.
\iffalse
\item {\color{red}{Each $ BS_{n} $ sends its associated row of the node-link incident matrix (i.e., $ G_{n,l}, ~ \forall l $) to the RRMS so that the RRMS can have the updated information about the topology of the network. (Not mention it?)}} Moreover, $ BS_{n} $ calculates the average rate of its outgoing links {\color{red}{under each DTX pattern}} %***, $ r_{l} \left(\boldsymbol{a}, \boldsymbol\rho^{\ast}  \left( \boldsymbol\omega \right) \right), ~ \forall \boldsymbol{a} \in \mathcal{A} $, 
and reports them to the RRMS. % for further central calculation and decision making.
%***\item Then, the RRMS updates the long-term control variables based on these average rates.
%***\item Finally, the RRMS broadcasts the updated long-term control variables and the weight vector $ \boldsymbol\omega $ to the BSs, for the next superframe.
\fi
\item After updating the long-term control variables based on these average rates, the RRMS sends the updated long-term control variables $ \boldsymbol{q}^{(i)}, \boldsymbol{x}_l^{(i)}, ~ \forall l \in \mathcal{T}(n) $, and the weights  $ \omega_l^{(i)}, ~ \forall l \in \mathcal{T}(n) $ for all outgoing links of $ \mathrm{BS}_n $ to it for the next superframe.

\end{enumerate}

\begin{figure}[]
\centering
\includegraphics[width=0.4\textwidth]{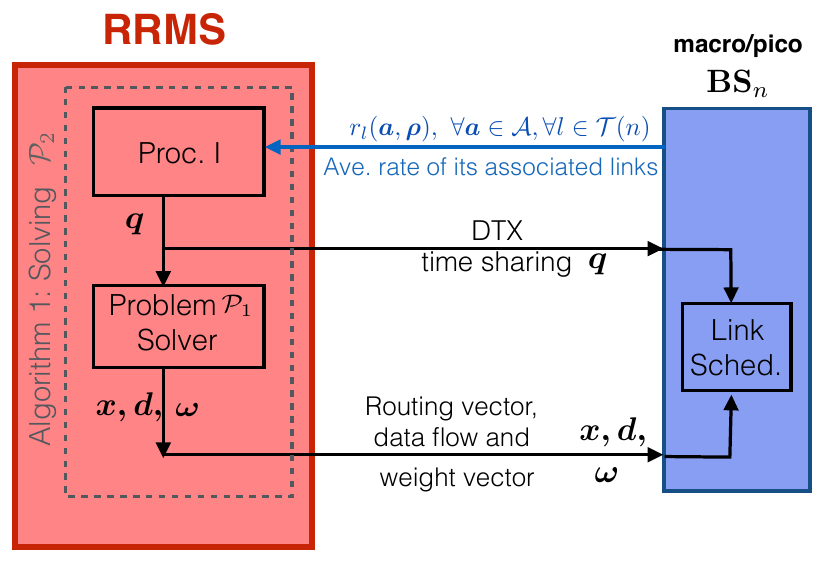} %.eps %Signaling_SuperFrame_v4.pdf
	\caption{Signalling flow at each superframe (i.e., slow signalling). }
	\label{fig:Signaling_SuperFrame}
\end{figure}

\begin{remark}
To provide robustness to the backhaul latency in practical implementations, we allow $ T_{d} $ subframes before the end of each superframe to start to do all of the above long-timescale signalling and calculations. The parameter $ T_{d} $ is chosen as a sufficient time for calculation and message passing. Moreover, since $ T_d $ subframes is a small portion of a superframe, it is much less than the coherence time of the channels statistics that are considered to be constant over a long time (i.e., over a large number of superframes). Accordingly, the delay imposed will have no effect on the accuracy of the algorithm.

%This timing configuration guarantees our design to be robust to the backhaul latency.

%Moreover, $ T_d $ subframes is a small portion of a superframe, and hence is much less than the coherence time of the channels statistics. Accordingly, the delay imposed will have no effect on the accuracy of the algorithm.

\end{remark}

%The RRMS starts to do all of this at T_d subframes before the end of each super frame, where T_d is chosen to allow sufficient time for calculation and signaling, This will also show that our design is not sensitive to the backhaul latency.

Fig. \ref{fig:Signaling_sub} shows the \textit{fast-timescale} message passing between the head and tail of each link at each subframe. Note that as we have considered downlink transmission, the head node is always a BS, while the tail node can be either an MU or another BS.  %According to the proposed algorithm and as illustrated in this figure as well, 
At each subframe, each tail node needs to feed back its %instantaneous CSI / 
received SINR over each subband to its transmitting BS so that the transmitting BS can update the short-term link scheduling of its corresponding links accordingly. %, using \eqref{equ: link sched}

%It is clear that the head node of each link is always a BS (macro or pico), while the tail node can be either a BS or an MU.

\begin{figure}[]
\centering
\includegraphics[width=0.35\textwidth]{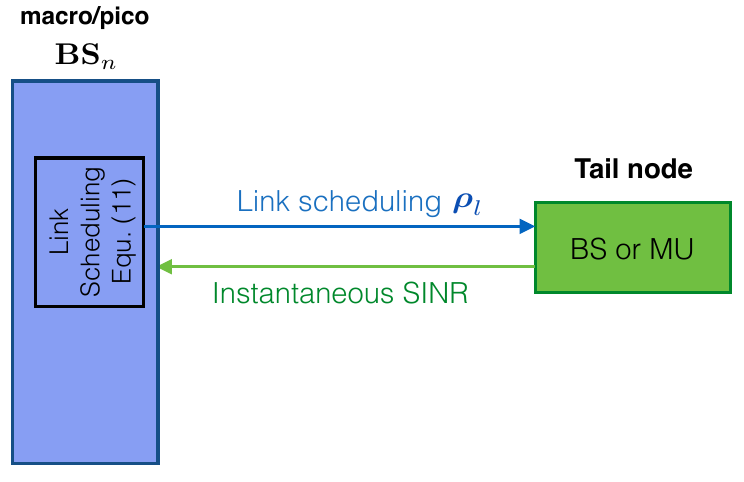} %.eps %Signaling_sub_v3.pdf
	\caption{Signalling flow at each subframe (i.e., fast signalling). }
	\label{fig:Signaling_sub}
\end{figure}

%\begin{figure}[t]
%\centering
%\includegraphics[width=0.7\textwidth]{pic/Signaling_beg_sup_v4.pdf}
%\caption{Signalling flow at the beginning of superframe \textit{(slow signalling)}.}
%\label{fig:Signaling_beg_sup}
%\end{figure}
%
%\begin{figure}[t]
%\centering
%\includegraphics[width=0.7\textwidth]{pic/Signaling_end_sup_v5.pdf}
%\caption{Signalling flow at the end of superframe \textit{(slow signalling)}.}
%\label{fig:Signaling_end_sup}
%\end{figure}

\iffalse
\begin{figure}
\centering
\includegraphics[width=0.38\textwidth]{pics/Signaling_sub_v3.pdf}
%\includegraphics[width=0.35\textwidth]{pics/Signaling_sub_v3.eps}
\caption{Signalling flow at each subframe \textit{(fast signalling)}.}
\label{fig:Signaling_sub}
\end{figure}
\fi

From the above analysis, the proposed hierarchical RRM has several advantages in terms of signalling overhead. First, the RRMS requires only global statistical information, which can be provided and fed back by the BSs in long-timescale signalling. This makes the proposed solution robust to the backhaul latency. On the other hand, in the short timescale, BSs only require the local CSI of their outgoing direct links, which can be provided by local message passing between each BS and its associated MUs. As a result, the proposed hierarchical algorithm benefits from low signalling overhead and message passing among different nodes and  good scalability of the complexity. %*****Therefore, it can be efficiently implemented in real applications and networks. 
In the next subsection, we provide rigorous analysis of the computational complexity and signalling overhead per iteration of the proposed algorithm. %as well.

}}

\iffalse ***
 the BSs and their associated MUs communicate locally with each other and each BS only requires local CSI of its outgoing direct links. %, which can be fed back by the receiving node (receiving BS or MU). 
Therefore, our proposed algorithm has low signalling overhead and message passing among different types of network components, as well as good scalability of the complexity. Hence, it can be efficiently implemented in real applications and networks.
\fi

{\color{black}{

\subsection{Signalling Overhead and Computational Complexity}\label{sec: signalling_complexity} %of the Proposed Algorithm per each Iteration}

In this section, we elaborate the signalling overhead (in terms of the number of bits exchanged among nodes) and the complexity of the proposed algorithm for each step in each iteration (superframe). For simplicity, we assume that each real number is quantized by $ B $ bits.

%Note that each iteration of the proposed algorithm includes {\color{red}{three steps}}:
%*** In the following, we discuss the signalling overhead and the complexity of each of them:
%***For investigating the complexity of the proposed algorithm, in this section we analyse its per-iteration complexity. Moreover, later in the next section, the overall complexity of the proposed algorithm (in terms of the CPU time) will be discussed by simulation results and comparisons to the baselines.

%***Then, we show the overall complexity (in terms of the CPU time) of the proposed algorithm with respect to the size of the problem, i.e., the number of nodes. %{\color{red}{(what about in the simulations? w.r.t. \textit{the number of BSs} (as asked by the reviewer) instead of \textit{the number of all nodes} is ok?)}}

\iffalse ***
\textbf{The complexity of each iteration of the proposed approach: }%per-iteration complexity of the proposed approach, 
Note that each iteration of Algorithm 1 includes four major steps:
\fi

%\subsubsection{Step 1 (link scheduling at each subframe)}

\begin{enumerate}
\item \textbf{Step 1 (link scheduling at each subframe)}: 
%\begin{itemize}

%\item 
\textbf{Signalling Overhead:} For each link $ l=1,\ldots, L $, the received SINR over each subband $ m=1,\ldots, M $ of the tail node should be reported to the head BS. Therefore, the total number of feedback bits per subframe is $ L \times M \times B $, and hence, the total signalling overhead per iteration (i.e., superframe) will be $ L \times M  \times B \times T_s  $ bits.
%the total number of feedbacks is $ L \times M $ per subframe, and hence, the total signalling overhead per iteration (i.e., superframe) would be $ O \left( L \times M \times T_s \right) $.

%\item 
\textbf{Computational Complexity:} %Having the reported received SINR of each outgoing links over each subband, each BS updates their link scheduling. For this purpose, 
\iffalse ***
As shown in \eqref{equ: link sched}, for each subband $ m $, each BS needs to calculate the weighted rates of each outgoing link {\color{blue}{by evaluating the log function of the SINR (as given in \eqref{equ: link sched}) and then multiplying to the corresponding weight (i.e., $ w_l $ for each outgoing link $ l $), and finally, comparing the resulted weighted rates together to find the largest one. Therefore, in total, it takes a complexity of {\color{blue}{$ O \left( \log(B) \times L \times N_{BS} \right) $}}

Evaluating the log function takes $ O \left( \log(B) \times L \times N_{BS} \right) $
This takes a complexity of {\color{blue}{$ O \left( L \times N_{BS} \right) $}}, in total.
}}
\fi
For the link scheduling over each subband $ m $, each BS needs to calculate the weighted rates of its outgoing links by \eqref{equ: link sched} and then comparing the resulting weighted rates to find the largest one. Considering all the BSs, this has a complexity of $ O \left( L \times M  \right) $ per iteration, in total \cite{anderson2005bit}.

\item \textbf{Step 2a (DTX time-sharing update):}
%\begin{itemize}

%\item 
\textbf{Signalling Overhead:} In this step, the average rate of each link $ l=1,\ldots, L $ under each DTX pattern $ j=1,\ldots, |\mathcal{A}| $ is fed back by the associated head BS of the link to the RRMS. Moreover, the RRMS sends the updated DTX time-sharing vector $ \boldsymbol{q} $ to the BSs. Therefore, the required  signalling overhead is $  ( L + N_{BS} ) \times  | \mathcal{A} | \times B $ bits per iteration. %OR: $  O \left( L \times | \mathcal{A} | \times B \right) $ 

%\item 
\textbf{Computational Complexity:} %*****Calculating the aforementioned average rates by  \eqref{ave_rate_DTXpattern} {\color{red}{has a complexity of $ O \left( ??? \right) $}}. Moreover, 
Procedure I that updates DTX time-sharing control $ \boldsymbol{q}^{(i)} $ includes solving a convex optimization problem as in \eqref{q_Proc} with the number of variables equal to $ |\mathcal{A}  | $. Therefore, using primal-dual interior point methods (that are well known to be very efficient for solving convex problems%\cite[Chapter 11]{boyd2009convex}
), this problem can be solved in  $ O ( { |\mathcal{A}  |}^3 ) $ \cite{skiena1998algorithm}, \cite[Chapter 11]{boyd2009convex}. %*****{\color{red}{ $ O \left(  |\mathcal{A}  |^2.373 \right) $}} [REF???]

\item \textbf{Step 2b (routing, flow control and the weights update):}
%\begin{itemize}

%\item 
\textbf{Signalling Overhead:} The RRMS needs to send the updated long-term variables $ \boldsymbol{x}^{(i)} $ and $ \boldsymbol\omega^{(i)} $ to the BSs. Therefore, the signalling overhead of this step is $ (K + 1 ) \times L \times B  $ bits per iteration.

%In order to update the RRMS with the latest topology of the network,  each $ BS_n $ needs to feed back its associated row (i.e., row $ \boldsymbol{g}_n $) of the node-link incident matrix to the RRMS, if there is any change in its connectivity. Therefore, the total number of feedbacks at each subframe is $ N_{BS} \times L $, and hence, the total signalling overhead per iteration  would be $ O \left( N_{BS} \times L \right) $.

%\item 
\textbf{Computational Complexity:} 
Solving $ \mathcal{P}_1 $ by primal-dual interior point methods (as mentioned in Remark \ref{remark: solving P1}) to update the routing, flow control and the weights has a complexity of $ O \left( K^3 L^3 \right) $ \cite{skiena1998algorithm}, \cite[Chapter 11]{boyd2009convex}. 

\iffalse *****
Having the %latest topology of the network and 
average rates $ \bar{\boldsymbol{r}} \left( \boldsymbol{q}^{(i)} , \boldsymbol{\rho}^{(i)}\right) $ (obtained in the previous step), the RRMS solves the slave subproblem $ \mathcal{P}_1 $ using primal-dual methods (as mentioned in Remark \ref{remark: solving P1}) to update the routing, flow control and the Lagrangian (dual) variables. {\color{red}{Therefore, the complexity of this step is $ O \left( ??? \right) $.}}
%polynomial in terms of the number of data flows ($ K $) and the number of links ($ | \mathcal{L}| $) in the network.
\fi

%Calculating the aforementioned average rates by  \eqref{ave_rate_DTXpattern} {\color{red}{has a complexity of $ O \left( ??? \right) $}}. Moreover, Procedure I that updates DTX time-sharing control $ \boldsymbol{q}^{(i)} $ includes solving a convex optimization problem as in \eqref{q_Proc} with the number of variables equal to $ |\mathcal{A}  | $. Therefore, using primal-dual interior point methods that are well known to be very efficient for solving convex problems \cite{boyd2009convex} , this problem can be solved in  $ O \left( ??? \right) $ [REF???].
 %can be solved in polynomial time in terms of the number of admissible patterns $ | \mathcal{A} |  $.

%\end{itemize}

\end{enumerate} 

To conclude the above discussion, it can be seen that the per-iteration complexity of the overall proposed algorithm is $ O \left( LM + |\mathcal{A}|^3 + K^3L^3 \right) $, which is at most polynomial in terms of the key system parameters, such as the number of links, data flows (active users), subbands and DTX patterns. This indicates that the proposed algorithm scales very well with the size of the problem. Moreover the signalling overhead of the proposed scheme is much lower than the signalling overhead of the fast-timescale RRM baseline (which has the signalling overhead of $ (N_{BS}+|\mathcal{A}| + L(M+K+2)) \times B \times T_s $  bits per superframe).

 }}

{\color{black}{

\section{Simulation Results} \label{sec: sim}

% In this section, we consider a single-cell HetNet with flexible backhaul. In the macro cell, there is a macro BS and 6 uniformly distributed pico BSs (constituting pico cells) and 6 active mobile users. Fig. \ref{fig: Simulation_Topology} illustrates the associated network graph. The black node is the macro BS and the green nodes and small red nodes stand for pico BSs and MUs, respectively. Moreover, existing of a line between two nodes in this graph, shows that those nodes are in the coverage area of each other, i.e, they can communicate/interfere with each other. The solid lines show the connection links between two BSs, while the dashed lines show the links between BS and MU. The transmit power of macro BS and pico BSs are $ p_{macro} = 40 ~ dBm $ and $ p_{pico} = 29 \sim 35 ~ dBm $, respectively. There are 10 available subbands in the network, and 6 data flows are to be routed in the downlink, i.e., from some BSs connected to the backhaul to some destined users. The PFS utility (as in equation \eqref{Alpha-fair}) is considered. %The following table summarizes the simulation parameters. 

In this section, we consider a multi-cell HetNet% with flexible backhaul
, as shown in Fig. \ref{fig: Simulation_Topology}, where each macro cell consists of a macro BS at the center, and four and eight uniformly distributed pico BSs and MUs, respectively. 
%Each red box corresponds to a macro cell. In the macro cell, there is a macro BS and three uniformly distributed pico BSs (constituting pico cells) and three active MUs. Black nodes, green nodes and small red nodes stand for macro BSs, pico BSs and MUs, respectively. 
The existence of a line between two nodes in this graph shows that those nodes are in each other's coverage area, i.e., they can communicate/interfere with each other. %The solid lines show the links between two BSs, while the dashed lines show the links between a BS and MU. 
The typical values of $p_{macro}= 40 $ dBm and $p_{pico} = 29 \sim 35 $ dBm  have been considered for the transmit power of each macro BS and pico BS, respectively \cite{damnjanovic2011survey}. We consider $ T_s=500 $ subframes within each superframe, and assume 
there are ten available subbands in the network. %, and three data flows are to be routed from cell 1/2/3 to cell 2/3/1, respectively, i.e., from some BSs connected to the backhaul to some destined users.
 The PFS utility is considered as the network utility function \cite{PFS}. %*** The PFS utility is considered.
%The PFS utility function (as in equation \eqref{Alpha-fair}) is considered.

%
\begin{figure}
  \centering
    \includegraphics[width=0.39\textwidth ]{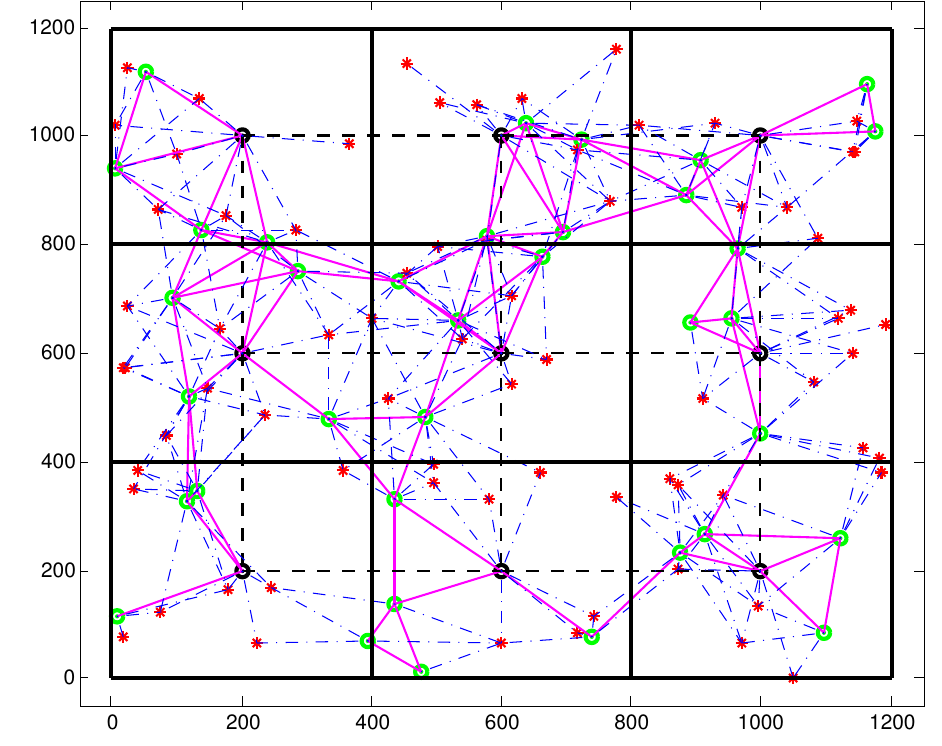} %width=0.39
  \caption{The considered HetNet topology which consists of nine macro cells. The black, green and red nodes represent macro BSs, pico BSs and MUs, respectively. 
 % Topology of a multi-cell HetNet. The source nodes and the destinations nodes have been labelled as $ s_1,\cdots,s_9 $ and $ d_1,\cdots,d_9 $, respectively. Moreover, $ (s_i,d_i), ~ \forall i=1,\cdots,9 $ indicates each source-destination pair.
 }
  \label{fig: Simulation_Topology} 
\end{figure}
%
%
%The path-loss model for BS-BS links is based on the free space model in \cite{xu2002spatial}. Moreover, for
For the BS-BS links and BS-MU links, the path loss models of \cite[Senario b5a]{ist2007deliverable} and \cite[Senario c2]{ist2007deliverable} (which are compatible with the LTE standard \cite{WinnerII_LTE}) are considered, respectively. %, as shown in Table \ref{tab: BS_MU link path loss model}. 
%***For the small-scale fading, a block fading channel model is considered, in which 
\iffalse
The small-scale fading is considered to be static during one subframe and i.i.d. over different subframes. %***(i.e., block fading channel model).
 Moreover, as mentioned before, the large-scale fading is considered to be a slow ergodic process which remains constant for a large number of %several 
 superframes. 
\fi 
 The backhaul delay for global CSI exchange and local CSI exchange are considered to be $ 2 ~\mathrm{ms} $ and $ 20 ~\mathrm{ms}$, respectively. 
 %The local CSI delay and global CSI delay  in the network are assumed  to be $ 2 ~ ms $ and $ 20 ~ ms  $, respectively {\color{green}{[LTE-REF???]}}.
  All the simulation results have been obtained by MATLAB R2014a on a simulation  platform with a Windows 7 x64, 2.6-GHz CPU and 8 GB RAM.

}}

We compare the  performance  of the proposed RRM scheme with  the following  baselines using numerical simulations. Note that %except the first baseline (which is for comparing to the conventional full backhaul case), 
all the baselines are considered with flexible backhaul deployment. Moreover, the first three baselines are two-timescale designs, while the last two baselines are single-timescale designs and are included to show the effectiveness of the proposed two-timescale approach over the conventional fast-timescale or slow-timescale RRM approaches.
\begin{itemize}

\iffalse *****
\item {\color{blue}{\textbf{Baseline 1, Full Backhaul Connection}: %Each pico BS is connected to the associated Macro BS in each cell.  
In this baseline, all the BSs have backhaul connectivity 
%***Note that obviously, there is no multi-hop routing in such scheme, %are connected to the backhaul.
 and the RRMS adopts our proposed Algorithm \ref{alg 1} to allocate the other resources. %This baseline is considered to compare the performance of RRM in HetNets with flexible 
 This baseline provides an upper bound for the performance of the HetNet under flexible backhaul deployment. %*****In fact, it shows how much performance we have to sacrifice in order to trade for the reduced backhaul cost when advanced flexible backhaul and dynamic resource allocation schemes are adopted. 
 }}
 \fi

\item \textbf{Baseline 1, Alternating Optimisation Approach \cite{bezdek2003convergence_AO}}: %*****The RRMS adopts the method of alternating optimization (AO) \cite{bezdek2003convergence_AO} to update the control variables. Under this method, 
The RRMS alternately updates each of the control variables by optimising the objective function with respect to that variable, while considering all the other variables to be fixed. This baseline is included in the simulations in order to show the importance of our considered joint optimization of the control variables and its advantage over alternating optimization. 

%***The RRMS adopts the method of alternating optimization (AO) [REF???] to update the control variables. Under this method, at each superframe, the RRMS alternatively updates each of the long-term control variables by optimising the objective function with respect to that variable, while considering all the other variables to be fixed, and updates the link scheduling at each subframe.

%***Under this method, for updating each control variable, the objective function is optimised with respect to that control variable, considering all the other control variables to be fixed.

\item \textbf{Baseline 2, Fixed Routing with Dynamic DTX and Link Scheduling Controls}: In this scheme, we consider fixed routing where each MU selects the nearest BS for its last hop communication link. The other control variables are updated in the same way as in our proposed algorithm.

\item \textbf{Baseline 3, Fixed DTX Control with Dynamic Routing, Flow Control and Link Scheduling}: Under this scheme, the DTX patterns are selected for each subband with equal probabilities (i.e., $ q_j = \dfrac{1}{ | \mathcal{A} | }, ~ \forall j=1,\cdots,| \mathcal{A} | $) at each superframe, and the other control variables are determined based on our proposed algorithm. 
%the subband allocation is  determined by the proposed link scheduling scheme in \eqref{equ: link sched}.
This baseline is included in our comparison in order to show the effectiveness of our proposed DTX time-sharing control scheme.

\item \textbf{Baseline 4, Fast-Timescale RRM Adaptive to the  Global Instantaneous CSI}: In this baseline, all the RRM control variables (either the long-term or short-term controls) are updated at each subframes, based on the global instantaneous CSI. %The RRMS uses our Algorithm \ref{alg 1} to update them at each subframe. 
%***Under this scheme, all the short-term and long-term controls are considered to be adaptive to the short-term global CSI in the network and  the RRMS updates them based on our proposed Algorithm \ref{alg 1} at each subframe. 
%*****{\color{red}{The purpose of comparing with this baseline is to show the effectiveness of the proposed two-timescale approach over the short-timescale instantaneous approach.}}

{\color{black}{\item \textbf{Baseline 5, Slow-Timescale RRM Adaptive to the Statistical CSI}:
In this baseline, 
all the RRM control variables are updated at a the slower timescale, i.e., at each superframe, based on the global long-term channel statistics. %The RRMS uses the proposed algorithm to update them at each superframe.
%***Under this scheme, all the short-term and long-term controls are considered to be adaptive to the long-term statistical CSI in the network and the RRMS updates them at each superframe, using the proposed algorithm.  
%*****{\color{red}{The purpose of comparing with this baseline is to show the effectiveness of the proposed two-timescale approach over the long-timescale statistical approach.}}
}}

%\item \textbf{Baseline 2, Fixed DTX and Dynamic Subband Allocation (FDDSA) \cite{tse2005fundamentals}}: The RRMS selects the DTX patterns with equal probabilities (i.e., $ q_j = \dfrac{1}{ | \mathcal{A} | }, ~ \forall j=1,\cdots,| \mathcal{A} | $) at each superframe, and the subband allocation is  determined by the proposed link scheduling scheme in \eqref{equ: link sched}.

%\item \textbf{Baseline 3, One-timescale RRM adapting to Statistical CSI (OTRSC) \cite{papandriopoulos2008optimal}}: The RRMS solves problem $P_{org}$ with the link scheduling policy depending on the large-scale instead of the small scale fading gain.
\end{itemize}

\iffalse
\begin{figure}
\centering
\includegraphics[width=0.38\textwidth]{pics/Sim_utility_new.pdf}
%\includegraphics[width=0.4\textwidth]{pics/Sim_utility_new.eps}
\caption{Utility function comparison.}
\label{fig: Simulation_obj_value_comparison}
\end{figure}
\fi

%***Fig. \ref{fig: Simulation_obj_value_comparison}  shows the utility function value versus the transmit power of the pico BSs for the proposed algorithms and the aforementioned baselines, when only $ 20 \% $ of the BSs are connected to the backhaul (except in Baseline 1 which is for full backhaul connection). It can be seen from this figure that the proposed scheme outperforms all the other baselines with flexible backhaul deployment. Moreover, Baseline 6 has the worst performance, since it does not exploit the instantaneous CSI that is locally available at each node to achieve multi-user diversity gain.  In addition, the performance of Baseline 5 shows that if the DTX time-sharing control is not optimised, the performance of the RRM algorithm will be highly degraded compared to our proposed algorithm that considers DTX time-sharing optimization as well. %worse than the performance of our proposed approach.  
%***Finally, the performance of the other three baselines (i.e., Baselines 2-4) is close to each other, but still worse than our proposed scheme. For example, our proposed scheme outperforms Baseline 4 (the fast-timescale RRM design), since the proposed two-timescale RRM design ???

\subsection{Performance Evaluation and Comparison}

{\color{black}{
Fig. \ref{fig: Simulation_obj_value_comparison}  shows the utility function value versus the transmit power of the pico BSs for the proposed algorithm and the aforementioned baselines, when $ 40 \% $ of the BSs are connected to the backhaul. %***** (except in Baseline 1 which is for full backhaul connection).  Note that as expected, the performance of Baseline 1 which has full backhaul connection is the upper bound for all the other baselines with flexible backhaul connection. 
It can be seen from Fig.~\ref{fig: Simulation_obj_value_comparison} that the proposed scheme outperforms all the other baselines, due to various advantages of the proposed design over the considered baselines.  For example, as shown by our theoretical analysis, the proposed algorithm achieves the optimal solution, while the alternating optimization (AO) algorithm (Baseline 1) does not converge to the optimal solution (as the problem is non-convex, the AO method cannot guarantee global optimality) and hence, it cannot achieve the same performance as ours. 
%although Baseline 2 uses alternating optimization algorithm to update each control variable, it still cannot achieve the same performance as our proposed algorithm, since the AO algorithm does not converge to the optimal solution (unlike the proposed algorithm, as also shown by theoretical proofs).  
Moreover, the proposed two-timescale RRM design outperforms the fast-timescale design (Baseline 4), because, due to the considered two-timescale design, the long-term controls, which require global coordination and signalling, are updated at the slow timescale, and hence the proposed design is less sensitive to the backhaul signalling delay. On the other hand, since the proposed two-timescale approach can exploit multi-user diversity gain by updating the local short-term controls (i.e., link scheduling) in the fast timescale, it also outperforms the slow-timescale design (i.e., Baseline 5). Furthermore, since we also perform dynamic routing in our design, the proposed RRM solution achieves higher performance than Baseline 2, which has fixed routing. 
In addition, the performance of Baseline 3 shows that if the DTX time-sharing control is not optimised, the performance of the RRM algorithm will be highly degraded compared to our proposed RRM design that optimises the time-sharing of different DTX patterns. Finally, Baseline 5 has the worst performance, since it does not exploit the instantaneous CSI that is locally available at each node to achieve multi-user diversity gain. 

%***Moreover, Baseline 6 has the worst performance, since it does not exploit the instantaneous CSI that is locally available at each node to achieve multi-user diversity gain.  In addition, the performance of Baseline 5 shows that if the DTX time-sharing control is not optimised, the performance of the RRM algorithm will be highly degraded compared to our proposed algorithm that considers DTX time-sharing optimization as well. %worse than the performance of our proposed approach. Finally, the performance of the other three baselines (i.e., Baselines 2-4) is close to each other, but still worse than our proposed scheme. For example, our proposed scheme outperforms Baseline 4 (the fast-timescale RRM design), since the proposed two-timescale RRM design ???

}}

\begin{figure*}[t!]
\centering
\begin{minipage}[t]{.48\textwidth}
  \centering
	%***\includegraphics[width=0.8\textwidth]{pics/Sim_utility_new.pdf}
	\includegraphics[width=0.8\textwidth]{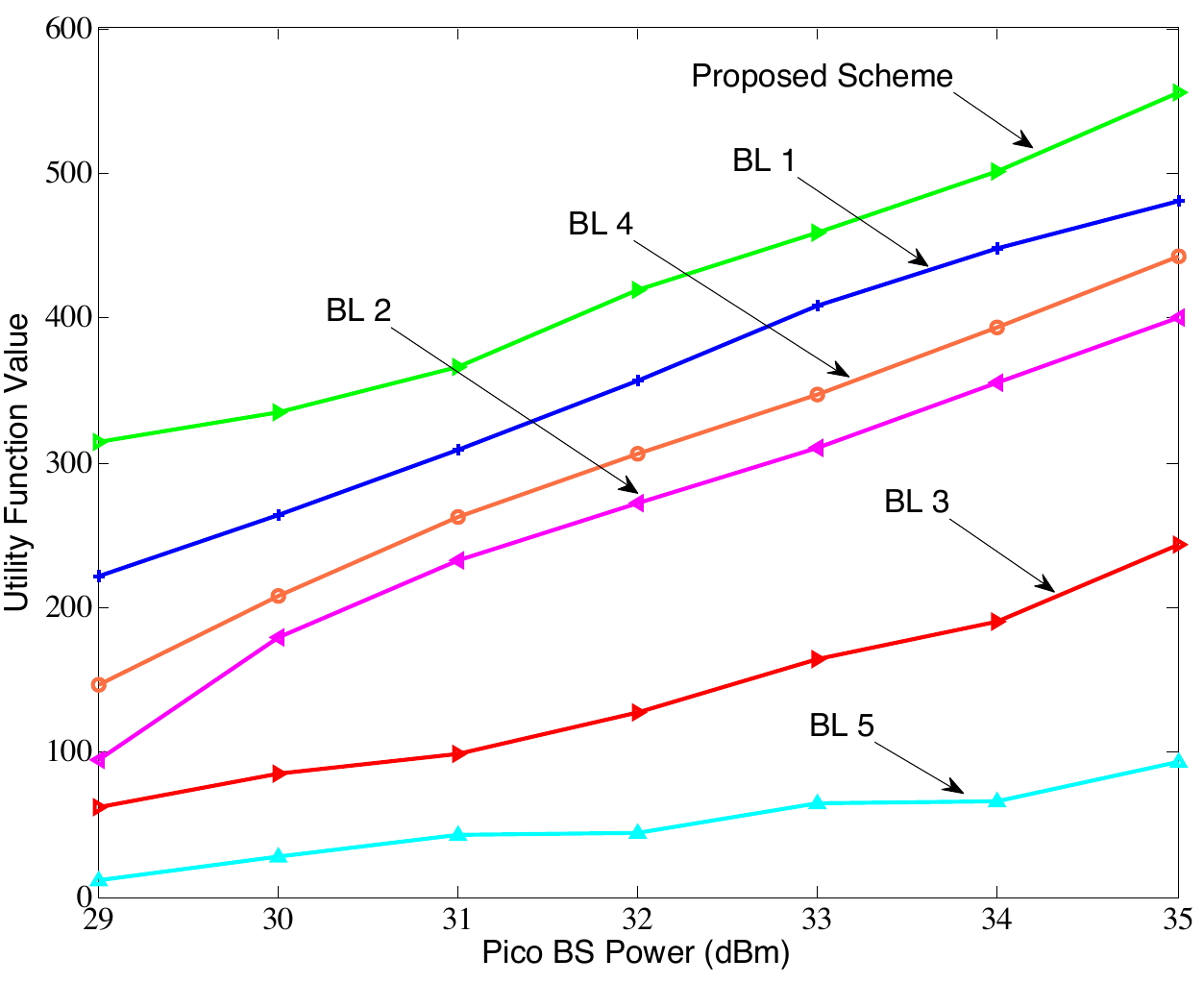}
	\caption{Network utility comparison.}%\vspace{-20 pt}
	\label{fig: Simulation_obj_value_comparison}
\end{minipage}%
~~
\begin{minipage}[t]{.48\textwidth}
\iffalse
  \centering
  \includegraphics[width=0.8\textwidth]{pics/Sim_throughput_new.pdf}
	%\includegraphics[width=0.35\textwidth]{pics/Signaling_sub_v3.eps}
	\caption{Average per-cell sum throughput comparison.}%\vspace{-20 pt}
	\label{fig: Simulation_throughput_comparison}
	\fi
	\centering
	\includegraphics[width=0.80\textwidth]{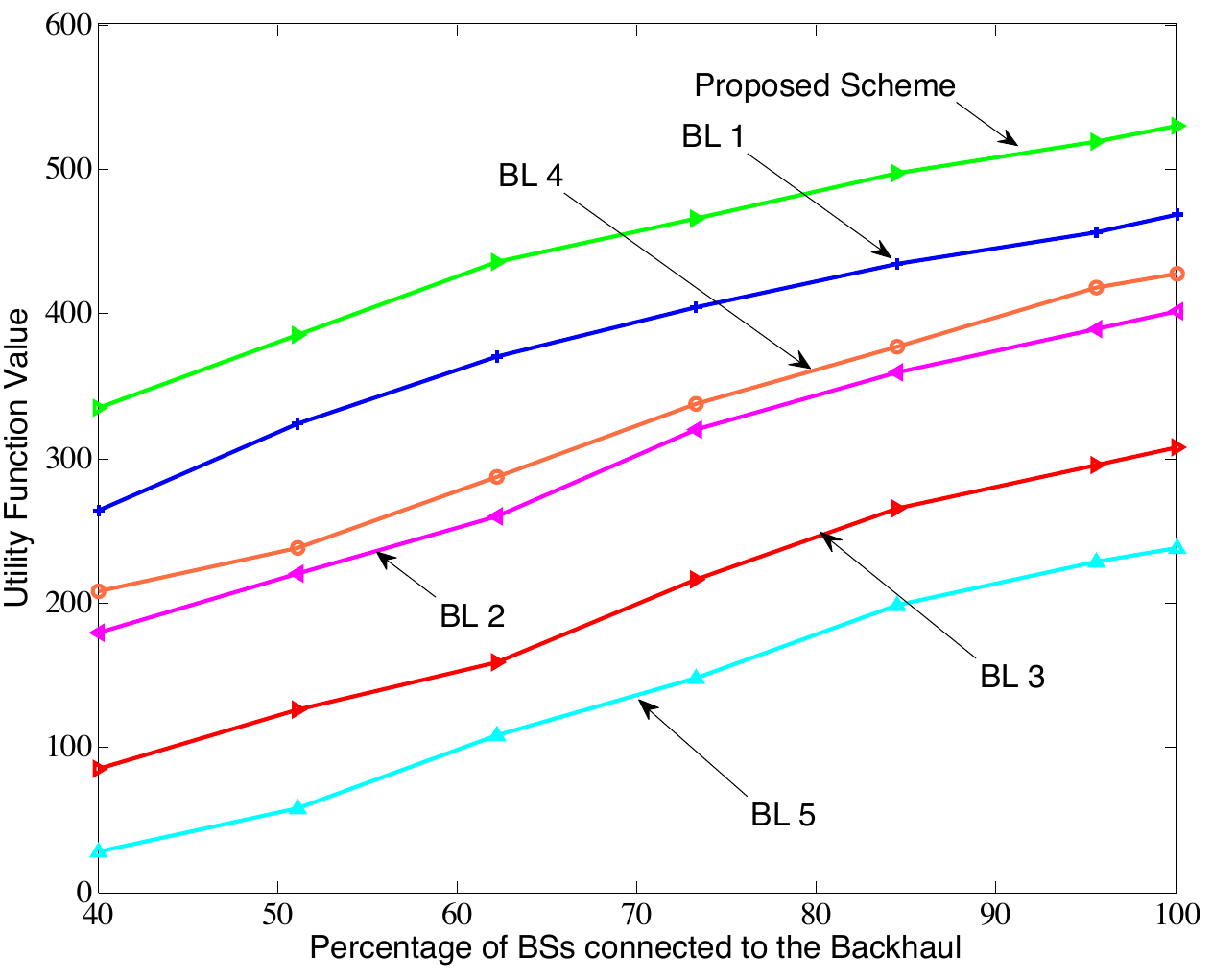}
	\caption{Performance comparison %of the proposed scheme and the baselines  
	under various percentages of the BSs with backhaul connection ($ p_{pico} = 30~ dBm $).}
	\label{fig: Perf_vs_BH}
\end{minipage}
%\vspace{-10 pt}
\end{figure*}

%from the simulation results, the performance of the proposed scheme is very close to BL1 with little performance loss due to the flexible backhaul design!!!!!!

Next, Fig. \ref{fig: Perf_vs_BH} compares the performance of the proposed RRM solution to the baselines, under different portions of BSs connected to the backhaul ranging from $ 40 \% $ to $ 100 \% $ (i.e., the conventional case with full backhaul connection). 
 %This baseline provides an upper bound for the performance of the HetNet under flexible backhaul deployment. 
 %*****Note that in Baseline 1 (i.e., full backhaul connection), this portion is always $ 100 \% $. %***For this simulation, the transmit power of pico BSs on each subband is fixed at $ p_{pico}= 30 ~dBm $. 
As expected, when the percentage of the BSs with backhaul connection varies from $ 40 \% $ to $ 100 \% $, the performance of all the baselines %*****with flexible backhaul connection (i.e., our proposed scheme as well as Baselines 2-6)
 is increasing. Specifically in our proposed scheme, the performance first increases quickly (almost linearly) with the portion of BSs with backhaul connection, and then increases more slowly for larger portions. %*****Finally, as expected, when this portion approaches $ 100 \% $ (i.e., full backhaul connection), the performance of our proposed scheme approaches the performance of Baseline 1 with full backhaul connection. 

\iffalse *****
\begin{figure}
\centering
\includegraphics[width=0.40\textwidth]{pics/perfvsbh.pdf}
%\includegraphics[width=0.40\textwidth]{pics/Sim_throughput_new.eps}
\caption{Performance comparison %of the proposed scheme and the baselines 
under various portions of BSs with backhaul connectivity.}
\label{fig: Perf_vs_BH}
\end{figure}
\fi

%***\subsection{The Overall Complexity}
\subsection{The Average Computational Time and Signalling Overhead}

Fig. \ref{fig: CPU_time} shows the computational complexity of the proposed algorithm and the baselines in terms of the average CPU time per subframe (for a fair comparison) for different network sizes. We have considered that $ 40\% $ of the BSs are connected to the backhaul and the pico BS power is $ p_{pico} = 30 ~dBm $. 
\iffalse ***
As can be seen from this figure, % the overall complexity of the proposed algorithm increases almost linearly with the size of the network.
%as the network size becomes larger, 
the overall computational complexity of the proposed scheme increases almost linearly with the network size. %Therefore, although the denser network will increase the convergence time (as is always expected), the complexity of the proposed algorithm is almost linear with the size of the problem, i.e., the number of BSs. 
This shows that the proposed algorithm scales very well with the size of the problem. 
\fi
As can be seen from this figure, the computational time (the average CPU time per subframe) of the proposed method grows polynomially with the network size, which is consistent with our analytical results in Section \ref{sec: signalling_complexity}. Moreover, 
the computational time %(the average CPU time per subframe) 
of Baseline 4 (fast-timescale RRM) is the highest among all the baselines. This is because this baseline needs to update all the control variables at each subframe. %Therefore the whole RRM algorithm is performed more frequently (i.e., 
Furthermore, Baseline 1 (the AO approach) also has high computational time, as it  
%is the highest among all the baselines. This is because the AO method 
needs to optimise each control variable alternatively with other variables fixed, and hence, for updating each control variable at each time, it needs to run an iterative algorithm to solve the optimization problem with respect to the corresponding variable, which takes more time. 
%Therefore, the overall complexity of the AO baseline is significantly high. 
Finally, the figure shows that the average computational time of the proposed scheme is similar to the other baselines (Baselines 2, 3 and 5) (while the proposed scheme achieves higher performance than them). Specifically, the proposed two-timescale scheme has similar computational time as the slow-timescale RRM scheme (Baseline 5), which is known for having low computational complexity.

 %Moreover, the computational time of all other schemes are similar. 
\iffalse ***
Finally, as can be seen from this figure, in the  proposed scheme, as the network size becomes larger, the overall CPU time will increase almost linearly. %Therefore, although the denser network will increase the convergence time (as is always expected), the complexity of the proposed algorithm is almost linear with the size of the problem, i.e., the number of BSs. 
This shows that the proposed algorithm scales very well with the size of the problem. 

optimising each variable alternatively with other variables fixed 

However, as shown earlier, the performance of this baseline is worse than the proposed scheme. Moreover, the computational time of all other schemes are similar. Finally, in our proposed scheme, as the network becomes denser, the overall CPU time will almost linearly increase. Therefore, although the denser network will increase the convergence time (as is always expected), the complexity of the proposed algorithm is almost linear with the size of the problem, i.e., the number of BSs. This indicates that the proposed algorithm scales very well with the size of the problem. 
\fi

\begin{figure*}[t!]
\centering
\begin{minipage}[t]{.48\textwidth}
  \centering
	%***\includegraphics[width=0.8\textwidth]{pics/Sim_utility_new.pdf}
	%\includegraphics[width=0.8\textwidth]{pics/CompvsSize_3.pdf}
	\includegraphics[width=0.8\textwidth]{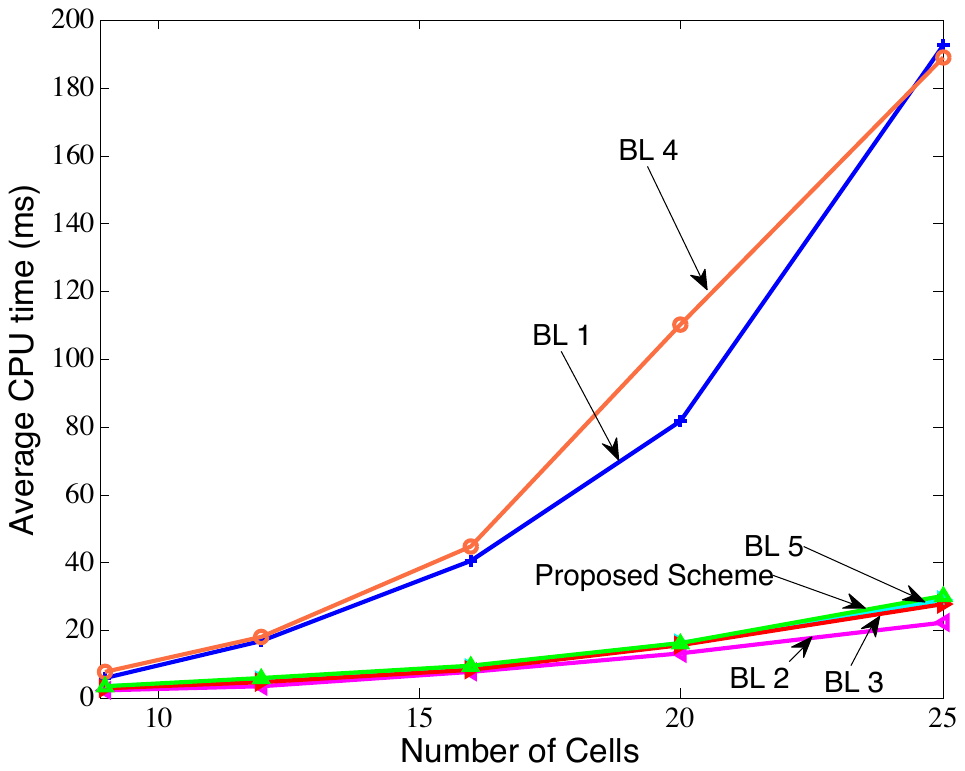}
	\caption{Average CPU time per subframe under different network sizes.}%\vspace{-20 pt}    
	\label{fig: CPU_time}
\end{minipage}%
~~
\begin{minipage}[t]{.48\textwidth}
	\centering
	\includegraphics[width=0.80\textwidth]{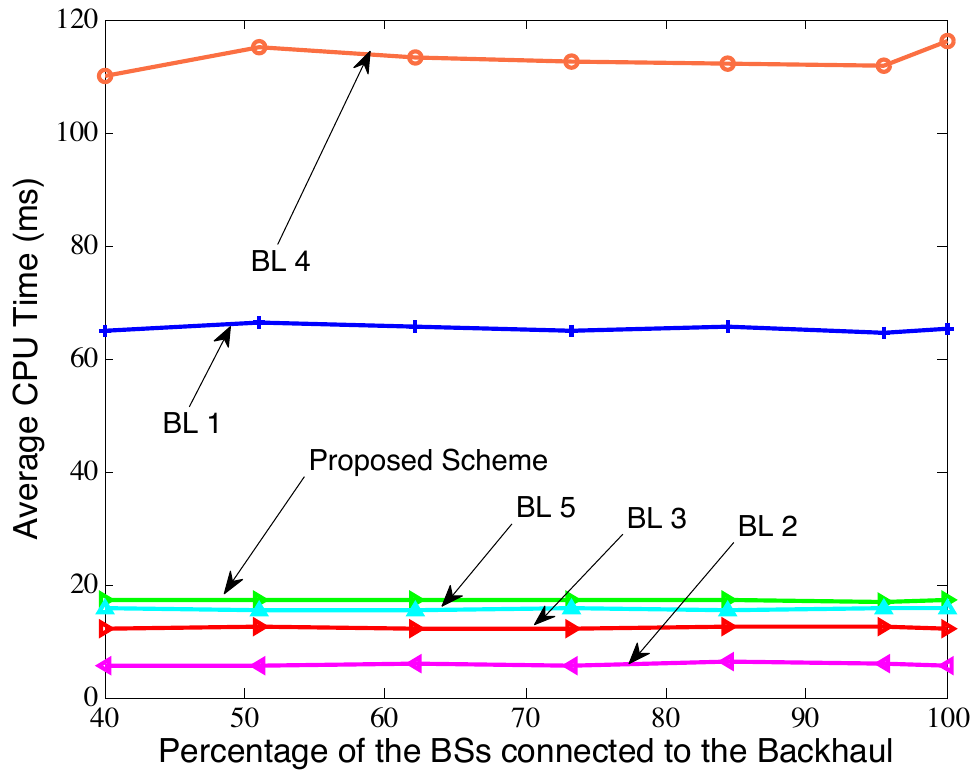}
	\caption{Average CPU time per subframe under different portions of BSs connected to the backhaul.}
	\label{fig: complexityfig}
\end{minipage}
%\vspace{-10 pt}
\end{figure*}

Furthermore, Fig. \ref{fig: complexityfig} shows the computational complexity of our proposed scheme and the baselines in terms of the average CPU time per subframe, under different portions of the BSs connected to the backhaul, when the pico BS power on each subband is $ p_{pico} = 30 ~ dBm $. It can be verified from this figure that under our proposed scheme or any of the baselines, the average CPU time per subframe is similar for different portions of the BSs with backhaul connection. 
Moreover, it can be seen that the average computational time of our proposed scheme is similar to Baselines 2, 3 and 5. This shows that while the performance of our proposed RRM scheme for flexible
backhaul is better than these baselines, its computational complexity is still similar to those baselines with low computational complexity. Furthermore, the average computational time of Baseline 2  per subframe is the smallest, since this baseline does not update the routing controls. However, as shown earlier, the this baseline worse than our proposed method, as it does not utilise dynamic routing. Finally, expected, Baselines 1 and 4 have the highest computational times.

{\color{black}{

%The signalling overhead of the proposed scheme and the baselines has been compared in Table ???.

Table \ref{signalling overhead} shows the signalling overhead of the proposed scheme and the baselines per BS per subband at each subframe when $ 40 \% $ of the BSs are connected to the backhaul. The signalling overhead %per BS per subband per one subframe 
is calculated based on the average number of bits that needs to be fed back to/from each BS per one subframe. We have considered $ B=6 $ bits as the average number of bits used to quantify the each real signalling.  %For this purpose, the signalling overhead of our proposed scheme is obtained by the formulation in Section ???, i.e., equation ???, and the signalling overhead for the other baselines are calculated similarly.  
As can be seen from this table, %Table \ref{table: timescales_RRM_comparison}, 
the signalling overhead of the proposed two-timescale design is similar to the simple slow timescale design and is much smaller than the fast timescale design. This is mainly due to the proposed two-timescale structure for updating the RRM control variables, in which, the global control variables are updated in a less frequent manner (once per each superframe) than the local control variables. Therefore, while the performance of the proposed scheme highly outperforms the slow timescale RRM scheme (as previously seen in Fig.s \ref{fig: Simulation_obj_value_comparison} and \ref{fig: Perf_vs_BH}), 
its signalling overhead is  similar to the slow timescale RRM design which is well-known for having low signalling overhead. 
%Therefore the signalling overhead of such two-timescale RRM scheme is comparable to the slow-timescale scheme. 
Moreover, the signalling overhead of Baseline 4 (the short-timescale RRM scheme) is significantly higher than the other schemes, since in this baseline, the iterations are done at the short-timescale (i.e., per each subframe), and hence, the required signalling and message passing need to be done significantly more frequently (at each subframe). Furthermore, Baseline 2 (RRM with fixed routing) has the lowest signalling overhead, which is due to the fact that this baseline does not need to update or report the routing variables (Note that its performance is worse than the proposed scheme). Finally, the signalling overhead of Baseline 1 is less than the proposed scheme. However, as seen in the previous figures, its performance as well as computational complexity are worse than the proposed scheme.

%*****The other baselines, including our scheme, have similar level of signalling overhead. Specifically, it should be noted that while the performance of the proposed scheme highly outperforms the slow timescale RRM scheme (as previously seen in Fig.s \ref{fig: Simulation_obj_value_comparison} and \ref{fig: Perf_vs_BH}), its signalling overhead is  similar to the slow timescale RRM design which is well-known for having low signalling overhead. This is due to the fact that in the proposed scheme, we have considered a two-timescale structure to update the RRM control variables, and hence, the global control variables are updated in a less frequent manner (once per each superframe). Therefore the signalling overhead of such two-timescale RRM scheme is comparable to the slow-timescale scheme.

%Therefore, while our proposed scheme have {\color{red}{higher performance gain}} than the other baselines, its  signalling overhead is still similar to them. This is mainly because the two-timescale 

\begin{small}
\begin{table*}[t]
	\centering
\begin{tabular}{| p{3cm} | p{1cm} p{2cm} p{1cm} p{1cm} p{1cm} p{1cm} |}
	\hline
		  {Baselines}& \multicolumn{1}{c|}{Proposed Scheme}  & \multicolumn{1}{c|}{Baseline 1} & \multicolumn{1}{c|}{Baseline 2}  & \multicolumn{1}{c|}{Baseline 3}  & \multicolumn{1}{c|}{Baseline 4}  & \multicolumn{1}{c|}{Baseline 5}   \\
	\hline
		{Signalling Overhead (bits)} &     \multicolumn{1}{c|}{43.92} &    \multicolumn{1}{c|}{21.72}  &    \multicolumn{1}{c|}{11.22}  &    \multicolumn{1}{c|}{44.52}  &  \multicolumn{1}{c|}{17110.92}  &     \multicolumn{1}{c|}{34.81}       \\
	\hline
\end{tabular}
	\caption{ \small{The average signalling overhead per BS per subband per one subframe.}\vspace{-10 pt}}
		\label{signalling overhead}   %***{table: complexity comparison}
%\vspace{-25 pt}
\end{table*}
\end{small}

\iffalse *****
%\begin{small}
\begin{table}[t]
	\centering
\begin{tabular}{| c | c c c c c c c |}
	\hline
		  {Baselines}& \multicolumn{1}{c|}{Proposed Scheme}  & \multicolumn{1}{c|}{Baseline 1} & \multicolumn{1}{c|}{Baseline 2}  & \multicolumn{1}{c|}{Baseline 3}  & \multicolumn{1}{c|}{Baseline 4}  & \multicolumn{1}{c|}{Baseline 5}  & \multicolumn{1}{c|}{Baseline 6}  \\
	\hline
		{{\color{red}{Signalling Overhead (bits)}}} &     \multicolumn{1}{c|}{73.2} &   \multicolumn{1}{c|}{120.7}  &    \multicolumn{1}{c|}{36.2}  &    \multicolumn{1}{c|}{28518.2}  &    \multicolumn{1}{c|}{18.7}  &    \multicolumn{1}{c|}{74.2}  &    \multicolumn{1}{c|}{58.1}       \\
	\hline
\end{tabular}
	\caption{ \small{The average signalling overhead per BS per one subframe.}}	
		\label{signalling overhead}   %***{table: complexity comparison}
\end{table}
%\end{small}
\fi

%Figure shows the cost saving against performance loss in HetNets with flexible backhaul, by reducing the backhaul connections.

%analysis of cost saving against performance loss, by reducing the backhaul connection

\subsection{Trade-off between Cost Saving and Performance Loss under Flexible Backhaul}

Fig. \ref{fig: BHcostPerf} shows  the trade-off between cost saving and performance loss by reducing the backhaul connections in the network with flexible backhaul. We have assumed 1 unit cost as the average cost of connecting each BS to the backhaul, and have changed the portion of BSs with backhaul connection from $ 40 \% $ to $ 95\% $. For each backhaul connection  portion, the backhaul cost saving is calculated as the difference between the cost of the backhaul connections for the portion of the BSs connected to the backhaul and the cost of full backhaul connection case, normalised to the cost of full backhaul connection case (which is the maximum cost). 
 As can be seen from this figure, by reducing the backhaul connections, the performance is reduced (as always expected), but at the same time a significant cost in deploying the network will be saved, which can be huge in the emerging densely deployed HetNets with a large number of pico BSs. As a result the considered flexible backhaul can help to save significant share of infrastructure costs for future HetNets.
%for different portions of the BSs without backhaul connectivity.

%The trade-off between

\subsection{Convergence of the Proposed Algorithm}

%*** Finally, Fig. \ref{fig: Convergence} shows the objective value %$ \widetilde{U} \left( \Omega \right) $ of $ \mathcal{P}_{2} $ versus the number of iterations. As can be verified from this figure, the proposed algorithm converges very fast.

Finally, Fig. \ref{fig: Convergence} shows the convergence result of the proposed scheme and baselines, when pico BS power is $ 30 ~ \mathrm{dBm} $. 
%*****As expected, Baseline 1 (with full backhaul connection) converges to {\color{red}{a higher value than all the other baselines OR the highest value among ...}}. {\color{red}{Note that this is because the backhaul scenario of this baseline is different from all the other baselines, in which only a portion of the BSs ($ 40 \% $ in the simulations for this figure) are connected to the backhaul. Consequently, Baseline 1 can achieve higher network utility. Putting Baseline 1, which has a different backhaul topology, aside, 
As can be seen in this figure, our proposed scheme achieves the best network utility among all the baselines, with a significant gap compared to them. In fact, the proposed algorithm accomplishes the optimal RRM solution (as previously shown by the analytical proofs), but the other baselines fail to reach the optimal solution and converge to solutions with pretty low network utility values.
Moreover, although Baseline 1 (AO approach) achieves a higher network utility than the other baselines, it is still sub-optimal and and the convergence speed of this baseline is very slow, compared to our proposed algorithm that converges to the optimal value very fast (as can be easily seen in this figure). Compared to the other baselines, the proposed algorithm has similar convergence speed, but much higher utility.

%Moreover, although Baseline 2 achieves a higher network utility than all other baselines, it is still sub-optimal and the convergence speed of this baseline is very slow, compared to our proposed algorithm that converges to the optimal value very fast (as can be easily seen in this figure).  These results show the advantage of the proposed RRM scheme over the other possible baselines in significantly better utilising the network resources in a HetNet with flexible backhaul. 

\begin{figure*}[t!]
\centering
\begin{minipage}[t]{.47\textwidth}
  \centering
	%***\includegraphics[width=0.8\textwidth]{pics/Sim_utility_new.pdf}
	\includegraphics[width=0.8\textwidth]{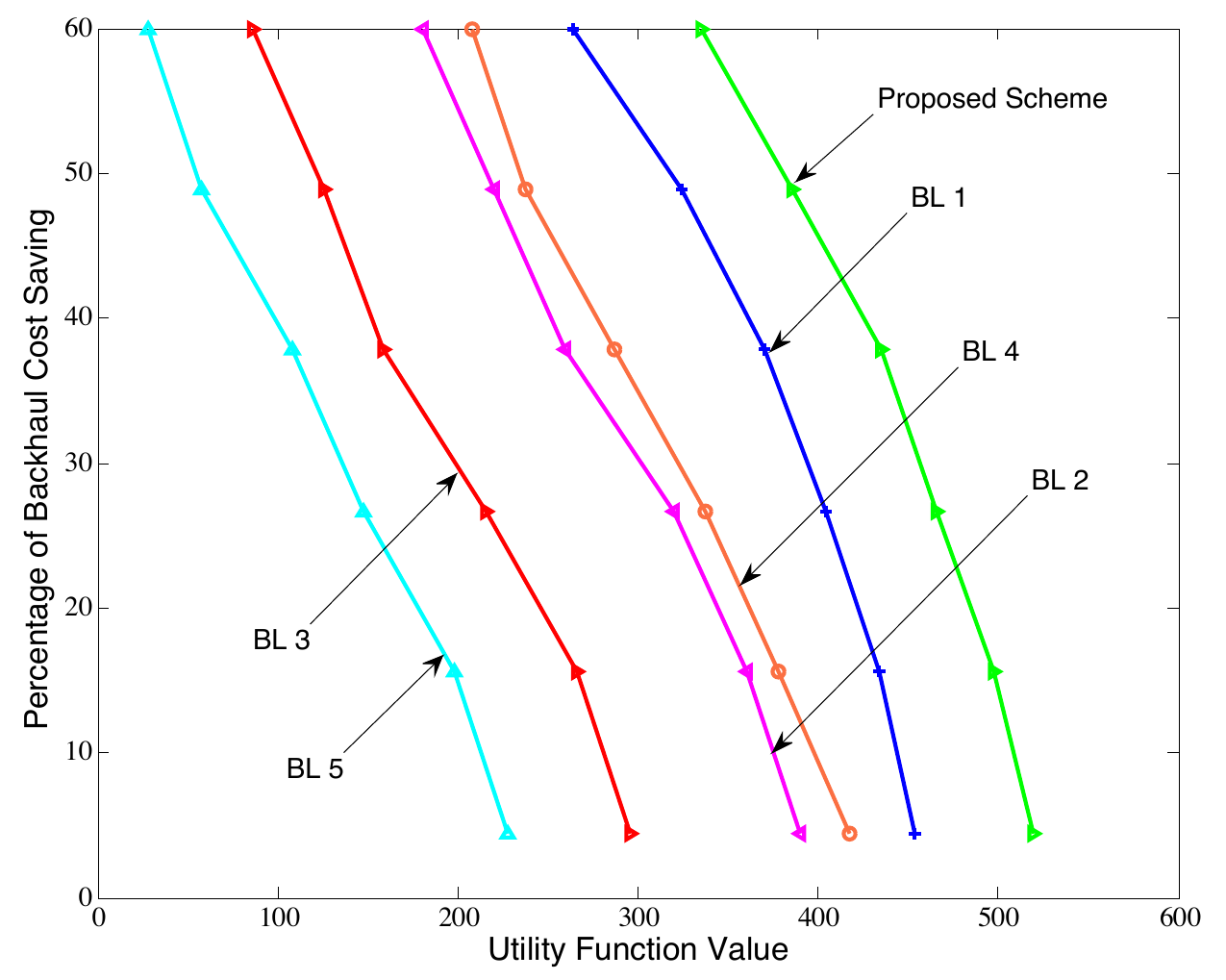}
	\caption{\small{Backhaul cost saving versus performance loss in flexible backhaul.}}%\vspace{-20 pt}    
	\label{fig: BHcostPerf}
\end{minipage}%
~~~
\begin{minipage}[t]{.47\textwidth}
	\centering
	\includegraphics[width=0.8\textwidth]{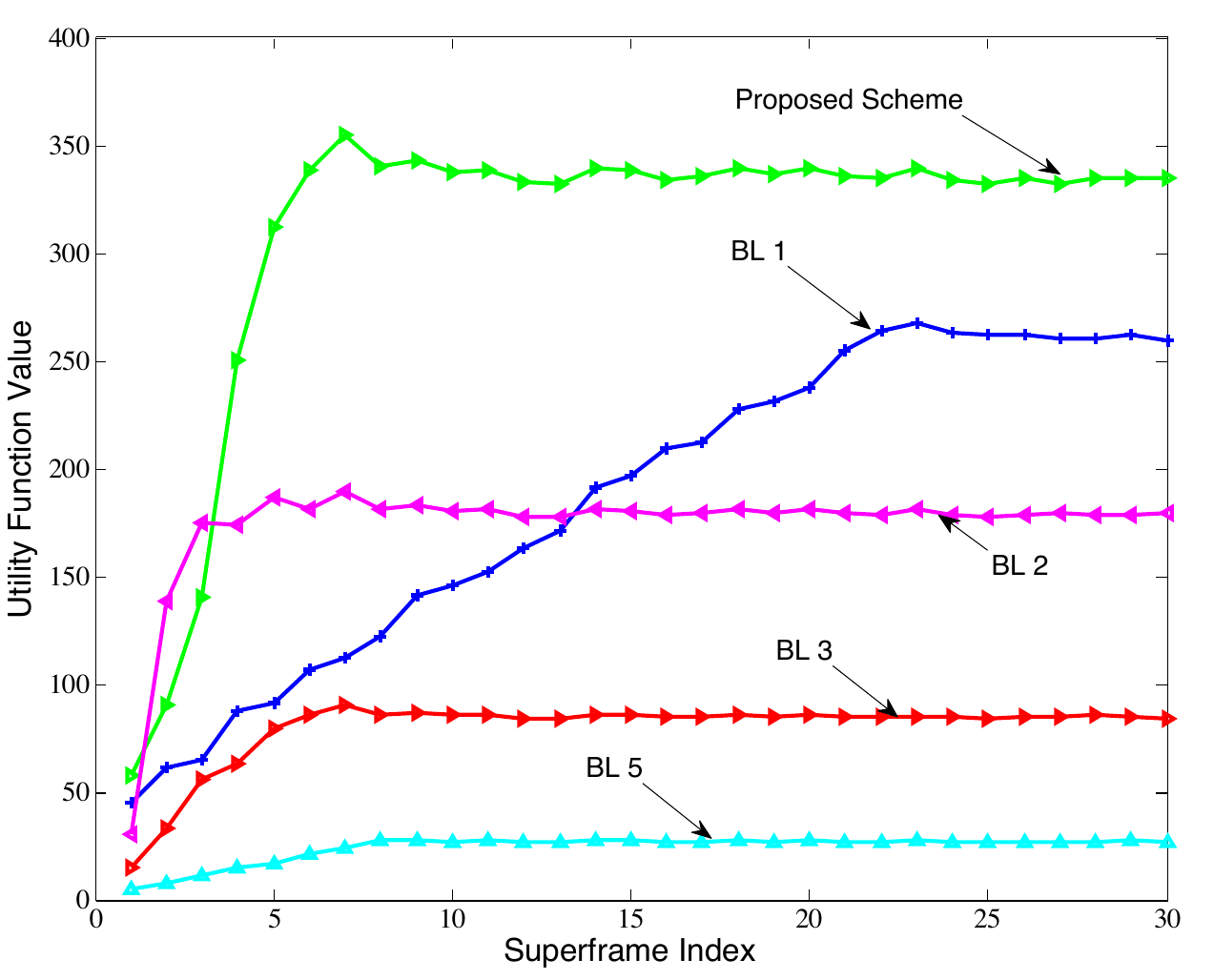}
	\caption{\small{Convergence results comparison.}}
	\label{fig: Convergence}
\end{minipage}
\vspace{-15 pt}
\end{figure*}

\iffalse
\begin{figure}
\centering
\includegraphics[width=0.4\textwidth]{pics/convergencefig.pdf}
%\includegraphics[width=0.40\textwidth]{pics/Sim_fig3.eps}
\caption{Convergence results comparison.}
\label{fig: Convergence}
\end{figure}
\fi

}}

{\color{black}{
\section{Conclusion} \label{sec: conclusion}

In this paper, we considered the problem of RRM for HetNets with flexible backhaul. We formulated the problem with a two-timescale stochastic optimization problem and considered the cross-layer design of flow control, routing control, interference control and link scheduling. Deriving a sufficient condition for the global optimality of a solution, we then proposed an iterative algorithm to reach the global optimal solution.  
The proposed two-timescale hierarchical design combines the benefits of both fast timescale RRM design and slow timescale RRM design and provides a significant performance gain while low signalling overhead and complexity and robustness to the backhaul latency. Simulation results show that the proposed RRM design achieves significant performance gains over various baselines.

\appendix 

\section{}
\subsection{Proof of Lemma \ref{theorem: P_2=P_E}} \label{ap: proof of theorem: P_2=P_E}
If $  \left( \boldsymbol{q}^\ast, \boldsymbol{\rho}^\ast \right) $ is the global optimal solution of $ \mathcal{P}_{2} $, then from the definition of problems $ \mathcal{P}_{E} $ and $ \mathcal{P}_{2} $ it follows that $ \overline{\boldsymbol{r}} \left( \boldsymbol{q}^\ast, \boldsymbol{\rho}^\ast \right) $ would be the optimal solution of $ \mathcal{P}_{E} $. On the other hand, suppose that $ \boldsymbol{r}^{\ast} $ is the optimal solution to $ \mathcal{P}_{E} $, and $  \left( \boldsymbol{q}^\ast, \boldsymbol{\rho}^\ast \right) $ satisfies $ \overline{\boldsymbol{r}} \left( \boldsymbol{q}^\ast, \boldsymbol{\rho}^\ast \right)=\boldsymbol{r}^{\ast} $. %By contradiction, 
If $  \left( \boldsymbol{q}^\ast, \boldsymbol{\rho}^\ast \right) $ is not the global optimal solution of $ \mathcal{P}_{2} $, then there exists another  $  \left( \boldsymbol{q}, \boldsymbol{\rho} \right) \in \Lambda $ such that $ \widetilde{U} \left( \overline{\boldsymbol{r}} \left( \boldsymbol{q}, \boldsymbol{\rho} \right) \right)  > \widetilde{U} \left( \overline{\boldsymbol{r}} \left( \boldsymbol{q}^\ast, \boldsymbol{\rho}^\ast \right) \right) $. As a consequence, $ \overline{\boldsymbol{r}} \left( \boldsymbol{q}^\ast,\boldsymbol{\rho}^\ast \right) =\boldsymbol{r}^{\ast} $ would not be the optimal solution of problem $ \mathcal{P}_{E} $, which is a contradiction.
%and this contradicts with our first assumption at the beginning.

\subsection{Proof of Proposition \ref{th: Convexity of P_E}}  \label{ap: proof of th: Convexity of P_E}

To  prove the concavity of the objective function $ \widetilde{U}(\boldsymbol{r}) $, using the concept of epigraph for convex functions \cite{boyd2009convex}, it is sufficient to show that the following set is convex:
\begin{equation}%\small 
\label{epi U_tilde}
epi \widetilde{U}  \triangleq \left\{ \boldsymbol{z} = \left[ \begin{smallmatrix} \boldsymbol{r} \\ y \end{smallmatrix} \right]: ~  \boldsymbol{r} \in \mathcal{R} , ~ y \leq \widetilde{U}( \boldsymbol{r} ) \right\}.
\end{equation}

\iffalse ***
First, to  prove the concavity of $ \widetilde{U}(\boldsymbol{r}) $, it is sufficient to show that the epigraph of function $ - \widetilde{U}(\boldsymbol{r}) $ is a  convex set. 

In order to prove the convexity of problem $ \mathcal{P}_{E} $, we first prove that its objective function $ \widetilde{U}(\boldsymbol{r}) $ is concave. Then, we prove that its feasible set is convex.

For the first part, it is sufficient to show that $ - \widetilde{U}(\boldsymbol{r}) $ is convex. Note that a function is convex if and only if its epigraph is a convex set \cite{boyd2009convex}. The epigraph of function $ - \widetilde{U}(\boldsymbol{r}) $ can be defined as $ \left\{ \boldsymbol{z} = \left[ \begin{smallmatrix} \boldsymbol{r} \\ y \end{smallmatrix} \right]: ~  \boldsymbol{r} \in \mathcal{R} , ~ y \geq - \widetilde{U}( \boldsymbol{r} ) \right\} $. Equivalently, it is sufficient to prove that the following set is convex:
\begin{equation} \label{epi U_tilde}
epi \widetilde{U}  \triangleq \left\{ \boldsymbol{z} = \left[ \begin{smallmatrix} \boldsymbol{r} \\ y \end{smallmatrix} \right]: ~  \boldsymbol{r} \in \mathcal{R} , ~ y \leq \widetilde{U}( \boldsymbol{r} ) \right\}.
\end{equation}

\fi

\begin{lemma}\label{lem: epi convex}
The set $ epi \widetilde{U} $ is a convex set.  %{\color{red}{Try to summarise it}}

\proof 

% \left( \boldsymbol\nu , y \right)

%In order to prove the first part, it is sufficient to prove that $ epi( \widetilde{U}(\overline{\boldsymbol{r}}) ) $ is a convex set, according to the following Lemma.

First note that %from Lemma \ref{}, !!!
at any arbitrary $ \boldsymbol{r} \in \mathcal{R} $, the Lagrangian multipliers vector $ \boldsymbol\lambda \left( \boldsymbol{r} \right) $ is the gradient of function $ \widetilde{U} $ \cite[Proposition 3.3.3]{bertsekas1999nonlinear} at point $ \boldsymbol{r} $. Hence, the $ (L+1) $-dimensional vector
$ \left[ \begin{smallmatrix}
\boldsymbol\lambda \left( \boldsymbol{r} \right) \\ -1 \end{smallmatrix} \right] $
defines a non-vertical supporting hyperplane to $ epi \widetilde{U} $ at any point $ \boldsymbol{z} = \left[ \begin{smallmatrix} \boldsymbol{r} \\ \widetilde{U}( \boldsymbol{r} ) \end{smallmatrix} \right], ~ \forall  \boldsymbol{r} \in \mathcal{R} $ \cite{boyd2009convex}, i.e.,
\begin{equation}\label{equ: supporting hyperplane}\small
( \boldsymbol{r'} - \boldsymbol{r} ) ^{T} \cdot \boldsymbol\lambda \left( \boldsymbol{r} \right)  + ( \widetilde{U} \left( \boldsymbol{r} \right) - y' ) \geq 0, ~ \forall \boldsymbol{r} \in \mathcal{R}, \forall \boldsymbol{z'} =   \left[ \begin{smallmatrix} \boldsymbol{r'} \\ y' \end{smallmatrix} \right]   \in epi \widetilde{U}.
\end{equation}

%Note that this hyperplane divides the closed set 

We prove Lemma \ref{lem: epi convex} by contradiction: Let's assume that $ epi \widetilde{U} $ is NOT a convex set. This implies that there exists some $ \boldsymbol{z_{1}} = \left[ \begin{smallmatrix} \boldsymbol{r_{1}} \\ y_{1} \end{smallmatrix} \right] , \boldsymbol{z_{2}} = \left[ \begin{smallmatrix} \boldsymbol{r_{2}} \\ y_{2} \end{smallmatrix} \right] \in epi \widetilde{U} $ and some $ 0 < \alpha < 1 $, such that
\begin{equation}\label{z0}
\boldsymbol{z_{0}} = \left[ \begin{smallmatrix} \boldsymbol{r_{0}} \\ y_{0} \end{smallmatrix} \right] \triangleq \alpha \boldsymbol{z_{1}} + \left( 1 - \alpha \right) \boldsymbol{z_{2}} \notin epi \widetilde{U}  .
\end{equation}

%\begin{align} 
%& \exists \boldsymbol{z_{1}} , \boldsymbol{z_{2}} , 0 < \alpha < 1: \\\notag & \boldsymbol{z_{1}}, \boldsymbol{z_{2}} \in epi \widetilde{U}, \alpha \boldsymbol{z_{1}} + \left( 1 - \alpha \right) \boldsymbol{z_{1}} not \in 
%\end{align}

According to the definition of $ epi \widetilde{U} $, $ \boldsymbol{r_{1}} $ and $ \boldsymbol{r_{2}} $ belong to the set $ \mathcal{R} $, and since it is a convex set, $ \alpha \boldsymbol{r_{1}} + \left( 1 - \alpha \right) \boldsymbol{r_{2}} \in epi \widetilde{U} $ as well. This, along with \eqref{z0}, implies that 
\begin{equation}\label{equ: y0 < U-tilde}
y_{0} > \widetilde{U} \left( \boldsymbol{r_{0}} \right).
\end{equation}

%Define $ \hat{\boldsymbol{z}} \triangleq \left[ \begin{smallmatrix} \boldsymbol{r_{0}} \\ \widetilde{U} \left( \boldsymbol{r_{0}} \right) \end{smallmatrix} \right] $. It is obvious that $ \hat{\boldsymbol{z}} \in epi \widetilde{U}$ and hence, from Lemma 0 ???, 

%Define another point $ \hat{\boldsymbol{z}} \triangleq \left[ \begin{smallmatrix} \boldsymbol{r_{0}} \\ \widetilde{U} \left( \boldsymbol{r_{0}} \right) \end{smallmatrix} \right] $. Using \eqref{equ: supporting hyperplane}, the supporting hyperplane to $ epi \widetilde{U} $ at this point would be as follows. 

Now consider point $ \hat{\boldsymbol{z}} \triangleq \left[ \begin{smallmatrix} \boldsymbol{r_{0}} \\ \widetilde{U} \left( \boldsymbol{r_{0}} \right) \end{smallmatrix} \right] $. The supporting hyperplane of $ epi \widetilde{U} $ at this point, so called $ SHP^{~epi \widetilde{U}} _{\boldsymbol{\hat{z}}} $, would be as follows:
%\vspace{-5 pt}
%{\small{
\begin{align}\label{equ: supporting hyperplane @ z-hat }\small
SHP^{~epi \widetilde{U}} _{\boldsymbol{\hat{z}}} =  \bigg\lbrace & \boldsymbol{z'}=\left[ \begin{smallmatrix} \boldsymbol{r'} \\ y' \end{smallmatrix} \right] \in  \mathbb{R}^{L+1} \Bigg|  \\
 & \left( \boldsymbol{r'} - \boldsymbol{r_{0}} \right) ^{T} \cdot \boldsymbol\lambda \left( \boldsymbol{r_{0}} \right)  + ( \widetilde{U} \left( \boldsymbol{r_{0}} \right) - y' ) = 0 \bigg\rbrace . \notag 
\end{align}
%}}
Substituting $ \boldsymbol{r} = \boldsymbol{r_{0}} $ into \eqref{equ: supporting hyperplane}, along with \eqref{equ: y0 < U-tilde} and \eqref{equ: supporting hyperplane @ z-hat }, concludes that $ \boldsymbol{z_{0}} $ and any $ \boldsymbol{z} \in epi \widetilde{U} $ are not located at the same side of the aforementioned hyperplane. In other words, we have
\begin{align}\label{equ: two half spaces}
\left( \boldsymbol{r} - \boldsymbol{r_{0}} \right) ^{T} \cdot \boldsymbol\lambda \left( \boldsymbol{r_{0}} \right)  + \left( y_0 - y \right) > 0, ~ \forall \boldsymbol{z}=\left[ \begin{smallmatrix} \boldsymbol{r} \\ y \end{smallmatrix} \right] \in epi \widetilde{U}.
\end{align}

Substituting $ \boldsymbol{z} = \boldsymbol{z_{1}}, \boldsymbol{z_{2}} \in epi \widetilde{U} $ into \eqref{equ: two half spaces} and noting that $ \boldsymbol{r_{0}} = \alpha \boldsymbol{r_{1}} + \left( 1 - \alpha \right) \boldsymbol{r_{2}} $, the following two inequalities are concluded 
%{\small{
\begin{align}
\left( 1 - \alpha \right) \left( \boldsymbol{r_{1}} - \boldsymbol{r_{2}} \right) ^{T} \cdot \boldsymbol\lambda \left( \boldsymbol{r_{0}} \right)  + \left( y_0 - y_{1} \right) &> 0, \label{equ : z=z1} \\
\label{equ : z=z2}
- \alpha \left( \boldsymbol{r_{1}} - \boldsymbol{r_{2}} \right) ^{T} \cdot \boldsymbol\lambda \left( \boldsymbol{r_{0}} \right)  + \left( y_0 - y_{2} \right) &> 0.
\end{align}
%}}
%Now, remember that  $ \boldsymbol{r_{0}} = \alpha \boldsymbol{r_{1}} + \left( 1 - \alpha \right) \boldsymbol{r_{2}} $ and substitute it in \eqref{equ : z=z1} and \eqref{equ : z=z2}:
%\hspace{8 pt}
Multiplying \eqref{equ : z=z1} and \eqref{equ : z=z2} by $ \alpha $ and $ \left( 1 - \alpha \right) $, respectively, and then summing them up together, it is obtained that $ y_{0} > \alpha y_{1} + \left( 1 - \alpha \right) y_{2} $. However, according to \eqref{z0}, the right-hand side of the above expression equals to $ y_{0} $. This is obviously a contradiction, and therefore Lemma \ref{lem: epi convex} is proven. \hfill$ \blacksquare $

%Now we multiply \eqref{equ : z=z1} and \eqref{equ : z=z2} by $ \alpha $ and $ \left( 1 - \alpha \right) $, respectively, and then sum up the obtained inequalities. It follows that $ y_{0} > \alpha y_{1} + \left( 1 - \alpha \right) y_{2} $, 
%
%\begin{equation}\label{eq: sum of two ineqs}
%\alpha \widetilde{U} \left( \boldsymbol{r_{1}} \right) + \left( 1 - \alpha \right) \widetilde{U} \left( \boldsymbol{r_{2}} \right) > y_{0}.
%\end{equation}
%
%On the other hand, due to the definition of $ epi \widetilde{U} $, $ \boldsymbol{z_{1}}, \boldsymbol{z_{2}} \in epi \widetilde{U} $ implies that $ y_{1} \geq \widetilde{U} \left( \boldsymbol{r_{1}} \right) $ and $ y_{2} \geq \widetilde{U} \left( \boldsymbol{r_{2}} \right) $. Substituting these two inequalities in \eqref{eq: sum of two ineqs}, it is concluded that:
%
\iffalse
\begin{equation}\label{equ: paradox}
y_{0} > \alpha y_{1} + \left( 1 - \alpha \right) y_{2},
\end{equation}
\fi
%while the right-hand side expression is indeed equal to $ y_{0} $, according to \eqref{z0}. This is obviously a contradiction, and therefore Lemma \ref{lem: epi convex} is proven. \hfill$ \blacksquare $
%*** Consequently, the epigraph of $ -\widetilde{U} $ is convex, and therefore, the objective function $ \widetilde{U} \left( \boldsymbol{r} \right) $ of the optimization problem $ \mathcal{P}_{E} $ is a concave function as well.\hfill$ \blacksquare $

%As both $ \boldsymbol{z_{1}} $ and $ \boldsymbol{z_{2}} $ belong to the set $ epi \widetilde{U} $, 

\end{lemma}

Next, as in the second part of Proposition \ref{th: Convexity of P_E}, we prove that the feasible set of $ \mathcal{P}_{E} $ is convex, too.

%\vspace{-5pt}

\begin{lemma}\label{lem: R is convex}
$ \mathcal{R} $ is a convex set.
\end{lemma}

%\vspace{-15pt}

\begin{proof}
In order to prove the convexity of $ \mathcal{R} $, we show that $ \mathcal{R} $ is the convex hull of the following set, i.e., $ \mathcal{R} = Conv\left( \mathcal{R'} \right) $, where $ \mathcal{R}' =  \left\lbrace \boldsymbol{r} \left( \boldsymbol{a} , \boldsymbol\rho \right) %= \left[ r_{1} \left( \boldsymbol{a} , \boldsymbol\rho  \right), \ldots , r_{L} \left( \boldsymbol{a} , \boldsymbol\rho  \right) \right]^{T} \hspace{- 5 pt} 
: 
\forall \boldsymbol{a} \in \mathcal{A} , \forall \boldsymbol\rho \in \Lambda_{\boldsymbol{\rho}} 
%\left( \boldsymbol{a} \right)  \right\rbrace \left( \boldsymbol{q} , \boldsymbol\rho \right) \in \Lambda 
\right\rbrace  
$.
%SHORTEN TWC
\iffalse
\begin{align}
\mathcal{R}' =  \left\lbrace \boldsymbol{r} \left( \boldsymbol{a} , \boldsymbol\rho \right) = \left[ r_{1} \left( \boldsymbol{a} , \boldsymbol\rho  \right), \ldots , r_{L} \left( \boldsymbol{a} , \boldsymbol\rho  \right) \right]^{T} \hspace{- 5 pt} : \boldsymbol{a} \in \mathcal{A} , \boldsymbol\rho \in \Lambda_{\boldsymbol{\rho}} \left( \boldsymbol{a} \right)  \right\rbrace. 
\end{align}
\fi
%\begin{equation}\label{def_R 2}
%\mathcal{R} =  \bigcup_{\Omega \in \Lambda} \left\{\boldsymbol{\nu} \in R_{+}^{L}: \boldsymbol{\nu} \leq \overline{\boldsymbol{r}}(\Omega)  \right\} , ~ \forall l=1,\ldots,L
%\end{equation}
Note that obviously,  $ \mathcal{R} \subset Conv\left( \mathcal{R'} \right) $. Hence, it is sufficient to show that $ Conv\left( \mathcal{R'} \right) \subset \mathcal{R}$. According to the definition of $ \mathcal{R} $ in \eqref{def_R}, if any point $ \boldsymbol{r} $ lies in $ \mathcal{R} $, then every point $ \boldsymbol\nu \in R^{L}_{+} $, where $ \boldsymbol\nu \leq \boldsymbol{r} $, lies in this region as well. Therefore, in order to prove $ Conv\left( \mathcal{R'} \right) \subset \mathcal{R}$, it is sufficient to show that any Pareto boundary point $ \boldsymbol{r'} $ of $ Conv\left( \mathcal{R'} \right) $ lies in $ \mathcal{R} $.

As $ \boldsymbol{r'} $ is an $ L $-dimensional Pareto boundary point, it can be expressed as a convex combination of $ L + 1 $ points in $ \mathcal{R'} $, i.e., $ \boldsymbol{r'} = \sum_{j=1}^{L+1} q_{j} \boldsymbol{r} \left( \boldsymbol{a_{j}}  , \boldsymbol\rho \right) $, where $ \sum_{j=1}^{L+1} q_{j} = 1 $, $ q_{j} \in \left[ 0,1 \right] $, $ \boldsymbol\rho \in \Lambda_{\boldsymbol{\rho}}  $ and $ \boldsymbol{a_{j}} \in \mathcal{A}, ~ \forall j $. Furthermore, as $ \boldsymbol{r'} $ is a Pareto boundary point of $ Conv\left( \mathcal{R'} \right) $, it follows that the point $ \boldsymbol{r} \left( \boldsymbol{a_{j}} , \boldsymbol\rho  \right) , ~ \forall \boldsymbol\rho,  \forall j: ~ q_{j} > 0 $ lies in the supporting hyperplane of $ Conv\left( \mathcal{R'} \right) $ at the Pareto boundary point $ \boldsymbol{r'} $. This fact implies that we can express $ \boldsymbol{r'} $ as a convex combination of $ L $ points in the set $ \left\lbrace  \boldsymbol{r} \left( \boldsymbol{a_{1}} , \boldsymbol\rho  \right), \ldots , \boldsymbol{r} \left( \boldsymbol{a_{L+1}} , \boldsymbol\rho  \right)  \right\rbrace $. Hence, $  \boldsymbol{r'} = \sum_{j=1}^{L} q'_{j} \boldsymbol{r} ( \boldsymbol{a'_{j}} , \boldsymbol\rho  ) $, where $ \sum_{j=1}^{L} q'_{j} = 1 $, $ q'_{j} \in \left[ 0,1 \right] $, $\boldsymbol\rho \in \Lambda_{\boldsymbol{\rho}} $ and $ \boldsymbol{a'_{j}} \in \left\lbrace \boldsymbol{a'_{1}} , \ldots , \boldsymbol{a'_{L}} \right\rbrace , ~ \forall j=1, \ldots , L $. Consequently, $ \boldsymbol{r'} $ lies in  $ \mathcal{R} $, which completes the proof of Lemma \ref{lem: R is convex}.
\end{proof}
% \hfill$ \blacksquare $

Having Lemma \ref{lem: epi convex} and Lemma \ref{lem: R is convex}, the proof of  Proposition \ref{th: Convexity of P_E} is completed.

\vspace{-10pt}

\subsection{Proof of Theorem \ref{theorem: Global Optimality Condition of P_2}} \label{ap: prrof of theorem: Global Optimality Condition of P_2}

%{\color{red}{Maybe omit it.}}

For proving the sufficient global optimality condition in Theorem \ref{theorem: Global Optimality Condition of P_2}, we first state the first order optimality condition of problem $ \mathcal{P}_{E} $ by Lemma \ref{lem: First Order Optimality Condition of P_E}. %, as summarized in the following lemma. %Next, using this lemma, Theorem \ref{theorem: Global Optimality Condition of P_2} derives a 
Next, using this lemma, we derive the sufficient global optimality condition for problem $ \mathcal{P}_{2} $, as stated in Theorem \ref{theorem: Global Optimality Condition of P_2}.

\begin{lemma}{(First Order Optimality Condition of Problem $ \mathcal{P}_{E} $)}\label{lem: First Order Optimality Condition of P_E}
The vector $ \boldsymbol{r^{\ast}} = [ r^{\ast}_{1},\ldots,r^{\ast}_{L} ]^{T} \in \mathcal{R} $ is the optimal solution for problem $ \mathcal{P}_{E} $ if %and only if 
\begin{equation}\label{opt_cond_PE}%\small
%***\at{\boldsymbol{g} ^{T} _{ \widetilde{U} } 
\at{{\nabla \widetilde{U}}^{T} \left( \boldsymbol{r} \right) }{\boldsymbol{r}=\boldsymbol{r^{\ast}}} \cdot \left(\boldsymbol{r^{\ast}} - \boldsymbol{r} \right) \geq 0, \hspace{5pt} \forall \boldsymbol{r} \in \mathcal{R},
\end{equation}
where $ %\boldsymbol{g}_{ \widetilde{U} } 
{\nabla \widetilde{U}} \left( \boldsymbol{r} \right) $ is the gradient of $ \widetilde{U} $ at point $ \boldsymbol{r} $.

\end{lemma}

%***\proof Refer to Appendix \ref{ap: proof of lem: First Order Optimality Condition of P_E}. %for the detailed proof.
\proof

Suppose $ \boldsymbol{r^{\ast}} = [ \overline{r}^{\ast}_{1},\ldots,\overline{r}^{\ast}_{L} ]^{T}$ satisfies \eqref{opt_cond_PE}, and define a function $ f \left( \boldsymbol{\overline{r}} \right) \triangleq \widetilde{U} \left( \boldsymbol{\overline{r}^{\ast}} \right) + {\nabla \widetilde{U}}^{T} \left( \boldsymbol{\overline{r}^{\ast}} \right) \cdot \left( \boldsymbol{\overline{r}} - \boldsymbol{\overline{r}^{\ast}} \right)  $ over the domain $ \mathcal{R} $. From the concavity of function $ \widetilde{U} \left( \boldsymbol{\overline{r}} \right) $, it follows that $ f \left( \boldsymbol{\overline{r}} \right) \geq \widetilde{U} \left( \boldsymbol{\overline{r}} \right) , ~ \forall \boldsymbol{\overline{r}} \in \mathcal{R} $.  Moreover, from \eqref{opt_cond_PE}, it is easy to see that $ \widetilde{U} \left( \boldsymbol{\overline{r}^{\ast}} \right) ~ \geq ~ f \left( \boldsymbol{\overline{r}} \right), ~ \forall \boldsymbol{\overline{r}} \in \mathcal{R} $. Therefore, combining the aforementioned two inequalities results that $ \widetilde{U} \left( \boldsymbol{\overline{r}^{\ast}} \right) \geq \widetilde{U} \left( \boldsymbol{\overline{r}} \right), ~ \forall \boldsymbol{\overline{r}} \in \mathcal{R} $. Hence, $ \boldsymbol{\overline{r}^{\ast}} $ is the global optimal solution for problem $ \mathcal{P}_{E} $.  \hfill$ \blacksquare $

Now suppose $ \left( \boldsymbol{q}^\ast, \boldsymbol{\rho}^\ast \right) $ satisfies condition \eqref{opt_cond_P1} in Theorem \ref{theorem: Global Optimality Condition of P_2}. It easily follows from \eqref{opt_cond_P1} along with the definition of $ \mathcal{R} $ in \eqref{def_R} that %{\small{
\begin{equation}\label{equ: proof The 3 _ eq1}
\at{ {\nabla  \widetilde{U} } ^T  \left( \boldsymbol{\overline{r}} \right) }{ \boldsymbol{\overline{r}} = \boldsymbol{\overline{r}} \left( \boldsymbol{q}^\ast, \boldsymbol{\rho}^\ast \right)  } \cdot \left( \boldsymbol{\overline{r}} \left( \boldsymbol{q}^{\ast} , \boldsymbol\rho^{\ast} \right) - \boldsymbol{\overline{r}}  \right) \geq 0, \quad \forall \boldsymbol{\overline{r}} \in \mathcal{R}.
\end{equation}%}}
Using Lemma \ref{lem: First Order Optimality Condition of P_E}, the above inequality results that $ \boldsymbol{\overline{r}} \left( \boldsymbol{q}^{\ast} , \boldsymbol\rho^{\ast} \right) $ satisfies the first-order global optimality condition for $ \mathcal{P}_E $, and hence, it is the global optimal solution to problem $ \mathcal{P}_E $. Consequently, according to Lemma \ref{theorem: P_2=P_E}, it is the optimal solution to problem $ \mathcal{P}_2 $ as well. % which completes the proof. %$ \hfill \blacksquare $

\vspace{-10pt}

\subsection{Proof of Theorem \ref{theorem: Global Optimality of Algorithm 1}}\label{App: proof of theorem: Global Optimality of Algorithm 1}

Assume that $ \left(\boldsymbol{q}^{(i+1)}, \boldsymbol{\rho}^{(i+1)} \right) $ is the DTX time-sharing and link scheduling control policy obtained in the $ {(i+1)}^{th} $ iteration of Algorithm \ref{alg 1}. According to \eqref{equ: link sched} and \eqref{q_Proc}, clearly $ \widetilde{U} \left(\boldsymbol{q}^{(i+1)}, \boldsymbol{\rho}^{(i+1)} \right) \geq \widetilde{U} \left(\boldsymbol{q}^{(i)}, \boldsymbol{\rho}^{(i)} \right) $, which implies that Algorithm \ref{alg 1} is increasing. This, along with the fact that the objective value is upper bounded, concludes that the algorithm is convergent, i.e., $ \lim_{i \rightarrow \infty} \widetilde{U} \left( \Omega^{(i)} \right) = \widetilde{U}^{\infty} $, 
\iffalse *****
\begin{equation}
\lim_{i \rightarrow \infty} \widetilde{U} \left( \Omega^{(i)} \right) = \widetilde{U}^{\infty}, %_{alg1},
\end{equation}
***** \fi
where $ \widetilde{U}^{\infty} $ is the convergence value of Algorithm \ref{alg 1}.

Now, let $ \left(\boldsymbol{q}^\infty, \boldsymbol{\rho}^\infty \right) $ be any limiting point of the sequence $ \left\lbrace \left(\boldsymbol{q}^{(i)}, \boldsymbol{\rho}^{(i)} \right) \right\rbrace $ generated by the algorithm and define $ {\nabla \widetilde{U}}_{\infty}^T \triangleq \at{{\nabla \widetilde{U}}^{T} \left( \bar{\boldsymbol{r}} \right) }{\bar{\boldsymbol{r}}=\bar{\boldsymbol{r}}\left(\boldsymbol{q}^\infty, \boldsymbol{\rho}^\infty \right)} $ and %define
 $ \boldsymbol{a}^{\infty} $ as the optimal solution of the following optimization problem: %\vspace{-8 pt}
\begin{equation}\label{equ: opt_problem of Proc I}%\small
\max_{ \boldsymbol{a} \in \mathcal{A} } ~  \at{{\nabla \widetilde{U}}^{T} \left( \bar{\boldsymbol{r}} \right) }{\bar{\boldsymbol{r}}=\bar{\boldsymbol{r}}\left(\boldsymbol{q}^\infty, \boldsymbol{\rho}^\infty \right)} \cdot \boldsymbol{r} \left( \boldsymbol{a} ,  \boldsymbol\rho \left( \boldsymbol{a}, \boldsymbol{H} \right)   \right),
\end{equation}
where $ \boldsymbol\rho^\infty \left( \boldsymbol{a}, \boldsymbol{H} \right)  $ is determined by \eqref{equ: link sched}. 
Obviously, we have %$ \at{{\nabla \widetilde{U}}^{T} \left( \bar{\boldsymbol{r}} \right) }{\bar{\boldsymbol{r}}=\bar{\boldsymbol{r}}\left(\boldsymbol{q}^\infty, \boldsymbol{\rho}^\infty \right)} \cdot \boldsymbol{r} \left( \boldsymbol{a}^\infty , \boldsymbol\rho^\infty \left( \boldsymbol{a}^\infty, \boldsymbol{H} \right)  \right) \geq \at{{\nabla \widetilde{U}}^{T} \left( \bar{\boldsymbol{r}} \right) }{\bar{\boldsymbol{r}}=\bar{\boldsymbol{r}}\left(\boldsymbol{q}^\infty, \boldsymbol{\rho}^\infty \right)} \cdot \boldsymbol{r} \left( \boldsymbol{a} , \boldsymbol\rho^\infty \left( \boldsymbol{a} , \boldsymbol{H} \right)  \right), \quad \forall \boldsymbol{a} \in \mathcal{A}. $ 
%Suppose that $ \boldsymbol{a^{\ast}} \left( \boldsymbol{\omega} \right) $ is the solution of the above problem. Hence, it follows that 
%\iffalse
\begin{align}\small %\label{3_equ: property of Proc I} 
&\notag \at{{\nabla \widetilde{U}}^{T} \left( \bar{\boldsymbol{r}} \right) }{\bar{\boldsymbol{r}}=\bar{\boldsymbol{r}}\left(\boldsymbol{q}^\infty, \boldsymbol{\rho}^\infty \right)} \cdot \boldsymbol{r} \left( \boldsymbol{a}^\infty , \boldsymbol\rho^\infty \left( \boldsymbol{a}^\infty, \boldsymbol{H} \right)  \right)  \notag \\
& \quad \geq \at{{\nabla \widetilde{U}}^{T} \left( \bar{\boldsymbol{r}} \right) }{\bar{\boldsymbol{r}}=\bar{\boldsymbol{r}}\left(\boldsymbol{q}^\infty, \boldsymbol{\rho}^\infty \right)} \cdot \boldsymbol{r} \left( \boldsymbol{a} , \boldsymbol\rho^\infty \left( \boldsymbol{a} , \boldsymbol{H} \right)  \right), \quad \forall \boldsymbol{a} \in \mathcal{A}. \notag  %\boldsymbol\rho \in \Lambda_{\boldsymbol{\rho}}. 
\end{align}
%\fi
 Moreover, according to \eqref{equ: link sched}, $ \boldsymbol\rho^\infty \left( \boldsymbol{a}, \boldsymbol{H} \right)  $ is maximising $ \at{{\nabla \widetilde{U}}^{T} \left( \bar{\boldsymbol{r}} \right) }{\bar{\boldsymbol{r}}=\bar{\boldsymbol{r}}\left(\boldsymbol{q}^\infty, \boldsymbol{\rho}^\infty \right)} \cdot \boldsymbol{r} \left( \boldsymbol{a} , \boldsymbol\rho   \right) $ with respect to $ \boldsymbol\rho  $ over the domain $  \Lambda_{\boldsymbol{\rho}} $. Therefore, it is concluded that
\begin{align}\label{eq: property_infty_1}\small
&\at{{\nabla \widetilde{U}}^{T} \left( \bar{\boldsymbol{r}} \right) }{\bar{\boldsymbol{r}}=\bar{\boldsymbol{r}}\left(\boldsymbol{q}^\infty, \boldsymbol{\rho}^\infty \right)} \cdot \boldsymbol{r} \left( \boldsymbol{a}^\infty , \boldsymbol\rho^\infty \left( \boldsymbol{a}^\infty, \boldsymbol{H} \right)  \right) \notag \\
&\quad \geq \at{{\nabla \widetilde{U}}^{T} \left( \bar{\boldsymbol{r}} \right) }{\bar{\boldsymbol{r}}=\bar{\boldsymbol{r}}\left(\boldsymbol{q}^\infty, \boldsymbol{\rho}^\infty \right)} \cdot \boldsymbol{r} \left( \boldsymbol{a} , \boldsymbol\rho  \right), \quad \forall \boldsymbol{a} \in \mathcal{A}, \forall \boldsymbol\rho \in \Lambda_{\boldsymbol{\rho}}, \notag \\ \notag
& \quad \geq  \at{{\nabla \widetilde{U}}^{T} \left( \bar{\boldsymbol{r}} \right) }{\bar{\boldsymbol{r}}=\bar{\boldsymbol{r}}\left(\boldsymbol{q}^\infty, \boldsymbol{\rho}^\infty \right)} \cdot \bar{\boldsymbol{r}} , \quad \forall \bar{\boldsymbol{r}} \in \mathcal{R},  \\ 
& \quad \geq  \at{{\nabla \widetilde{U}}^{T} \left( \bar{\boldsymbol{r}} \right) }{\bar{\boldsymbol{r}}=\bar{\boldsymbol{r}}\left(\boldsymbol{q}^\infty, \boldsymbol{\rho}^\infty \right)} \cdot \bar{\boldsymbol{r}}^\ast,
\end{align}
where the second inequality is due to the definition of $ \mathcal{R} $ and the last one is due to the fact that $ \bar{\boldsymbol{r}}^\ast \in \mathcal{R} $. Combining \eqref{eq: property_infty_1} and the concavity of $ \widetilde{U} $ results in
%{\small{
\begin{align}\label{eq: gap_ineq}\small
%\widetilde{U}^{\ast}_{alg1}  + \boldsymbol{g}^ {\ast T} \hspace{-5 pt} \cdot \left( \boldsymbol{r} \left( a^\ast \right) - \bar{\boldsymbol{r}} \left( \Omega^\ast \right) \right)  \geq  
%
& \widetilde{U}^{\infty} + \notag \\
 & ~\at{{\nabla \widetilde{U}}^{T} \hspace{-4 pt} \left( \bar{\boldsymbol{r}} \right) }{\bar{\boldsymbol{r}}=\bar{\boldsymbol{r}}\left(\boldsymbol{q}^\infty, \boldsymbol{\rho}^\infty \right)} \hspace{-10 pt} \cdot \left( \boldsymbol{r} \left( \boldsymbol{a}^\infty , \boldsymbol\rho^\infty \left( \boldsymbol{a}^\infty, \boldsymbol{H} \right)  \right) -\bar{\boldsymbol{r}} \left(\boldsymbol{q}^\infty, \boldsymbol{\rho}^\infty \right) \right) \nonumber\\
& ~ \geq 
\widetilde{U}^{\infty}+ \at{{\nabla \widetilde{U}}^{T} \left( \boldsymbol{r} \right) }{\boldsymbol{r}=\bar{\boldsymbol{r}}\left(\boldsymbol{q}^\infty, \boldsymbol{\rho}^\infty \right)}  \cdot \left( \bar{\boldsymbol{r}}^\ast - \bar{\boldsymbol{r}}\left(\boldsymbol{q}^\infty, \boldsymbol{\rho}^\infty \right) \right) \notag \\
& ~ \geq  \widetilde{U}^{\ast},
\end{align} %}}
where $ \bar{\boldsymbol{r}}^\ast $ is the global optimal solution of problem $ \mathcal{P}_2 $ and $ \widetilde{U}^{\ast} = \widetilde{U} \left(   \bar{\boldsymbol{r}}^\ast \right)$ is the optimal value. Consequently, it follows that
\begin{align}\label{eq: gap_ineq}\small
&\at{{\nabla \widetilde{U}}^{T} \left( \boldsymbol{r} \right) }{\boldsymbol{r}=\bar{\boldsymbol{r}}\left(\boldsymbol{q}^\infty, \boldsymbol{\rho}^\infty \right)} \hspace{-6 pt}  \cdot \left( \boldsymbol{r} \left( \boldsymbol{a}^\infty , \boldsymbol\rho^\infty \left( \boldsymbol{a}^\infty, \boldsymbol{H} \right)  \right) - \bar{\boldsymbol{r}}\left(\boldsymbol{q}^\infty, \boldsymbol{\rho}^\infty \right) \right) \notag \\
& \quad \quad \quad \geq  \widetilde{U}^{\ast} - \widetilde{U}^{\infty} \notag \\
& \quad \quad \quad \geq 0,
\end{align} 
where the last inequality follows from the fact that $  \widetilde{U}^{\ast} $ is the optimal (i.e., maximum) value of $ \mathcal{P}_2 $. 

To prove that the convergence value of Algorithm \ref{alg 1} equals to the global optimal solution of the problem, i.e., $ \widetilde{U}^{\infty} = \widetilde{U}^{\ast} $, it suffices to show that the left hand side of \eqref{eq: gap_ineq} is less than or equal to 0. In the following, we will prove it by contradiction: 
%
\iffalse
Now, it is sufficient to show that the left hand side of \eqref{eq: gap_ineq} is equal to 0. 

$ \at{{\nabla \widetilde{U}}^{T} \left( \boldsymbol{r} \right) }{\boldsymbol{r}=\bar{\boldsymbol{r}}\left(\boldsymbol{q}^\infty, \boldsymbol{\rho}^\infty \right)} \hspace{-6 pt}  \cdot \left( \bar{\boldsymbol{r}}^\ast - \bar{\boldsymbol{r}}\left(\boldsymbol{q}^\infty, \boldsymbol{\rho}^\infty \right) \right) = 0 $, which concludes that the convergence value of Algorithm \ref{alg 1} equals to the global optimal solution of the problem, i.e., $ \widetilde{U}^{\infty} = \widetilde{U}^{\ast} $. In the following, we will prove it by contradiction.
\fi
Assume that {\small $ \at{{\nabla \widetilde{U}}^{T} \left( \boldsymbol{r} \right) }{\boldsymbol{r}=\bar{\boldsymbol{r}}\left(\boldsymbol{q}^\infty, \boldsymbol{\rho}^\infty \right)} \hspace{-6 pt}  \cdot \left( \boldsymbol{r} \left( \boldsymbol{a}^\infty , \boldsymbol\rho^\infty \left( \boldsymbol{a}^\infty, \boldsymbol{H} \right)  \right) - \bar{\boldsymbol{r}}\left(\boldsymbol{q}^\infty, \boldsymbol{\rho}^\infty \right) \right)   > \varepsilon $}, %Hence, from \eqref{eq: gap_ineq_2} it is concluded that there exists a subsequence of the iterations of Algorithm 1 such that $ \boldsymbol{g}^ {\ast T} \cdot \left( \boldsymbol{r} \left( a^\ast \right) - \bar{\boldsymbol{r}} \left( \Omega^\ast \right) \right) \geq \epsilon $, 
for some $ \varepsilon > 0 $. Hence, using Taylor expansion at point $ \bar{\boldsymbol{r}} \left(\boldsymbol{q}^\infty, \boldsymbol{\rho}^\infty \right) $, we have
%{\small{
\begin{align}
 &\widetilde{U} \bigg( \bar{\boldsymbol{r}} \left(\boldsymbol{q}^\infty, \boldsymbol{\rho}^\infty \right) + \tau \Big[ \boldsymbol{r} \left( \boldsymbol{a}^\infty , \boldsymbol\rho^\infty \left( \boldsymbol{a}^\infty, \boldsymbol{H} \right)  \right) - \bar{\boldsymbol{r}}\left(\boldsymbol{q}^\infty, \boldsymbol{\rho}^\infty \right) \Big] \bigg)   \notag \\
 &  \quad = \widetilde{U} \left( \bar{\boldsymbol{r}} \left(\boldsymbol{q}^\infty, \boldsymbol{\rho}^\infty \right) \right) + \tau \at{{\nabla \widetilde{U}}^{T} \left( \boldsymbol{r} \right) }{\boldsymbol{r}=\bar{\boldsymbol{r}}\left(\boldsymbol{q}^\infty, \boldsymbol{\rho}^\infty \right)} \hspace{-6 pt}   \notag \\
 & ~~ \quad \quad   \quad  \quad \cdot \left( \boldsymbol{r} \left( \boldsymbol{a}^\infty , \boldsymbol\rho^\infty \left( \boldsymbol{a}^\infty, \boldsymbol{H} \right)  \right) - \bar{\boldsymbol{r}}\left(\boldsymbol{q}^\infty, \boldsymbol{\rho}^\infty \right) \right) + o \left( \tau ^2 \right) \notag \\
&  \quad > \widetilde{U} \left( \bar{\boldsymbol{r}} \left(\boldsymbol{q}^\infty, \boldsymbol{\rho}^\infty \right) \right) + \tau \varepsilon + o \left( \tau ^2 \right),
\end{align}%}} 
for some sufficiently small $ \tau > 0 $. Consequently, for a sufficiently small value of $ \tau > 0 $, we have $ \widetilde{U} \left( \bar{\boldsymbol{r}} \left(\boldsymbol{q}^\infty, \boldsymbol{\rho}^\infty \right) + \tau \left( \boldsymbol{r} \left( \boldsymbol{a}^\infty , \boldsymbol\rho^\infty \left( \boldsymbol{a}^\infty, \boldsymbol{H} \right)  \right) - \bar{\boldsymbol{r}}\left(\boldsymbol{q}^\infty, \boldsymbol{\rho}^\infty \right) \right) \right) > \widetilde{U} \left( \bar{\boldsymbol{r}} \left(\boldsymbol{q}^\infty, \boldsymbol{\rho}^\infty \right) \right). $
%*** This, along with the contradiction assumption, results in:
%{\small{
\iffalse
\begin{equation}
\widetilde{U} \left( \bar{\boldsymbol{r}} \left(\boldsymbol{q}^\infty, \boldsymbol{\rho}^\infty \right) + \tau \left( \boldsymbol{r} \left( \boldsymbol{a}^\infty , \boldsymbol\rho^\infty \left( \boldsymbol{a}^\infty, \boldsymbol{H} \right)  \right) - \bar{\boldsymbol{r}}\left(\boldsymbol{q}^\infty, \boldsymbol{\rho}^\infty \right) \right) \right) > \widetilde{U} \left( \bar{\boldsymbol{r}} \left(\boldsymbol{q}^\infty, \boldsymbol{\rho}^\infty \right) \right).
\end{equation}
\fi
%}}
 Therefore, we have found some point $ \left(1 - \tau \right) \bar{\boldsymbol{r}} \left(\boldsymbol{q}^\infty, \boldsymbol{\rho}^\infty \right) + \tau \left( \boldsymbol{r} \left( \boldsymbol{a}^\infty , \boldsymbol\rho^\infty \left( \boldsymbol{a}^\infty, \boldsymbol{H} \right) \right) \right)  $ with a strictly larger utility value than the utility value at point $ \bar{\boldsymbol{r}} \left(\boldsymbol{q}^\infty, \boldsymbol{\rho}^\infty \right)  $. Note that this point is in fact $ \bar{\boldsymbol{r}} \left( \boldsymbol{q}^\prime , \boldsymbol{\rho}^\infty \right) $, where 
\begin{equation} \notag %\small
\forall j = 1, \ldots,  |\mathcal{A}|: \quad 
q_j^\prime  = \left\lbrace
	\begin{array}{ll}
		\left( 1- \tau \right) q_j^\infty + \tau & \mbox{if $ \boldsymbol{a}^{(j)} = \boldsymbol{a}^\infty $,} \\
		\left( 1- \tau \right) q_j^\infty & \mbox{otherwise.} 
	\end{array}
\right.
\end{equation}
Hence, $ \widetilde{U} \left( \bar{\boldsymbol{r}} \left( \boldsymbol{q}^\prime , \boldsymbol{\rho}^\infty \right) \right) >   \widetilde{U} \left( \bar{\boldsymbol{r}} \left( \boldsymbol{q}^\infty , \boldsymbol{\rho}^\infty \right) \right) $, which contradicts with the fact that $ \boldsymbol{q}^\infty $ is the maximiser of $ \widetilde{U} \left( \bar{\boldsymbol{r}} \left( \boldsymbol{q} , \boldsymbol{\rho}^\infty \right) \right) $ with respect to $ \boldsymbol{q} $ (according to \eqref{q_Proc}). Therefore, the left hand side of \eqref{eq: gap_ineq} is equal to 0 and hence,  $ \widetilde{U}^{\infty} = \widetilde{U}^{\ast} $, which indicates that the proposed algorithm converges to the optimal solution of $ \mathcal{P}_2 $.  
Moreover, assume that $ \left( \boldsymbol{d}^\infty, \boldsymbol{x}^\infty \right) $ is the solution to subproblem $ \mathcal{P}_1 $ under $ \bar{\boldsymbol{r}}\left( \boldsymbol{q}^\infty, \boldsymbol{\rho}^\infty \right) $,  derived from Step 2b of Algorithm \ref{alg 1}. According to Lemma \ref{lem: primal decomposition}, we have that %if $ \left( \boldsymbol{d}^\ast, \boldsymbol{x}^\ast \right) $ and $ \left( \boldsymbol{q}^\ast, \boldsymbol{\rho}^\ast \right) $ are the optimal solutions of subproblems $ \mathcal{P}_1 $ and $ \mathcal{P}_2 $, respectively,}} then 
$ \left( \boldsymbol{d}^\infty, \boldsymbol{x}^\infty, \boldsymbol{q}^\infty, \boldsymbol{\rho}^\infty \right) $ is the optimal solution to the original problem $ \mathcal{P}_{org} $. Therefore, we have $ U \left( \boldsymbol{d}^\infty \right) = \widetilde{U}^{\ast} = {U}^{\ast} $% (where $ {U}^{\ast} $  is defined as the optimal solution to $ \mathcal{P}_{org} $)
, which completes the proof. $ \hfill \blacksquare $ }}
%\begin{equation}
%U \left( \boldsymbol{d}^\infty \right) = \widetilde{U}^{\ast}  = {U}^{\ast}
%\end{equation}
%which completes the proof. $ \hfill \blacksquare $

% and hence, $ U\left( \boldsymbol{d}^\infty \right) = \widetilde{U}^{\ast} $
}}

\ifCLASSOPTIONcaptionsoff
  \newpage
\fi



%\bibliographystyle{IEEEtranTCOM}

%\begin{thebibliography}{1}
%\end{thebibliography}

\vspace{-5pt}

%\nocite{*}
\bibliography{References}

\begin{thebibliography}{10}
\providecommand{\url}[1]{#1}
\csname url@samestyle\endcsname
\providecommand{\newblock}{\relax}
\providecommand{\bibinfo}[2]{#2}
\providecommand{\BIBentrySTDinterwordspacing}{\spaceskip=0pt\relax}
\providecommand{\BIBentryALTinterwordstretchfactor}{4}
\providecommand{\BIBentryALTinterwordspacing}{\spaceskip=\fontdimen2\font plus
\BIBentryALTinterwordstretchfactor\fontdimen3\font minus
  \fontdimen4\font\relax}
\providecommand{\BIBforeignlanguage}[2]{{%
\expandafter\ifx\csname l@#1\endcsname\relax
\typeout{** WARNING: IEEEtran.bst: No hyphenation pattern has been}%
\typeout{** loaded for the language `#1'. Using the pattern for}%
\typeout{** the default language instead.}%
\else
\language=\csname l@#1\endcsname
\fi
#2}}
\providecommand{\BIBdecl}{\relax}
\BIBdecl

\bibitem{omidvar2015Globecom}
N.~Omidvar, A.~Liu, V.~Lau, F.~Zhang, D.~Tsang, and M.~R. Pakravan,
  ``Two-timescale radio resource management for heterogeneous networks with
  flexible backhaul,'' in \emph{2015 IEEE Global Communications Conference
  (GLOBECOM)}.\hskip 1em plus 0.5em minus 0.4em\relax IEEE, 2015, pp. 1--6.

\bibitem{damnjanovic2011survey}
A.~Damnjanovic, J.~Montojo, Y.~Wei, T.~Ji, T.~Luo, M.~Vajapeyam, T.~Yoo,
  O.~Song, and D.~Malladi, ``A survey on {3GPP} heterogeneous networks,''
  \emph{Wireless Communications, IEEE}, vol.~18, no.~3, pp. 10--21, 2011.

\bibitem{paolini2011crucial}
M.~Paolini, ``Crucial economics for mobile data backhaul,'' \emph{White paper},
  2011.

\bibitem{tombaz2011impact}
S.~Tombaz, P.~Monti, K.~Wang, A.~Vastberg, M.~Forzati, and J.~Zander, ``Impact
  of backhauling power consumption on the deployment of heterogeneous mobile
  networks,'' in \emph{Global Telecommunications Conference (GLOBECOM
  2011)}.\hskip 1em plus 0.5em minus 0.4em\relax IEEE, 2011, pp. 1--5.

\bibitem{farias2013green}
F.~Farias, P.~Monti, A.~Vastberg, M.~Nilson, J.~Costa, and L.~Wosinska, ``Green
  backhauling for heterogeneous mobile access networks: {W}hat are the
  challenges?'' in \emph{9th International Conference on Information,
  Communications and Signal Processing (ICICS)}.\hskip 1em plus 0.5em minus
  0.4em\relax IEEE, 2013, pp. 1--5.

\bibitem{tornatore2017Fiber-WirelessConvergenceinNext-GenerationCommunicationNetworks}
M.~Tornatore, G.-K. Chang, and G.~Ellinas, \emph{Fiber-Wireless Convergence in
  Next-Generation Communication Networks: Systems, Architectures, and
  Management}.\hskip 1em plus 0.5em minus 0.4em\relax Springer, 2017.

\bibitem{relay2001integrated}
H.~Wu, C.~Qiao, S.~De, and O.~Tonguz, ``Integrated cellular and ad hoc relaying
  systems: {iCAR},'' \emph{Selected Areas in Communications, IEEE Journal on},
  vol.~19, no.~10, pp. 2105--2115, 2001.

\bibitem{sreng2003relayer}
V.~Sreng, H.~Yanikomeroglu, and D.~D. Falconer, ``Relayer selection strategies
  in cellular networks with peer-to-peer relaying,'' in \emph{58th Vehicular
  Technology Conference}, vol.~3.\hskip 1em plus 0.5em minus 0.4em\relax IEEE,
  2003, pp. 1949--1953.

\bibitem{relay2011flexible}
{\"O}.~Bulakci, A.~B. Saleh, S.~Redana, B.~Raaf, and J.~H{\"a}m{\"a}l{\"a}inen,
  ``Flexible backhaul resource sharing and uplink power control optimization in
  {LTE}-advanced relay networks,'' in \emph{Vehicular Technology Conference
  (VTC Fall)}.\hskip 1em plus 0.5em minus 0.4em\relax IEEE, 2011, pp. 1--6.

\bibitem{survey2004relay}
R.~Pabst, B.~H. Walke, D.~C. Schultz, P.~Herhold, H.~Yanikomeroglu,
  S.~Mukherjee, H.~Viswanathan, M.~Lott, W.~Zirwas, M.~Dohler \emph{et~al.},
  ``Relay-based deployment concepts for wireless and mobile broadband radio,''
  \emph{Communications {M}agazine, IEEE}, vol.~42, no.~9, pp. 80--89, 2004.

\bibitem{niyato2009relay}
D.~Niyato, E.~Hossain, D.~I. Kim, and Z.~Han, ``Relay-centric radio resource
  management and network planning in {IEEE} 802.16 j mobile multihop relay
  networks,'' \emph{IEEE Transactions on wireless communications}, vol.~8,
  no.~12, 2009.

\bibitem{wang2012performance}
Y.~Wang and K.~I. Pedersen, ``Performance analysis of enhanced inter-cell
  interference coordination in {LTE}-advanced heterogeneous networks,'' in
  \emph{75th Vehicular Technology Conference (VTC Spring)}.\hskip 1em plus
  0.5em minus 0.4em\relax IEEE, 2012, pp. 1--5.

\bibitem{pang2012optimized}
J.~Pang, J.~Wang, D.~Wang, G.~Shen, Q.~Jiang, and J.~Liu, ``Optimized
  time-domain resource partitioning for enhanced inter-cell interference
  coordination in heterogeneous networks,'' in \emph{Wireless Communications
  and Networking Conference (WCNC)}.\hskip 1em plus 0.5em minus 0.4em\relax
  IEEE, 2012, pp. 1613--1617.

\bibitem{liu2014hierarchical}
A.~Liu, V.~K. Lau, L.~Ruan, J.~Chen, and D.~Xiao, ``Hierarchical radio resource
  optimization for heterogeneous networks with enhanced inter-cell interference
  coordination (e{ICIC}),'' \emph{Signal Processing, IEEE Transactions on},
  vol.~62, no.~7, pp. 1684--1693, 2014.

\bibitem{omidvar2016cross}
N.~Omidvar, F.~Zhang, A.~Liu, V.~K. Lau, D.~H. Tsang, and M.~R. Pakravan,
  ``Cross-layer {QSI}-aware radio resource management for {HetN}ets with
  flexible backhaul,'' in \emph{Wireless Communications and Networking
  Conference (WCNC 2016), Doha, Qatar}.\hskip 1em plus 0.5em minus 0.4em\relax
  IEEE, 2016, pp. 945--950.

\bibitem{omidvar2015PIMRC}
N.~Omidvar, A.~Liu, V.~Lau, F.~Zhang, D.~H. Tsang, and M.~R. Pakravan,
  ``Two-timescale {QoS}-aware cross-layer optimisation for {HetNets} with
  flexible backhaul,'' in \emph{Personal, Indoor, and Mobile Radio
  Communications (PIMRC), 2015 IEEE 26th Annual International Symposium
  on}.\hskip 1em plus 0.5em minus 0.4em\relax IEEE, 2015, pp. 1072--1076.

\bibitem{khaled2007interpolation}
N.~Khaled, B.~Mondal, G.~Leus, R.~W. Heath, and F.~Petr{\'e},
  ``Interpolation-based multi-mode precoding for {MIMO-OFDM} systems with
  limited feedback,'' \emph{IEEE Transactions on Wireless Communications},
  vol.~6, no.~3, 2007.

\bibitem{sadek2008transmit}
A.~K. Sadek, W.~Su, and K.~R. Liu, ``Transmit beamforming for space-frequency
  coded {MIMO-OFDM} systems with spatial correlation feedback,'' \emph{IEEE
  Transactions on Communications}, vol.~56, no.~10, pp. 1647--1655, 2008.

\bibitem{hasan2017survey}
M.~Z. Hasan, H.~Al-Rizzo, and F.~Al-Turjman, ``A survey on multipath routing
  protocols for {QoS} assurances in real-time wireless multimedia sensor
  networks,'' \emph{IEEE Communications Surveys \& Tutorials}, 2017.

\bibitem{khandekar2010LTE}
A.~Khandekar, N.~Bhushan, J.~Tingfang, and V.~Vanghi, ``{LTE}-advanced:
  Heterogeneous networks,'' in \emph{European Wireless Conference (EW)}.\hskip
  1em plus 0.5em minus 0.4em\relax IEEE, 2010, pp. 978--982.

\bibitem{he2008toward}
J.~He and J.~Rexford, ``Toward internet-wide multipath routing,'' \emph{IEEE
  network}, vol.~22, no.~2, 2008.

\bibitem{kandula2007dynamic}
S.~Kandula, D.~Katabi, S.~Sinha, and A.~Berger, ``Dynamic load balancing
  without packet reordering,'' \emph{ACM SIGCOMM Computer Communication
  Review}, vol.~37, no.~2, pp. 51--62, 2007.

\bibitem{xu2006miro}
W.~Xu and J.~Rexford, \emph{{MIRO}: multi-path interdomain routing}.\hskip 1em
  plus 0.5em minus 0.4em\relax ACM, 2006, vol.~36, no.~4.

\bibitem{singh2015survey}
S.~K. Singh, T.~Das, and A.~Jukan, ``A survey on internet multipath routing and
  provisioning,'' \emph{IEEE Communications Surveys \& Tutorials}, vol.~17,
  no.~4, pp. 2157--2175, 2015.

\bibitem{SDN2015programming}
I.~Gasparis, U.~C. Kozat, and M.~O. Sunay, ``Programming flows in dense mobile
  environments: A multi-user diversity perspective,'' \emph{arXiv preprint
  arXiv:1506.07816}, 2015.

\bibitem{astely2009lte}
D.~Ast{\'e}ly, E.~Dahlman, A.~Furusk{\"a}r, Y.~Jading, M.~Lindstr{\"o}m, and
  S.~Parkvall, ``{LTE:} the evolution of mobile broadband,'' \emph{IEEE
  Communications magazine}, vol.~47, no.~4, 2009.

\bibitem{deb2014eICIC}
S.~Deb, P.~Monogioudis, J.~Miernik, and J.~P. Seymour, ``Algorithms for
  enhanced inter-cell interference coordination {(eICIC)} in {LTE} {HetNets},''
  \emph{IEEE/ACM transactions on networking}, vol.~22, no.~1, pp. 137--150,
  2014.

\bibitem{An_eICIC_Arxiv}
\BIBentryALTinterwordspacing
A.~Liu, V.~K.~N. Lau, L.~Ruan, J.~Chen, and D.~Xiao, ``Hierarchical radio
  resource optimization for heterogeneous networks with enhanced inter-cell
  interference coordination {(eICIC)},'' \emph{CoRR}, vol. abs/1305.5884, 2013.
  [Online]. Available: \url{http://arxiv.org/abs/1305.5884.}
\BIBentrySTDinterwordspacing

\bibitem{alpha_fair}
J.~Mo and J.~Walrand, ``Fair end-to-end window-based congestion control,''
  \emph{IEEE/ACM Transactions on Networking (ToN)}, vol.~8, no.~5, pp.
  556--567, 2000.

\bibitem{PFS}
F.~P. Kelly, A.~K. Maulloo, and D.~K. Tan, ``Rate control for communication
  networks: {S}hadow prices, proportional fairness and stability,''
  \emph{Journal of the {O}perational {R}esearch {S}ociety}, pp. 237--252, 1998.

\bibitem{palomar2006tutorial}
D.~P. Palomar and M.~Chiang, ``A tutorial on decomposition methods for network
  utility maximization,'' \emph{IEEE Journal on Selected Areas in
  Communications}, vol.~24, no.~8, pp. 1439--1451, 2006.

\bibitem{bertsekas1999nonlinear}
D.~P. Bertsekas, \emph{Nonlinear {P}rogramming}.\hskip 1em plus 0.5em minus
  0.4em\relax Athena Scientific, 1999.

\bibitem{boyd2009convex}
S.~Boyd and L.~Vandenberghe, \emph{Convex {O}ptimization}.\hskip 1em plus 0.5em
  minus 0.4em\relax Cambridge University Press, 2009.

\bibitem{anderson2005bit}
S.~E. Anderson, ``Bit twiddling hacks,'' \emph{URL:
  http://graphics.stanford.edu/\~{}seander/bithacks. html}, 2005.

\bibitem{skiena1998algorithm}
S.~S. Skiena, \emph{The algorithm design manual}.\hskip 1em plus 0.5em minus
  0.4em\relax Springer Science \& Business Media, 2008.

\bibitem{ist2007deliverable}
I.~IST-WINNER, ``Deliverable 1.1.2 v.1.2, “{WINNER II} channel models”,
  {IST-WINNER2},'' Tech. Rep., 2008 (http://projects.celti
  c-initiative.org/winner+/deliverables.html), Tech. Rep., 2007.

\bibitem{WinnerII_LTE}
M.~Hossain, R.~Adhikary, and N.~Yesmin, ``Performance evaluation of {WINNER-II}
  channel model for long term evolution {(LTE)},'' \emph{International Journal
  of Scientific \& Engineering Research}, vol.~4, no.~5, 2013.

\bibitem{bezdek2003convergence_AO}
J.~C. Bezdek and R.~J. Hathaway, ``Convergence of alternating optimization,''
  \emph{Neural, Parallel \& Scientific Computations}, vol.~11, no.~4, pp.
  351--368, 2003.

\end{thebibliography}
\bibliographystyle{IEEEtran}

\end{document}

%@inproceedings{chen2013convergence,
%  title={Convergence analysis of mixed timescale cross-layer stochastic optimization},
%  author={Chen, Junting and Lau, Vincent KN},
%  booktitle={Signals, Systems and Computers, 2013 Asilomar Conference on},
%  pages={225--229},
%  year={2013},
%  organization={IEEE}
%}

%
%@inproceedings{shi2014flexible,
%  title={A Flexible Wireless Backhaul Solution for Emerging Small Cells Networks},
%  author={Shi, Yi and Li, Mingchao and Xiong, Xin and Han, Guanglin},
%  booktitle={IEEE ICASSP},
%  year={2014},
%  organization={IEEE}
%}